\documentclass[a4paper]{aa}
\usepackage[english]{babel}  
\usepackage{graphicx}  
\usepackage[utf8]{inputenc}  
\usepackage{microtype}  
\usepackage{booktabs}  
\usepackage{rotating}  
\usepackage{lscape}
\usepackage[squaren]{SIunits}
\usepackage{natbib}
\usepackage{amsmath}
\usepackage{amssymb}
\usepackage{pstricks}
\usepackage{txfonts}
\usepackage{tikz}
\usepackage{color}
\usepackage[config, labelfont={bf}]{subfig}
\usepackage{hyperref}

\newcommand{\tex}{T_\mathrm{{ex}}}
\newcommand{\tmb}{T_\mathrm{{MB}}}

\newcommand{\taStar}{T_\mathrm{{A}}^{*}}

\newcommand{\td}{T_\mathrm{{d}}}

\newcommand{\pot}[1]{10^{#1}}

\newcommand{\vlsr}{V_\mathrm{{LSR}}}

\newcommand{\cm}{\usk\centi \metre}

\newcommand{\pc}{\usk\mathrm{pc}}

\newcommand{\kpc}{\usk\mathrm{kpc}}
\newcommand{\kpctab}{\mathrm{kpc}}
\newcommand{\hii}{H\textsc{ii}}

\newcommand{\msun}{\usk\mathrm{M_\odot}}

\newcommand{\lsun}{\usk\mathrm{L_\odot}}

\newcommand{\kms}{\usk\kilo\metre\usk\second^{-1}}
\newcommand{\kmstab}{\kilo\metre\usk\second^{-1}}
\newcommand{\kel}{\usk\kelvin}
\newcommand{\mum}{\usk\micro\metre}

\newcommand{\dgc}{D_{GC}}

\newcommand{\asymErr}[2]{\ensuremath{^{{#1}}_{{#2}}}}

\newcommand{\SigmaDust}{\ensuremath{\Sigma_\mathrm{dust}}}
\newcommand{\gtd}{\ensuremath{\gamma}}
\newcommand{\ghz}{\ensuremath{\usk\giga\hertz}}
\newcommand{\vs}{\ensuremath{\,vs.\,}}

\newcommand{\koenigOGText}{K\"onig et al. (in prep.)}

\renewcommand{\dgc}{\ensuremath{R_\mathrm{GC}}}

\title{Galactocentric variation of the gas-to-dust ratio and its relation with metallicity}
\author{A. Giannetti \inst{\ref{ira},\ref{mpi}}
\and S. Leurini \inst{\ref{oac},\ref{mpi}}
\and C. K\"onig \inst{\ref{mpi}}
\and J.~S. Urquhart \inst{\ref{ukent}}
\and T. Pillai \inst{\ref{mpi}}
\and J. Brand \inst{\ref{ira}}
\and J. Kauffmann \inst{\ref{mpi}}
\and F. Wyrowski \inst{\ref{mpi}}
\and K.~M. Menten \inst{\ref{mpi}}
}
\institute{
INAF - Istituto di Radioastronomia \& Italian ALMA Regional Centre, Via P. Gobetti 101, I-40129 Bologna, Italy\label{ira}
\and Max-Planck-Institut f\"ur Radioastronomie, auf dem H\"ugel 69, D-53121, Bonn, Germany \label{mpi}
\and INAF-Osservatorio Astronomico di Cagliari, Via della Scienza 5, I-09047, Selargius (CA), Italy \label{oac}
\and School of Physical Sciences, University of Kent, Ingram Building, Canterbury, Kent CT2\,7NH,\,UK \label{ukent}
}

\abstract{
	The assumption of a gas-to-dust mass ratio $\gtd$ is a common approach to estimate the basic properties of molecular clouds, such as total mass and column density of molecular hydrogen, from (sub)mm continuum observations of the dust. In the Milky Way a single value is used at all galactocentric radii, independently of the observed metallicity gradients. Both models and extragalactic observations suggest that this quantity increases for decreasing metallicity $Z$, typical of the outer regions in disks, where fewer heavy elements are available to form dust grains.
}
{
	We aim to investigate the variation of the gas-to-dust ratio as a function of galactocentric radius and metallicity, to allow a more accurate characterisation of the quantity of molecular gas across the galactic disk, as derived from observations of the dust.
}
{
	Observations of the optically thin C$^{18}$O (2--1) transition were obtained with the APEX telescope for a sample of 23 massive and dense star-forming regions in the far outer Galaxy (galactocentric distance greater than $14\kpc$). From the modelling of this line and of the spectral energy distribution of the selected clumps we computed the gas-to-dust ratio and compared it to that of well-studied sources from the ATLASGAL TOP100 sample in the inner galactic disk. 
}
{
	The gradient in $\gtd$ is found to be $0.087\asymErr{+0.047}{-0.025}\usk\mathrm{dex}\kpc^{-1}$ (or equivalently 
	$\gtd \propto Z^{-1.4\asymErr{+0.3}{-1.0}}$). The dust-to-metal ratio, decreases with galactocentric radius, which is the most common situation also for external late-type galaxies. This suggests that grain growth dominates over destruction. The predicted $\gtd$ is in excellent agreement with the estimates in Magellanic clouds, for the appropriate value of $Z$.
}
{
}

\offprints{A. Giannetti (agianne@ira.inaf.it)}

\begin{document}  

\maketitle

\section{Introduction}\label{sec:intro}
	In the past decade many surveys of the galactic plane have been carried out in the continuum, covering wavelengths from the millimetre regime to the infrared (IR).
	They provide a complete picture of the dust emission, tracing both very cold material (at millimetre, sub-mm and far-IR wavelengths; for example ATLASGAL, \citealt{Schuller+09_aap504_415}, and Hi-GAL,  \citealt{Molinari+10_pasp122_314}), and hot dust and PAHs (in the mid- and near-IR; for example MSX, \citealt{Egan+03_AFRL}, MIPSGAL, \citealt{Carey+09_pasp121_76}, WISE, \citealt{Wright+10_aj140_1868}). The temperature, mass and column density of the dust can be estimated by constructing and modelling the spectral energy distribution of the thermal dust emission \citep[SED; e.g.][]{Koenig+17_aap599_139}. The dust, however, constitutes only a minor fraction of the total mass of molecular clouds. One has to assume a gas-to-dust mass ratio ($\gtd$) to derive the mass and column density of molecular hydrogen. 
	A direct, local determination shows that the hydrogen-to-dust mass ratio is $\sim100$, corresponding to a gas-to-dust mass ratio $\gtd\approx136$, when accounting for helium \citep{Draine+07_apj663_866}. 
	Current research uses a constant value of the gas-to-dust ratio irrespective of the galactocentric distance of the cloud \citep[typically $100-150$, e.g.][]{Elia+13_apj772_45,Elia+17_mnras471_100,Koenig+17_aap599_139}, and while these values are reasonable within the solar circle they are not likely to be reliable for the outer parts of the disk, where the metallicity and average disk surface density might be substantially lower.
	
	Heavy elements are the main constituents of dust grains, and therefore when their abundance with respect to hydrogen changes, dust may be influenced too. Models combining chemical evolution of the Galaxy with dust evolution indeed suggest that $\gtd$ increases with decreasing metallicity $Z$ \citep[][]{Dwek98_apj501_643,Mattsson+12_mnras423_38,HirashitaHarada17_mnras467_699}. This is also supported by observations in nearby galaxies \citep[e.g.][]{Sandstrom+13_apj777_5}. 
	
	Except for a few cases, the data for external galaxies are averaged over the entire galaxy, and in all cases optically thick CO lines are used to obtain the mass of molecular gas. Moreover, in studies in which the gradient in $\gtd$ with $Z$ can be spatially resolved, the resolution is of the order of a kpc, introducing large uncertainties, for example, by assuming a uniform single temperature for dust or a specific calibration in deriving the metallicity \citep[e.g.][]{Sandstrom+13_apj777_5}. As \citet{Mattsson+12_mnras423_38} discuss, this could lead to a dust content which, in the central regions, often is larger than the amount of available metals in the interstellar medium (ISM).
	
	The study of the metallicity-$\gtd$ relation in the Milky Way not only opens the possibility to have, for the first time, more accurate estimates of the amount of molecular gas in clouds,
	but also provides the possibility to explore it on spatial scales and sensitivities that are extremely challenging to obtain, if not inaccessible, in galaxies other than our own. \citet{Issa+90_aap236_237} studied how the gradient in gas-to-dust ratio depends on the galactocentric radius, but for a limited range of $\dgc$ ($9-11\kpc$) and using optically thick CO lines to estimate the amount of molecular gas, via the integrated intensity of the CO (1--0) line-to-molecular mass conversion factor $X_\mathrm{CO}$. 
	
\begin{figure}[t]
	\centering
	\includegraphics[width=\columnwidth]{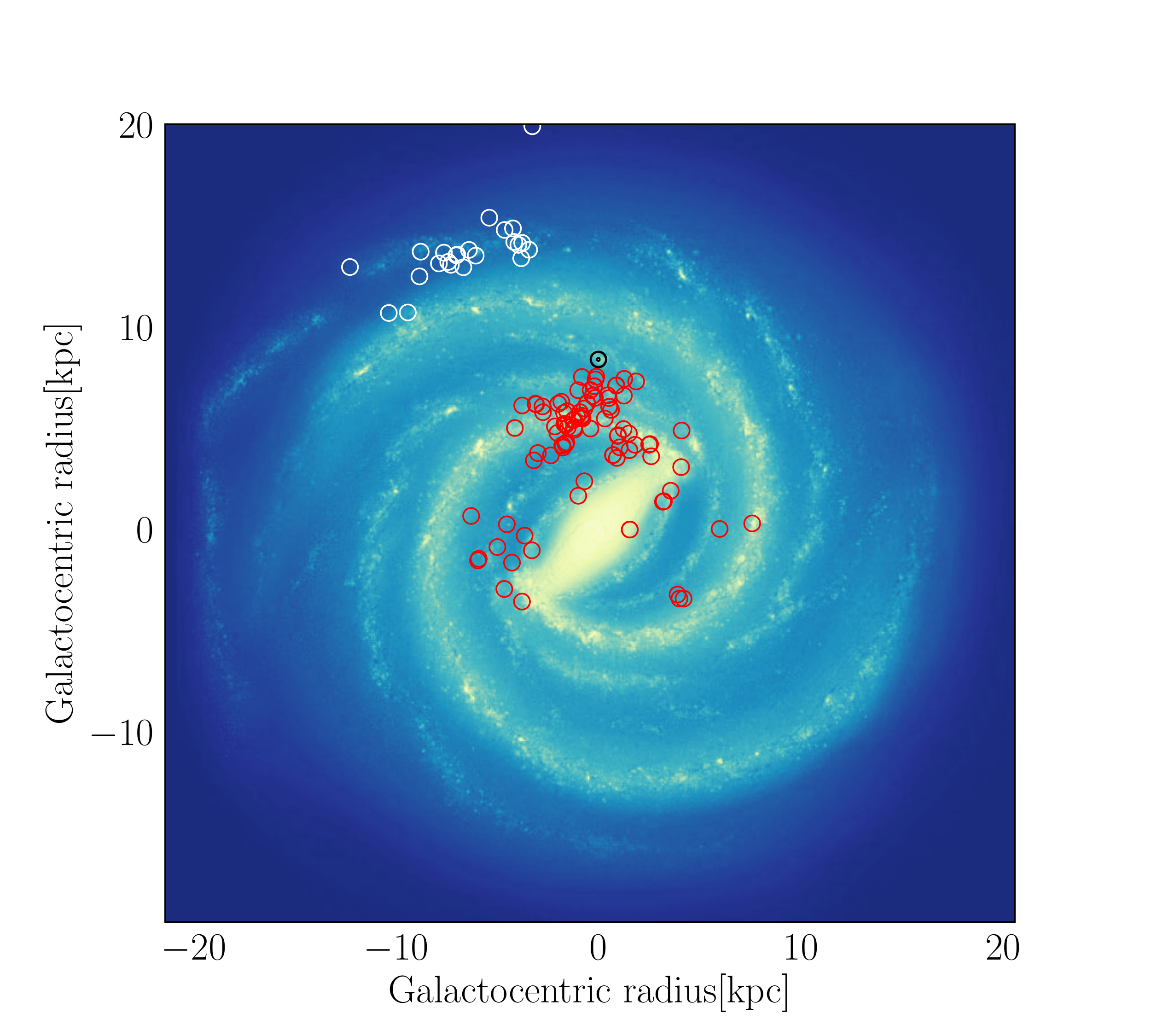}
	\caption{Distribution of the sources considered in this work. In white we show the sources in the FOG, and in red we show clumps from the TOP100 sample. The background image is an artist impression of the Milky Way as seen from the northern galactic pole (courtesy of NASA/JPL-Caltech/R. Hurt -- SSC/Caltech). The Sun in at (0, 8.34) kpc.}\label{fig:distribution_clumps}
    
    \end{figure}

	In this work we use a sample of 23 sources in the far outer Galaxy, complemented by 57 sources from the ATLASGAL TOP100 in the inner Galaxy (Fig.~\ref{fig:distribution_clumps}) to expand this pioneering work, exploring the variation of $\gtd$ across the entire disk of the Milky Way. 
	This opens up the possibility of using the appropriate value of the gas-to-dust ratio to obtain more precise estimates of the very basic properties of molecular clouds throughout the Milky Way from publicly available surveys, such as the total mass and H$_{2}$ column density. From these quantities it is possible to derive molecular abundances and, in combination with complete surveys of the galactic disk, a reliable distribution of mass of molecular gas in the Milky Way.

\section{Observations and sample selection}\label{sec:obs_and_sample}

	From the \citet{WouterlootBrand89_aaps80_149} IRAS/CO catalogue and that compiled by \koenigOGText\ using $^{12}$CO(2--1) and $^{13}$CO (2--1)\footnote{Observed with the APEX-1 receiver at the Atacama Pathfinder Experiment (APEX) 12-m telescope.}, we selected a sample of 23 sources in the far outer Galaxy ($\dgc > 14\kpc$; FOG) with the following criteria: i) the source must be associated with IR emission in WISE images ii) Herschel data must be available to estimate the dust content, and iii) the surface density of dust ($\SigmaDust$) at the emission peak must exceed $3\times10^{-5}\usk\gram\cm^{-2}$, or $N_\mathrm{H_{2}}=8.75\times10^{20}\cm^{-2}$ (i.e. $\Sigma_{gas}\sim19\msun\pc^{-2}$), assuming $\gtd=136$. 
	According to the model of \citet{HirashitaHarada17_mnras467_699}, the latter condition is sufficient to ensure that the vast majority of gas is in molecular form for $Z \gtrsim 0.2 \usk Z_{\odot}$. 
	In the FOG, in fact, the metallicity ranges from $\sim0.5\,Z_{\odot}$ at $\dgc\sim14\kpc$ to $\sim0.2\,Z_{\odot}$ at $\dgc\sim21\kpc$ \citep[using the results in][ see Eq.~\ref{eq:Z_grad_MW}]{LuckLambert11_aj142_136}.
	Observations of the Magellanic Clouds also provide support for this statement. 
	The metallicities in the Large and Small Magellanic Clouds (LMC, SMC) are $Z = 0.5 \, Z_{\odot}$ and $Z = 0.2\, Z_{\odot}$, respectively \citep{RussellDopita92_apj384_508}, encompassing the range of the far outer Galaxy. 
	Observations of the atomic and molecular gas in these galaxies by \citet{RomanDuval+14_apj797_86} demonstrate that the H\textsc{i}--H$_{2}$ transition occurs at $\approx30\msun\pc^{-2}$ in the LMC and $\approx80\msun\pc^{-2}$ in the SMC. Our criterion on the surface density of dust, when using the gas-to-dust ratios estimated by \citet{RomanDuval+14_apj797_86} in the Magellanic Clouds, exceeded these observed thresholds: $\Sigma_{gas}\approx70\msun\pc^{-2}$ for $Z=0.5\, Z_{\odot}$ and $\sim230\msun\pc^{-2}$ for $Z=0.2\, Z_{\odot}$.

	The selection of only IR-bright sources, 
	still associated with substantial molecular material, 
	implies that we are dealing with clumps in a relatively advanced stage of the star formation process, when CO is not significantly affected by depletion \citep{Giannetti+14_aap570_65}.
	The sources have been followed-up with single-pointing observations centred on the dust emission peak, as identified in Herschel images, carried out using the APEX-1 receiver at APEX tuned to $218.09\ghz$, a setup which includes C$^{18}$O(2--1). Here, we have used this transition to estimate the total amount of H$_{2}$ at the position of the dust emission peak. The angular resolution of APEX at this frequency is $\sim28\arcsec$.
	Observations were performed between September 29 and October 15 2015, and between December 3 and 11 2015. The typical rms noise on the $\taStar$-scale ranges between $10\usk\milli\kelvin$ and $20\usk\milli\kelvin$ at a spectral resolution of $0.4\kms$. 
	We converted the antenna temperature $\taStar$ to main beam brightness temperature, $\tmb$, using $\eta_{\mathrm{MB}} = 0.75$.
	
\section{Results}\label{sec:results}

	As a first step we constructed the SED for each of the sources in the FOG to obtain the peak mass surface density of the dust. 
	For the SED construction and fitting, we follow the procedure described in \citet{Koenig+17_aap599_139} and adopted in \citet{Giannetti+17_aap603_33} and \citet{Urquhart+17_ArXiv}, with minor changes due to the absence of ATLASGAL images for the outer Galaxy. We considered the five Hi-GAL \citep{Molinari+10_pasp122_314} bands (500, 350, 250, 160 and $70\mum$) from the SPIRE \citep{Griffin+10_aap518_3} and PACS \citep{Poglitsch+10_aap518_2} instruments, to reconstruct the cold dust component of the SED. The contribution from a hot embedded component is estimated from mid-IR continuum measurements, using MSX \citep{Egan+03_AFRL} and WISE \citep{Wright+10_aj140_1868} images at 21, 14, 12 and $8\mum$, and 24 and $12\mum$, respectively.
	
	The flux for each of the bands was calculated using an aperture-and-annulus scheme. The aperture is centred on the emission peak at $250\mum$ and its size was set to three times the FWHM of a Gaussian fitted to the $250\mum$ image. The background was calculated as the median flux over an annulus with inner and outer radii of 1.5 and 2.5 the aperture size, respectively. After being normalised to the area of the aperture, the background was subtracted from the flux within the aperture.
	The uncertainties on the background-corrected fluxes were calculated summing in quadrature the pixel noise of the images and a flux calibration uncertainty. 
	We adopted a calibration uncertainty of $20\%$ for the 350, 250, 160 and $70\mum$ fluxes, and of $30\%$ for the mid-IR bands. An uncertainty of $50\%$ is assumed for the $500\mum$ flux, due to the
	large pixel size, and for the $8\mum$ flux, due to contamination from PAHs.
	The grey-body plus black-body model was optimised via a $\chi^{2}$ minimisation, and the uncertainties on the parameters were estimated propagating numerically the errors on the observables.
	Differently from \citet{Koenig+17_aap599_139} and \citet{Urquhart+17_ArXiv}, we use the $350\mum$ Herschel flux measurement to calculate the peak dust surface density of the clump, because the sources were not covered in ATLASGAL and because this image has a comparable resolution to our molecular-line observations. The method is discussed in more detail in \citet{Koenig+17_aap599_139} and \citet{Urquhart+17_ArXiv}, and we refer the interested reader to these publications.

\begin{figure}
		\centering
		\includegraphics[width=0.5\textwidth]{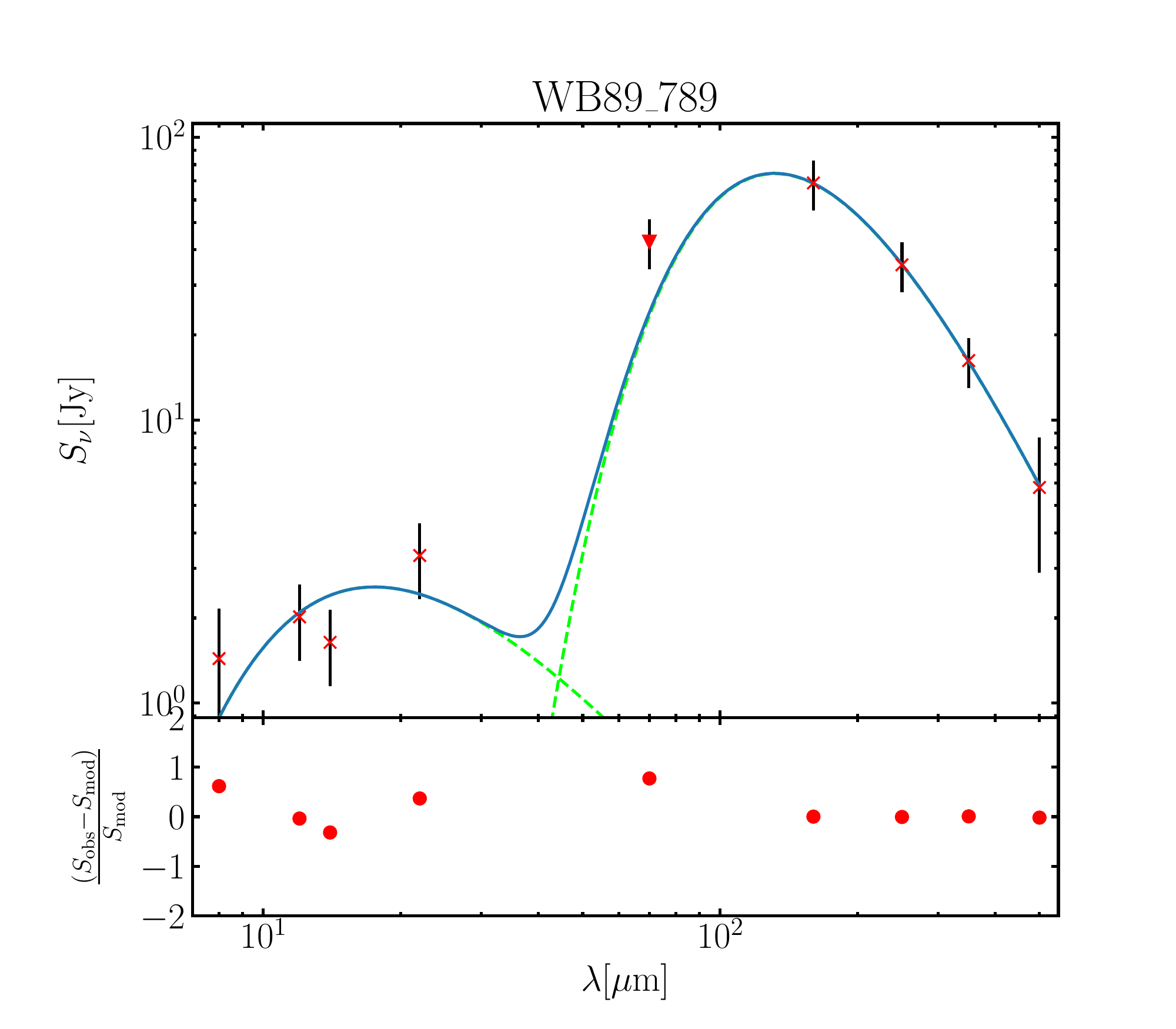}
		\caption{Top: example of the SED fit for WB$\_$789, hosting one of the the furthest clusters from the galactic centre yet detected \citep{BrandWouterloot07_aa464_909}. Extracted fluxes are indicated by the red crosses, and upper and lower limits are indicated by triangles pointing downwards and upwards, respectively. The best fit curve is indicated in blue, and the separate contributions of the grey and black bodies are shown by the green dashed lines. Bottom: Residuals calculated as $(S_{\mathrm{obs}} - S_{\mathrm{mod}}) / S_{\mathrm{obs}}$. The SED and their residuals for all the other sources are included in Fig.~\ref{fig:seds}.\label{fig:sed_example}}
	\end{figure}

	The dust opacity and emissivity used are the same as in \citet{Koenig+17_aap599_139}, that is, $\kappa_{870\mum}=1.85\cm^{2}\usk\gram^{-1}$ and $\beta=1.75$, respectively \citep[see e.g.][]{Kauffmann+08_aap487_993}. The SEDs for the entire sample can be found in Fig.~\ref{fig:seds}; an example is shown in Fig.~\ref{fig:sed_example}. In addition to the mass surface density of dust at the far-IR peak, we derived the bolometric luminosity, the dust temperature and mass of the sources, as measured within the apertures listed in Table~\ref{tab:sed_fit}, that contains the complete results of the SED fit.
	
\begin{figure}
		\centering
		\includegraphics[width=0.8\columnwidth]{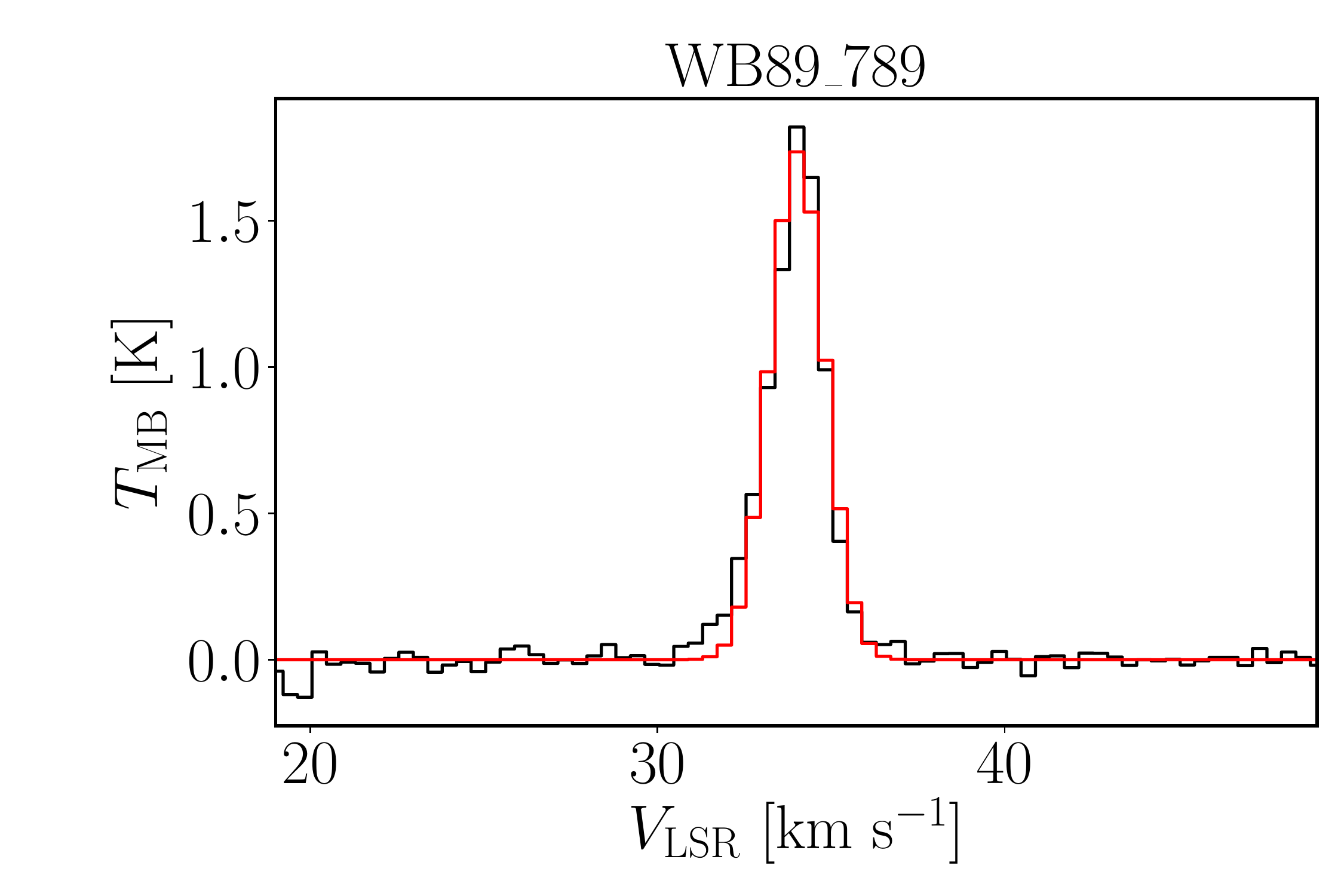}
		\caption{Example of the C$^{18}$O(2--1) observations for WB$\_$789. We indicate in red the best fit from MCWeeds. The spectra and their fits for the entire sample are shown in Fig.~\ref{fig:spectra}.\label{fig:fit_example}}
	\end{figure}

	We fitted the C$^{18}$O (2--1) line using MCWeeds \citep{Giannetti+17_aap603_33} with the algorithm that makes use of the Normal approximation \citep{gelman2003bayesian} to obtain the column density of carbon monoxide, under the assumption of LTE; the adopted partition function is reported in Table~\ref{tab:part_funct}. Using the relation between the dust temperature and the excitation temperature of CO isotopologues found in \citet[][see their Fig.~10]{Giannetti+17_aap603_33}, we estimated the excitation conditions for the sources in the FOG. We used this value of $\tex$ as the most probable one in the prior, with a value of $\sigma$ equal to the measured intrinsic scatter; all priors are fully described in Table~\ref{tab:priors} and the results are listed in Table~\ref{tab:mcweeds_fit_fog}. To exclude biases connected to the $\td\,vs.\tex$ relation, we compared the column densities with those computed using the unmodified values of the dust temperature from the SED; this has only a minor impact on the derived quantities. In Appendix~\ref{app:spectra} we show the fit results, superimposed on the observed spectra; an example is given in Fig.~\ref{fig:fit_example}.
	
	In order to study how the gas-to-dust ratio varies across the galactic disk, we complemented the FOG sample with sources selected from the TOP100 \citep{Giannetti+14_aap570_65, Koenig+17_aap599_139}, a representative and statistically significant sample of high-mass star-forming clumps covering a wide range of evolutionary phases \citep{Koenig+17_aap599_139, Giannetti+17_aap603_33}. These sources are among the brightest in their evolutionary class in the inner Galaxy.
	For the 57 sources classified as \hii\ and IRb in the TOP100, we used the column density determinations from \citet{Giannetti+14_aap570_65} to derive the H$_{2}$ column density. Among the isotopologues analysed in that work, we elected to use C$^{17}$O, because in these extreme sources C$^{18}$O can have a non-negligible optical depth, and because we have FLASH$^{+}$  \citep{Klein+14_TransThzSciTech4_588} observations of C$^{17}$O (3--2) for the entire subsample. 
	
	We have $\mathrm{C^{17}O(1-0)}$ for 17 of the selected sources in the TOP100, for which we were able to estimate the optical depth; $30\%$ of the sample has an optical depth $\approx0.1$, the remaining clumps have optical depths below this value.
	Assuming LTE, $T=30\kel$, and $\tau_\mathrm{C^{17}O(1-0)}=0.1$, the optical depth of $\mathrm{C^{17}O(3-2)}$ is about a factor of four to five higher than that of the (1--0) transition, leading to an underestimate of the carbon monoxide column density less than $\sim30\%$. Therefore the correction for opacity can be considered negligible (see Sect.~\ref{sec:discussion}). The column density of C$^{17}$O is then converted to C$^{18}$O using $^{18}\mathrm{O}/^{17}\mathrm{O} = 4$, according to \citet{Giannetti+14_aap570_65}, as determined from the same sample. The peak surface densities (and total masses) of dust for the clumps in the TOP100 were taken from the results of \citet{Koenig+17_aap599_139}. The properties of the sources extracted from the TOP100 are listed in Table~\ref{tab:mcweeds_fit}.
	
	We computed the mass surface density of the gas from the column density of C$^{18}$O, obtained via MCWeeds for the FOG sample, and of C$^{17}$O for the TOP100. When more than one velocity component was observed in the spectra of the CO isotopologues, the column densities were summed to obtain the total surface density of carbon monoxide along the line of sight, because all sources contribute to the observed continuum. This has the effect of introducing scatter in the value of $\gtd$ at a particular $\dgc$, as the clumps have different distances, but it only happens in a minor fraction of the sources (e.g. two in the FOG sample), and depends on the $\dgc$ of the sources. From the C$^{18}$O surface density, we derived the total mass surface density of the cloud ($\Sigma$), accounting for helium, and assuming that the expected abundance of the C$^{18}$O is described by:
	\begin{equation}
		\chi_\mathrm{C^{18}O}^\mathrm{E} = \frac{9.5 \times \pot{-5} \times 10^{\alpha(\dgc - R_\mathrm{GC,\odot})}}{^{16}\mathrm{O}/^{18}\mathrm{O}} , \label{eq:expected_ab}
	\end{equation}
	where $\dgc$ is expressed in $\kpctab$, $R_\mathrm{GC,\odot}=8.34 \kpc$ \citep{Reid+14_apj783_130}, and $\alpha$ describes the C/H gradient, taken to be $-0.08\usk \mathrm{dex}\kpc^{-1}$ from \citet{LuckLambert11_aj142_136}. We assumed that the CO abundance is controlled by the carbon abundance, because it is always less abundant than oxygen, and becomes progressively more so in the outer Galaxy.
	A smaller abundance of CO at lower $Z$ 
	is consistent with observations of low-metallicity galaxies, where the detectable CO-emitting region is significantly smaller than the H$_{2}$ envelope \citep{Elmegreen+13_nat495_487,Rubio+15_nat525_218}.
	The oxygen isotopic ratio is commonly described by the relation $^{16}\mathrm{O}/^{18}\mathrm{O} = 58.8 \dgc + 37.1$ \citep{WilsonRood94_araa32_191}. On the one hand, independent measurements of $^{16}\mathrm{OH}/^{18}\mathrm{OH}$ by \citet{Polehampton+05_aap437_957}, despite finding consistent results with the previous works, do not strongly support such a gradient. On the other hand, \citet{Wouterloot+08_aa487_237} find an even steeper gradient considering sources in the FOG, where C$^{18}$O is likely to be less abundant, if not for the oxygen isotopic ratio, then for selective photodissociation due to lower shielding of the dust, and self-shielding. 

\begin{figure*}[t]
	\centering
	\includegraphics[width=\textwidth]{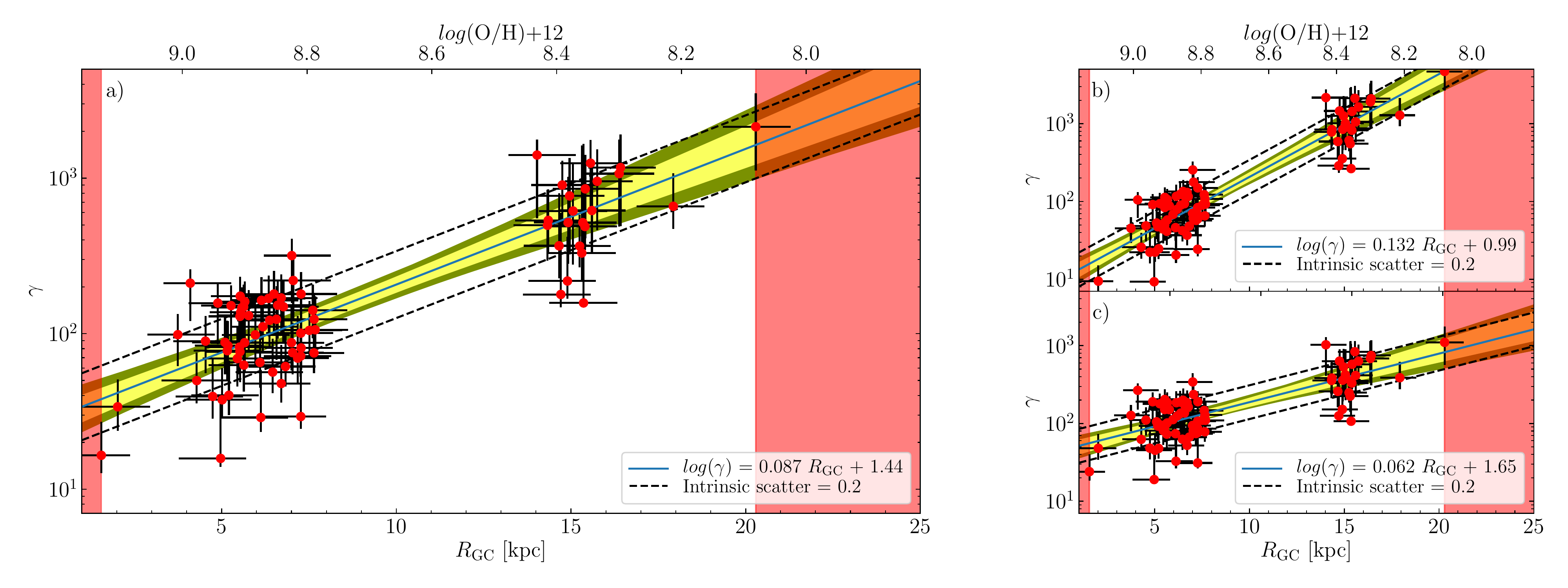}
	\caption{Variation of the gas-to-dust ratio with galactocentric radius, for our fiducial case (Panel a), considering a CO/C$^{18}$O galactocentric gradient (Panel b), and assuming that the abundance of CO follows the radial oxygen gradient, rather than the C/H (Panel c). The thick blue lines indicate the best fit, reported in the bottom right corner; the $68\%$ and $95\%$ highest probability density intervals of the fit parameters are indicated by the light and dark yellow-shaded regions, respectively. The intrinsic scatter is indicated by the dashed lines. 
	For comparison with external galaxies $log$(O/H) + 12 is shown on the top axis.}\label{fig:gtd_fit}
    \end{figure*}

	For the moment we ignore the effect of a gradient, and adopt the local CO/C$^{18}$O ratio. We used $\Sigma$, together with $\SigmaDust$, as observed data for a JAGS\footnote{\url{http://mcmc-jags.sourceforge.net/}} \citep[Just Another Gibbs Sampler,][]{Plummer2003_IWDSC} model which derives the gas-to-dust ratios as $\Sigma/\Sigma_{dust}$, and fits the points in a log-linear space, considering an intrinsic scatter. 
	Figure~\ref{fig:gtd_fit}a shows that the gas-to-dust ratio increases with galactocentric distance, with a gradient for $\gtd \vs\ \dgc$ described by:
	\begin{equation}
		log(\gtd) = \left( 0.087\left[ \asymErr{+0.045}{-0.025} \right]\pm0.007\right) \, \dgc + \left( 1.44 \left[ \asymErr{-0.45}{+0.21} \right]\pm0.03 \right), \label{eq:gtd_gradient}
	\end{equation}
	where $\dgc$ is expressed in $\kpctab$; we first indicate the systematic uncertainty between square brackets (discussed in the next section), and the statistical uncertainties afterwards. 
	This equation gives values for $\gtd$ at the Sun distance between $\approx130$ and $\approx 145$, well consistent with the local value of 136, considering the intrinsic scatter of the observed points (cf. Fig.~\ref{fig:gtd_fit}) and the uncertainties in the derived relation.
	As indicated in Fig.~\ref{fig:gtd_fit}a, our results are, in general, valid only between $\sim2\kpc$ and $\sim20\kpc$ from the galactic centre, the range spanned by the sources in our sample. 
	
	The slope of the gradient is very close to that used in Eq.~\ref{eq:expected_ab}, showing that C$^{18}$O behaves in a way comparable to the dust, with respect to metallicity. This implies that the results are closely linked to the assumed galactocentric carbon gradient. In the next section we discuss as limiting cases how the $\gtd$ gradient would change if the CO abundance follows the oxygen variation instead, and if C$^{18}$O is less abundant with respect to CO in the outer Galaxy, as a consequence of the $^{16}\mathrm{O}/^{18}\mathrm{O}$ gradient, or of selective photodissociation (see also Fig.~\ref{fig:gtd_fit}b, c). The effect of such systematic uncertainties causes the variations in the slope and intercept of Eq.~\ref{eq:gtd_gradient} indicated in the square brackets.

\section{Discussion}\label{sec:discussion}

	In this section, we discuss the uncertainties in the gas-to-dust ratio estimates and in its galactocentric gradient, why $\gtd$ has to be higher at large $\dgc$, and how it depends on metallicity. Estimates of the gas-to-dust ratio are difficult, resting on the derivation of surface densities of a tracer of H$_{2}$ (C$^{18}$O and C$^{17}$O in our case) and of the dust. Several assumptions introduce a systematic uncertainty in $\gtd$. For the surface density of molecular gas, the main sources of uncertainties are the canonical CO abundance, that can vary by a factor of two, the CO--C$^{18}$O conversion, discussed in Sect.~\ref{sec:results}, and the assumption of LTE, which is likely less important, given the results of the comparison between temperatures derived from CH$_{3}$CCH and CO isotopologues in the TOP100 sample in \citet{Giannetti+17_aap603_33}. Dust is more problematic, especially because its properties are poorly constrained. Opacity and emissivity are sensitive to the grain composition and size distribution:
	distinct models can induce discrepancies in the estimated mass surface density of dust up to a factor of approximately three \citep[see e.g.][]{OssenkopfHenning94_aap291_943, LiDraine01_apj554_778, Gordon+14_apj797_85, Gordon+17_apj837_98}.
	The simple SED model adopted is a crude approximation as well: 
	temperature varies along the line of sight, and in the extreme case where the representative grey-body temperature changes from $\approx20\kel$ (the median in our FOG sample is $\approx23\kel$) to $\approx 50\kel$, the dust surface density changes by a factor of approximately five.
	
	Propagating these to the gas-to-dust ratio implies a global uncertainty of nearly a factor of six on $\gtd$ for each target. It is therefore relevant to test whether a simpler model with a constant $\gtd$ ratio is to be favoured over the proposed gradient. A Bayesian model comparison, which automatically takes into account the Ockham's Razor principle \citep[e.g.][]{BolstadIBS}, shows that, in the unfavourable case that the CO abundance follows the oxygen gradient, the odds ratio is approximately eight in favour of the gradient model\footnote{Considering a flat prior on the slope and intercept of the $log(\gtd) \vs \dgc$ relation in the ranges $0-1$ and $0-4$, respectively.}, which is then to be preferred over a constant value of $\gtd$ across the entire disk.
	
	Factors that can change the slope of the $\gtd \vs \dgc$ relation are the molecular gas- and the CO-dark gas fractions, the CO abundance gradient, and the dust model. Larger quantities of gas in atomic form, as well as more CO-dark gas at lower metallicities (due to a lower shielding of dust and self-shielding) would cause the relation to be steeper. However, because we target exclusively dense molecular clouds, the vast majority of gas should be in molecular form (see Sect.~\ref{sec:obs_and_sample}).
	A larger fraction of CO-dark gas is evident for the low-metallicity galaxy WLM \citep{Elmegreen+13_nat495_487,Rubio+15_nat525_218}; a less extreme, but analogous situation is possible for clouds at the edge of the Milky Way disk \citep[in the FOG, $Z$ is larger than in WLM by a factor of between approximatley two and five, see][]{Leaman+12_apj750_33}. 

	On the other hand, the variation of dust composition and size distribution of grains tend to make the measured relation flatter. 
	Silicates are likely to be more common in the outer Galaxy \citep[e.g.][]{Carigi+05_apj623_213}; in this case the opacity would be lower, leading to an underestimate of the dust surface density. The models in \citet{OssenkopfHenning94_aap291_943} show that a variation in the silicate-to-carbon fraction has a much smaller impact than the change in size distribution due to coagulation. 
	In the extreme case in which no coagulation takes place in the FOG, and it is, on the contrary, efficient in the inner Galaxy, the dust opacity changes by a factor of approximately two in the far-IR and submm regimes, reducing the mass surface densities by the same quantity. This effect has an impact similar in magnitude, but opposite, to that of the CO/C$^{18}$O gradient. 

	In the outer Galaxy, where extinction is lower, dust grains can be more efficiently reprocessed. As a consequence, more carbon is present in the gas phase \citep[e.g.][]{Parvathi+12_apj760_36}, effectively making the gradient in CO abundance shallower than the C/H one, if it follows the gas-phase abundance of carbon. A limiting case is obtained by using the slope of the oxygen gradient in Eq.~\ref{eq:expected_ab}, in which case the slope in Eq.~\ref{eq:gtd_gradient} can be as shallow as $0.062 \usk\mathrm{dex}\kpc^{-1}$. It is, however, unlikely that the increased abundance of C in the gas phase and the less efficient coagulation have such an important effect. Conversely, if we neglect these effects, but consider the CO/C$^{18}$O gradient, we can obtain an upper limit for the gradient slope. Under these conditions, $\gtd$ varies by $0.132 \usk\mathrm{dex}\kpc^{-1}$.

	The CO/C$^{18}$O abundance gradient, and larger fractions of CO-dark and atomic gas at lower metallicity \citep[see e.g.][]{Elmegreen+13_nat495_487,Rubio+15_nat525_218} contrast the effects of the increased fraction of C in the gas phase, of dust size distribution and composition. 
	For simplicity, as a fiducial value, we have therefore considered the relation for which the variation in grain size distribution and the increased fraction of CO-dark gas cancel out the impact of the CO/C$^{18}$O abundance gradient (Fig.~\ref{fig:gtd_fit}a).
	
	\medskip
	
	Theoretical considerations also indicate that the gas-to-dust ratio has to be higher in the FOG.
	In the following, we show that at a distance of $\sim15\kpc$ the fraction of heavy elements locked into dust grains must be $80\%$ to maintain $\gtd$ at the local value of $136$.
	Following \citet{Mattsson+12_mnras423_38} we can conservatively use the O/H gradient to obtain an approximation of the galactocentric metallicity behaviour. 
	The radial $Z$ gradient for our Galaxy can be reliably obtained via measurements of the abundance of heavy elements in Cepheids, which are young enough to represent the present-day composition. 
	We use the results from \citet{LuckLambert11_aj142_136}, who consider a large number of Cepheids with $5\kpc\lesssim\dgc\lesssim17\kpc$, deriving for oxygen the gradient $d\mathrm{[O/H]}/d\dgc = -0.056 \,\mathrm{dex}\kpc^{-1}$. We obtain for $Z$:
	\begin{equation}
		log(Z) = -0.056 \dgc - 1.176\label{eq:Z_grad_MW},
	\end{equation} 
	which gives, at the location of the Sun, an H-to-metal mass ratio $\sim44$. If approximately $40\%$ of the heavy elements are locked into dust grains \citep{Dwek98_apj501_643}, this implies $\gtd=110$, which is in very good agreement with the locally-estimated value of $136$.
		
	The dust-to-gas mass ratio $Z_{d}$ is the inverse of $\gtd$, and the fraction of mass in heavy elements locked in dust grains, the dust-to-metal ratio, can be expressed as the ratio of $Z_{d}$ and the gas metallicity, that is, $Z_{d}/Z$. A dust-to-metal ratio of one would imply that dust grains contain all elements heavier than helium.
	If we were to assume that the gas-to-dust ratio remains constant to $\gtd \equiv Z_{d}^{-1}=136$ (implying that progressively more heavy elements end up in dust grains), using Eq.~\ref{eq:Z_grad_MW} we would see that the dust metallicity $Z_{d}/Z$ reaches $80\%$ at $\dgc\approx15\kpc$. In addition, the metallicity gradient is most likely steeper than the oxygen gradient \citep[e.g.][]{Mattsson+12_mnras423_38}, moving this limit inwards, thus indicating that in the FOG the gas-to-dust ratio is bound to be higher.
	
	Now using our results for the increase of the gas-to-dust ratio with $\dgc$, the dust metallicity can be derived from Eqs.~\ref{eq:gtd_gradient} and \ref{eq:Z_grad_MW}:
	\begin{equation}
		log\left( \frac{Z_{d}}{Z} \right) = \left( -0.031 \left[\asymErr{+0.025}{-0.047} \right] \right) \dgc - \left( 0.26 \left[ \asymErr{-0.21}{+0.45} \right] \right), \label{eq:Zdust}
	\end{equation}
	which shows that the dust-to-metal ratio decreases with galactocentric radius.  
	A decrease of the the dust-to-metal ratio is the most common situation in late-type galaxies and indicates that grain growth in the dense ISM dominates over dust destruction \citep[e.g.][]{Mattsson+12_mnras423_38,Mattsson+14_mnras444_797}. This strongly reinforces the previous argument that a constant $\gtd=136$ cannot be sustained in the far outer Galaxy, because the metal-to-dust ratio virtually always decreases moving outwards in the disk, for Milky-Way type galaxies.

	A good test bench for Eq.~\ref{eq:gtd_gradient} is represented by the Magellanic Clouds. Combining Eqs.~\ref{eq:gtd_gradient} and \ref{eq:Z_grad_MW}, and using the appropriate metallicity \citep[$Z = 0.5 \, Z_{\odot}$ and $Z = 0.2\, Z_{\odot}$ for the Large and Small Magellanic Clouds, respectively;][]{RussellDopita92_apj384_508}, we obtain $\gtd \sim 420\asymErr{+250}{-110}$ and $\gtd = 1750\asymErr{+4100}{-900}$, in excellent agreement with the results of \citet{RomanDuval+14_apj797_86}.
	    
\section{Summary and conclusions}

	We combined our molecular-line surveys towards dense and massive molecular clouds in the inner- and far outer disk of the Milky Way to study how the gas-to-dust ratio $\gtd$ varies with galactocentric distance and metallicity.
	We estimated conservative limits for the galactocentric gradient of gas-to-dust mass ratio, by considering multiple factors that influence its slope (see Sect.~\ref{sec:discussion}), and defined, for simplicity, the fiducial value as the case where dust coagulation and the larger fraction of carbon in the gas phase in the FOG balance the CO/C$^{18}$O abundance gradient.
	The gas-to-dust mass ratio is shown to increase with $\dgc$ according to Eq.~\ref{eq:gtd_gradient}, and this gradient is compared with that of metallicity, as obtained from Cepheids by \citet{LuckLambert11_aj142_136}.
	The variation in gas-to-dust ratio is steeper than that of $Z$ ($\gtd \propto Z^{-1.4\asymErr{+0.3}{-1.0}}$),
	implying that the the dust-to-metal ratio decreases with distance from the galactic centre. This indicates that dust condensation in the dense ISM dominates over dust destruction, which is typical of late-type galaxies like ours \citep{Mattsson+12_mnras423_38,Mattsson+14_mnras444_797}. 
	The predictions obtained combining Eq.~\ref{eq:gtd_gradient} and \ref{eq:Z_grad_MW} for the metallicities of the Magellanic Clouds are in excellent agreement with the results on $\gtd$ in these galaxies by \citet{RomanDuval+14_apj797_86}.

	The use of Eq.~\ref{eq:gtd_gradient} to calculate the appropriate value of $\gtd$ at each galactocentric radius is fundamental for the study of individual objects, allowing us to derive accurate H$_{2}$ column densities and total masses from dust continuum observations, as well as for any study that compares the properties of molecular clouds in the inner and outer Galaxy. This opens the way for a complete view of the galactic disk and of the influence of $Z$ on the physics and chemistry of molecular clouds. 

\begin{acknowledgements}
We are thankful to Frank Israel for a discussion on the uncertainties involved in deriving the gas-to-dust ratio and to the anonymous referee that both helped to improve the quality and the clarity of this paper. This work was partly carried out within the Collaborative Research Council 956, sub-project A6, funded by the Deut\-sche For\-schungs\-ge\-mein\-schaft (DFG). This paper is based on data acquired with the Atacama Pathfinder EXperiment (APEX). APEX is a collaboration between the Max Planck Institute for Radioastronomy, the European Southern Observatory, and the Onsala Space Observatory. 
This research made use of Astropy, a community-developed core Python package for Astronomy \citep[][\url{http://www.astropy.org}]{astropy_2013}, of NASA's Astrophysics Data System, and of Matplotlib \citep{Hunter_2007_matplotlib}. MCWeeds makes use of the PyMC package \citep{Patil+10_jstatsoft35_1}.
\end{acknowledgements}

\bibliographystyle{bibtex/aa}
\bibliography{bibtex/biblio.bib}

\AtEndDocument{
\Online
\clearpage
\begin{appendix}
		
	\section{Tables}
	\begin{table*}
	\centering
	\caption{Properties of the sources and results of the SED fit procedure for the far outer Galaxy sample.\label{tab:sed_fit}}
	\begin{tabular}{l*{9}{r}}
	\hline \hline
	Source & $l$ & $b$ & Dist.\tablefootmark{a} & $\dgc\tablefootmark{a}$ & Diam. & $\td$ & $L_{bol}$ & $M_{dust}$ & $\Sigma_{dust}$\tablefootmark{b} \\
	 & deg & deg & [kpc] & [kpc] & $[\arcsec]$ & [K] & $[10^x\lsun]$ & $[10^x\msun]$ & $[10^x\usk\gram\cm^{-2}]$ \\
	\hline
	WB89\_898 & $217.604$ & $-2.617$ & 8.8 & 16.4 & 109 & $24.1 \pm 5.7$ & $3.65 \pm 0.24$ & $0.15 \pm 0.19$ & $-3.84 \pm 0.19$ \\
	WB89\_986 & $229.772$ & $0.060$ & 7.9 & 14.9 & 120 & $22.0 \pm 4.7$ & $3.48 \pm 0.28$ & $0.35 \pm 0.18$ & $-4.04 \pm 0.18$ \\
	WB89\_909 & $217.331$ & $-1.369$ & 6.3 & 14.0 & 120 & $20.6 \pm 2.6$ & $3.72 \pm 0.20$ & $0.77 \pm 0.13$ & $-3.42 \pm 0.13$ \\
	WB89\_1024 & $238.959$ & $-1.684$ & 9.2 & 15.4 & 107 & $24.3 \pm 4.4$ & $3.58 \pm 0.18$ & $0.04 \pm 0.15$ & $-4.04 \pm 0.15$ \\
	WB89\_890 & $212.282$ & $-0.617$ & 6.4 & 14.3 & 120 & $27.2 \pm 2.2$ & $3.69 \pm 0.13$ & $0.02 \pm 0.10$ & $-4.23 \pm 0.10$ \\
	WB89\_858 & $213.192$ & $-3.325$ & 6.9 & 14.8 & 91 & $25.5 \pm 0.4$ & $3.37 \pm 0.05$ & $-0.22 \pm 0.09$ & $-4.22 \pm 0.08$ \\
	WB89\_1125 & $256.151$ & $-1.375$ & 9.7 & 14.3 & 76 & $19.5 \pm 1.8$ & $3.09 \pm 0.13$ & $0.22 \pm 0.11$ & $-4.01 \pm 0.11$ \\
	WB89\_879 & $214.931$ & $-2.719$ & 6.9 & 14.7 & 99 & $26.5 \pm 5.9$ & $3.42 \pm 0.29$ & $-0.19 \pm 0.17$ & $-4.26 \pm 0.17$ \\
	WB89\_896 & $215.888$ & $-2.010$ & 7.9 & 15.6 & 108 & $23.7 \pm 4.4$ & $3.50 \pm 0.16$ & $-0.07 \pm 0.16$ & $-4.18 \pm 0.16$ \\
	WB89\_856 & $213.098$ & $-3.561$ & 7.7 & 15.6 & 114 & $26.3 \pm 2.3$ & $3.79 \pm 0.15$ & $0.24 \pm 0.11$ & $-4.07 \pm 0.10$ \\
	WB89\_1126 & $257.508$ & $-2.252$ & 10.6 & 15.0 & 53 & $24.4 \pm 6.8$ & $3.30 \pm 0.31$ & $-0.15 \pm 0.21$ & $-4.11 \pm 0.21$ \\
	WB89\_1006 & $235.686$ & $-1.246$ & 8.1 & 14.7 & 110 & $28.5 \pm 5.3$ & $3.99 \pm 0.27$ & $0.24 \pm 0.15$ & $-3.95 \pm 0.15$ \\
	WB89\_1008 & $236.999$ & $-1.838$ & 8.8 & 15.2 & 114 & $19.8 \pm 0.7$ & $3.19 \pm 0.06$ & $0.24 \pm 0.09$ & $-4.05 \pm 0.08$ \\
	WB89\_789 & $195.820$ & $-0.567$ & 12.0 & 20.3 & 109 & $23.2 \pm 3.1$ & $4.09 \pm 0.20$ & $0.81 \pm 0.13$ & $-3.56 \pm 0.13$ \\
	WB89\_1066 & $245.103$ & $-0.993$ & 9.8 & 15.4 & 87 & $16.9 \pm 0.2$ & $2.85 \pm 0.03$ & $0.35 \pm 0.08$ & $-3.91 \pm 0.08$ \\
	WB89\_1023 & $238.772$ & $-1.810$ & 10.3 & 16.4 & 54 & $24.6 \pm 5.0$ & $3.09 \pm 0.28$ & $-0.31 \pm 0.17$ & $-4.30 \pm 0.16$ \\
	WB89\_1080 & $249.599$ & $-2.075$ & 13.1 & 17.9 & 120 & $21.8 \pm 0.6$ & $3.81 \pm 0.05$ & $0.67 \pm 0.08$ & $-4.25 \pm 0.08$ \\
	WB89\_873 & $215.599$ & $-3.414$ & 7.1 & 14.9 & 114 & $21.2 \pm 1.4$ & $3.22 \pm 0.10$ & $0.12 \pm 0.10$ & $-4.18 \pm 0.10$ \\
	G237.32--1.28 & $237.320$ & $-1.280$ & 8.7 & 15.1 & 114 & $31.4 \pm 11.8$ & $4.51 \pm 0.42$ & $0.50 \pm 0.23$ & $-3.80 \pm 0.23$ \\
	G233.38--1.60 & $233.380$ & $-1.600$ & 8.7 & 15.4 & 120 & $12.1 \pm 1.0$ & $1.82 \pm 0.09$ & $-0.04 \pm 0.13$ & $-4.53 \pm 0.13$ \\
	G229.76--0.44 & $229.760$ & $-0.440$ & 8.4 & 15.3 & 120 & $13.9 \pm 0.4$ & $2.13 \pm 0.05$ & $-0.01 \pm 0.09$ & $-4.54 \pm 0.09$ \\
	G235.35--1.74 & $235.350$ & $-1.690$ & 9.3 & 15.8 & 120 & $18.1 \pm 0.3$ & $2.61 \pm 0.04$ & $-0.04 \pm 0.08$ & $-4.53 \pm 0.08$ \\
	G233.76--1.25 & $233.760$ & $-1.250$ & 8.7 & 15.3 & 117 & $13.9 \pm 3.7$ & $2.38 \pm 0.15$ & $-0.03 \pm 0.27$ & $-4.49 \pm 0.27$ \\
	\hline
	\end{tabular}
	\tablefoot{\tablefoottext{a}{Calculated using the \citet{BrandBlitz93_aap275_67} rotation curve.} \tablefoottext{b}{Derived from the $350\mum$ peak flux.}}
	\end{table*}
	\def\arraystretch{1}
	\clearpage
	\begin{table*}
		\caption{Measurements of the C$^{18}$O rotational partition function.}\label{tab:part_funct}
		\centering
		\begin{tabular}{lrrrrrrr}
			\hline
			\hline
			Temperatures [K]                 & 9.375  & 18.75   & 37.5   & 75      & 150     & 225     &  300    \\
			\hline 
			Q(C$^{18}$O)                     & 3.9    & 7.5     & 14.6   & 28.8    & 57.3    & 85.8    & 114.3   \\
			\hline
		\end{tabular}
		\tablefoot{The partition function was obtained from the JPL database \citep{Pickett+98_jqsrt60_883}.}
	\end{table*}

	\begin{table*}
		\caption{Priors used in MCWeeds.}\label{tab:priors}
		\centering
		\begin{tabular}{lcccc}
			\hline
			\hline
			C$^{18}$O         & Temperature                            & Column density   & Linewidth            & Velocity     \\
							  & [K]                                    & [log(cm$^{-2}$)] & [$\kmstab$]          & [$\kmstab$]  \\
			\hline
			Prior             & Truncated normal                       & Normal           & Truncated normal     & Normal       \\
			Parameters (cool) & $\mu = T_{ex,CO}\tablefootmark{a}$     & $\mu = 14$       & $\mu = 1$            & $\mu=\vlsr\tablefootmark{b}$ \\
							  & $\sigma = 15$                          & $\sigma = 1.5$   & $\sigma = 1$         & $\sigma = 2$ \\
							  & $low = 7, high=100$                    &                  & $low = 0.3, high=6$  &              \\
			\hline
		\end{tabular}
		\tablefoot{\tablefoottext{a}{The mean for the C$^{18}$O excitation temperature is derived from dust temperature, using the relation in \citet{Giannetti+17_aap603_33}} \tablefoottext{b}{The mean for the radial velocity is obtained from previous observations of CO and isotopologues from \koenigOGText\ and from \citet{WouterlootBrand89_aaps80_149}.}
	}
	\end{table*}

	\def\arraystretch{1.5}
	\begin{table*}
	\centering
	\caption{Column densities, expected C$^{18}$O abundances, gas-to-dust ratios and radial velocities for the far outer Galaxy sample.\label{tab:mcweeds_fit_fog}}
	\begin{tabular}{l*{6}{r}}
	\hline \hline
	Source & $N_\mathrm{C^{18}O}$ & $\chi^{E\,}_{C^{18}O}$ & $\gtd$ & $\gtd_{bf}$ & $N_\mathrm{H_2}\tablefootmark{b}$ & $\vlsr$ \\
	 & $[10^x\cm^{-2}]$ & [$10^{-8}$] &  &  & $[10^{22}\cm^{-2}]$ & $[\kmstab]$ \\
	\hline
	WB89\_898 & $15.14 \pm 0.06$ & $3.8 \asymErr{+2.2}{-1.7}$ & $1168 \asymErr{+923}{-421}$ & $739 \asymErr{+811}{-273}$ & $1.7 \asymErr{+1.8}{-0.6}$ & 63.35 \\
	WB89\_986 & $14.71 \pm 0.04$ & $5.0 \asymErr{+2.2}{-2.0}$ & $518 \asymErr{+328}{-158}$ & $547 \asymErr{+429}{-171}$ & $0.8 \asymErr{+0.6}{-0.2}$ & 70.57 \\
	WB89\_909 & $15.83 \pm 0.08$ & $5.9 \asymErr{+2.2}{-2.1}$ & $1405 \asymErr{+758}{-379}$ & $458 \asymErr{+284}{-127}$ & $2.7 \asymErr{+1.7}{-0.8}$ & 50.06/52.40$^{a}$ \\
	WB89\_1024 & $14.88 \pm 0.08$ & $4.6 \asymErr{+2.2}{-1.9}$ & $852 \asymErr{+583}{-276}$ & $603 \asymErr{+531}{-200}$ & $0.9 \asymErr{+0.8}{-0.3}$ & 82.42 \\
	WB89\_890 & $14.55 \pm 0.09$ & $5.6 \asymErr{+2.2}{-2.0}$ & $497 \asymErr{+284}{-140}$ & $486 \asymErr{+327}{-141}$ & $0.5 \asymErr{+0.3}{-0.1}$ & 46.29/47.59$^{a}$ \\
	WB89\_858 & $14.78 \pm 0.09$ & $5.2 \asymErr{+2.2}{-2.0}$ & $902 \asymErr{+555}{-269}$ & $529 \asymErr{+397}{-162}$ & $0.5 \asymErr{+0.4}{-0.2}$ & 49.34 \\
	WB89\_1125 & $14.79 \pm 0.04$ & $5.6 \asymErr{+2.2}{-2.0}$ & $535 \asymErr{+306}{-151}$ & $488 \asymErr{+330}{-142}$ & $0.7 \asymErr{+0.5}{-0.2}$ & 86.08 \\
	WB89\_879 & $14.04 \pm 0.12$ & $5.2 \asymErr{+2.2}{-2.0}$ & $179 \asymErr{+109}{-53}$ & $525 \asymErr{+390}{-160}$ & $0.4 \asymErr{+0.3}{-0.1}$ & 51.98 \\
	WB89\_896 & $14.59 \pm 0.08$ & $4.5 \asymErr{+2.2}{-1.8}$ & $618 \asymErr{+435}{-204}$ & $627 \asymErr{+576}{-213}$ & $0.6 \asymErr{+0.6}{-0.2}$ & 57.55 \\
	WB89\_856 & $15.01 \pm 0.09$ & $4.5 \asymErr{+2.2}{-1.8}$ & $1246 \asymErr{+872}{-410}$ & $622 \asymErr{+567}{-210}$ & $0.8 \asymErr{+0.8}{-0.3}$ & 52.63 \\
	WB89\_1126 & $14.80 \pm 0.08$ & $5.0 \asymErr{+2.2}{-2.0}$ & $768 \asymErr{+489}{-235}$ & $551 \asymErr{+436}{-173}$ & $0.7 \asymErr{+0.5}{-0.2}$ & 90.41 \\
	WB89\_1006 & $14.67 \pm 0.10$ & $5.3 \asymErr{+2.2}{-2.0}$ & $367 \asymErr{+222}{-108}$ & $519 \asymErr{+381}{-157}$ & $0.9 \asymErr{+0.7}{-0.3}$ & 74.81 \\
	WB89\_1008 & $14.52 \pm 0.04$ & $4.8 \asymErr{+2.2}{-1.9}$ & $365 \asymErr{+244}{-116}$ & $584 \asymErr{+496}{-190}$ & $0.8 \asymErr{+0.7}{-0.3}$ & 80.41 \\
	WB89\_789 & $15.38 \pm 0.07$ & $1.9 \asymErr{+1.8}{-1.0}$ & $2135 \asymErr{+2555}{-1032}$ & $1604 \asymErr{+3511}{-795}$ & $7.0 \asymErr{+15.3}{-3.5}$ & 34.24 \\
	WB89\_1066 & $14.77 \pm 0.04$ & $4.6 \asymErr{+2.2}{-1.9}$ & $488 \asymErr{+333}{-158}$ & $602 \asymErr{+529}{-200}$ & $1.2 \asymErr{+1.0}{-0.4}$ & 87.01 \\
	WB89\_1023 & $14.65 \pm 0.08$ & $3.9 \asymErr{+2.2}{-1.7}$ & $1068 \asymErr{+839}{-383}$ & $733 \asymErr{+798}{-270}$ & $0.6 \asymErr{+0.6}{-0.2}$ & 88.74 \\
	WB89\_1080 & $14.36 \pm 0.06$ & $2.9 \asymErr{+2.0}{-1.4}$ & $658 \asymErr{+624}{-271}$ & $1000 \asymErr{+1470}{-422}$ & $0.9 \asymErr{+1.3}{-0.4}$ & 105.92 \\
	WB89\_873 & $14.20 \pm 0.06$ & $5.1 \asymErr{+2.2}{-2.0}$ & $218 \asymErr{+137}{-66}$ & $545 \asymErr{+425}{-170}$ & $0.6 \asymErr{+0.4}{-0.2}$ & 53.65 \\
	G237.32--1.28 & $15.01 \pm 0.10$ & $4.9 \asymErr{+2.2}{-1.9}$ & $614 \asymErr{+397}{-190}$ & $563 \asymErr{+456}{-179}$ & $1.4 \asymErr{+1.1}{-0.4}$ & 78.07 \\
	G233.38--1.60 & $13.67 \pm 1.30$ & $4.7 \asymErr{+2.2}{-1.9}$ & $157 \asymErr{+107}{-51}$ & $597 \asymErr{+520}{-197}$ & $0.3 \asymErr{+0.2}{-0.1}$ & 76.70 \\
	G229.76--0.44 & $13.98 \pm 0.93$ & $4.7 \asymErr{+2.2}{-1.9}$ & $330 \asymErr{+222}{-105}$ & $592 \asymErr{+509}{-194}$ & $0.3 \asymErr{+0.2}{-0.1}$ & 73.03 \\
	G235.35--1.74 & $14.42 \pm 0.14$ & $4.3 \asymErr{+2.2}{-1.8}$ & $953 \asymErr{+686}{-320}$ & $646 \asymErr{+614}{-223}$ & $0.3 \asymErr{+0.3}{-0.1}$ & 80.98 \\
	G233.76--1.25 & $14.23 \pm 0.65$ & $4.7 \asymErr{+2.2}{-1.9}$ & $517 \asymErr{+350}{-166}$ & $595 \asymErr{+516}{-196}$ & $0.3 \asymErr{+0.3}{-0.1}$ & 76.91 \\
	\hline
	\end{tabular}
	\tablefoot{The uncertainties reported for each quantity are calculated using the limiting cases for the C$^{18}$O abundance discussed in the text. The values of the gas-to-dust ratio obtained with the best-fit relation in Eq.~\ref{eq:gtd_gradient} are reported in the $\gtd_{bf}$ column. \tablefoottext{a}{Sources with multiple velocity components.} \tablefoottext{b}{The surface density of gas is obtained multiplying $N_\mathrm{H_{2}}$ by the mean molecular mass $\mu \, m_\mathrm{H}$.}}
	\end{table*}
	\longtab{
		\onecolumn
		\begin{longtable}{lrrrrr}
		\caption{Column densities, expected C$^{18}$O abundances and gas-to-dust ratios for the subsample of the TOP100 selected for this work.\label{tab:mcweeds_fit}}\\
		\hline\hline
		Source & $N_\mathrm{C^{18}O}\tablefootmark{a}$ & $\chi^{E\,}_{C^{18}O}$ & $\gtd$ & $\gtd_{bf}$ & $N_\mathrm{H_2}\tablefootmark{b}$ \\
		 & $[10^x\cm^{-2}]$ & [$10^{-8}$] &  &  & $[10^{22}\cm^{-2}]$ \\
		\hline
		\endfirsthead
		\caption{Continued.}\\
		\hline\hline
		Source & $N_\mathrm{C^{18}O}\tablefootmark{a}$ & $\chi^{E\,}_{C^{18}O}$ & $\gtd$ & $\gtd_{bf}$ & $N_\mathrm{H_2}\tablefootmark{b}$ \\
		 & $[10^x\cm^{-2}]$ & [$10^{-8}$] &  &  & $[10^{22}\cm^{-2}]$ \\
		\hline
		\endhead
		\hline
		\endfoot
		\hline
		\multicolumn{6}{p{13cm}}{\tablefoot{The uncertainties reported for each quantity are calculated using the limiting cases for the C$^{18}$O abundance discussed in the text. The values of the gas-to-dust ratio obtained with the best-fit relation in Eq.~\ref{eq:gtd_gradient} are reported in the $\gtd_{bf}$ column. \tablefoottext{a}{Calculated multiplying the measured C$^{17}$O column density by a factor of 4 \citep[see][]{Giannetti+14_aap570_65}.} \tablefoottext{b}{The surface density of gas is obtained multiplying $N_\mathrm{H_{2}}$ by the mean molecular mass $\mu \, m_\mathrm{H}$.}}}
		\endlastfoot
		AGAL010.472+00.027 & $16.66 \pm 0.04$ & $59.2 \asymErr{+198.5}{-18.5}$ & $17 \asymErr{+8}{-13}$ & $38 \asymErr{+18}{-22}$ & $12.9 \asymErr{+6.2}{-7.5}$ \\
		AGAL010.624--00.384 & $17.16 \pm 0.05$ & $39.5 \asymErr{+46.5}{-8.9}$ & $98 \asymErr{+28}{-53}$ & $58 \asymErr{+18}{-27}$ & $16.0 \asymErr{+4.9}{-7.5}$ \\
		AGAL012.804--00.199 & $17.01 \pm 0.04$ & $25.2 \asymErr{+10.0}{-2.8}$ & $111 \asymErr{+14}{-31}$ & $95 \asymErr{+13}{-29}$ & $25.4 \asymErr{+3.5}{-7.8}$ \\
		AGAL013.658--00.599 & $16.05 \pm 0.04$ & $35.8 \asymErr{+33.5}{-7.2}$ & $50 \asymErr{+13}{-24}$ & $65 \asymErr{+17}{-28}$ & $3.0 \asymErr{+0.8}{-1.3}$ \\
		AGAL015.029--00.669 & $16.87 \pm 0.07$ & $23.3 \asymErr{+7.4}{-2.1}$ & $152 \asymErr{+15}{-37}$ & $103 \asymErr{+11}{-28}$ & $15.9 \asymErr{+1.7}{-4.4}$ \\
		AGAL019.882--00.534 & $16.21 \pm 0.06$ & $30.2 \asymErr{+19.0}{-4.8}$ & $40 \asymErr{+8}{-16}$ & $78 \asymErr{+16}{-29}$ & $7.6 \asymErr{+1.5}{-2.9}$ \\
		AGAL028.861+00.066 & $16.50 \pm 0.04$ & $37.0 \asymErr{+37.5}{-7.7}$ & $211 \asymErr{+56}{-106}$ & $63 \asymErr{+18}{-28}$ & $1.9 \asymErr{+0.5}{-0.8}$ \\
		AGAL030.818--00.056 & $16.39 \pm 0.10$ & $31.5 \asymErr{+22.1}{-5.4}$ & $16 \asymErr{+3}{-6}$ & $75 \asymErr{+16}{-29}$ & $26.9 \asymErr{+5.9}{-10.6}$ \\
		AGAL031.412+00.307 & $16.64 \pm 0.04$ & $31.3 \asymErr{+21.5}{-5.2}$ & $38 \asymErr{+8}{-15}$ & $75 \asymErr{+16}{-29}$ & $20.3 \asymErr{+4.4}{-7.9}$ \\
		AGAL034.258+00.154 & $17.02 \pm 0.05$ & $20.7 \asymErr{+4.3}{-1.2}$ & $72 \asymErr{+4}{-12}$ & $118 \asymErr{+8}{-26}$ & $60.9 \asymErr{+4.1}{-13.5}$ \\
		AGAL034.401+00.226 & $16.43 \pm 0.05$ & $20.7 \asymErr{+4.3}{-1.2}$ & $101 \asymErr{+6}{-17}$ & $118 \asymErr{+8}{-26}$ & $11.3 \asymErr{+0.8}{-2.5}$ \\
		AGAL034.411+00.234 & $16.03 \pm 0.04$ & $20.7 \asymErr{+4.3}{-1.2}$ & $29 \asymErr{+2}{-5}$ & $118 \asymErr{+8}{-26}$ & $15.4 \asymErr{+1.0}{-3.4}$ \\
		AGAL034.821+00.351 & $16.24 \pm 0.05$ & $20.6 \asymErr{+4.2}{-1.2}$ & $180 \asymErr{+11}{-31}$ & $118 \asymErr{+8}{-26}$ & $4.1 \asymErr{+0.3}{-0.9}$ \\
		AGAL035.197--00.742 & $16.35 \pm 0.04$ & $22.4 \asymErr{+6.2}{-1.8}$ & $61 \asymErr{+5}{-13}$ & $108 \asymErr{+10}{-28}$ & $12.8 \asymErr{+1.2}{-3.3}$ \\
		AGAL037.554+00.201 & $16.14 \pm 0.04$ & $30.4 \asymErr{+19.4}{-4.9}$ & $84 \asymErr{+16}{-33}$ & $78 \asymErr{+16}{-29}$ & $3.1 \asymErr{+0.6}{-1.2}$ \\
		AGAL043.166+00.011 & $17.28 \pm 0.05$ & $19.4 \asymErr{+3.0}{-0.8}$ & $141 \asymErr{+6}{-19}$ & $127 \asymErr{+6}{-24}$ & $64.3 \asymErr{+3.0}{-12.3}$ \\
		AGAL049.489--00.389 & $17.15 \pm 0.06$ & $24.0 \asymErr{+8.2}{-2.4}$ & $57 \asymErr{+6}{-14}$ & $101 \asymErr{+12}{-29}$ & $77.6 \asymErr{+9.2}{-22.2}$ \\
		AGAL053.141+00.069 & $16.27 \pm 0.04$ & $19.3 \asymErr{+2.9}{-0.7}$ & $124 \asymErr{+5}{-16}$ & $127 \asymErr{+6}{-24}$ & $7.4 \asymErr{+0.3}{-1.4}$ \\
		AGAL059.782+00.066 & $16.03 \pm 0.04$ & $19.3 \asymErr{+2.9}{-0.7}$ & $75 \asymErr{+3}{-10}$ & $127 \asymErr{+6}{-24}$ & $6.9 \asymErr{+0.3}{-1.3}$ \\
		AGAL301.136--00.226 & $16.63 \pm 0.04$ & $20.6 \asymErr{+4.2}{-1.2}$ & $81 \asymErr{+5}{-14}$ & $118 \asymErr{+8}{-26}$ & $22.3 \asymErr{+1.5}{-4.9}$ \\
		AGAL305.209+00.206 & $16.50 \pm 0.05$ & $21.6 \asymErr{+5.2}{-1.5}$ & $75 \asymErr{+6}{-15}$ & $113 \asymErr{+9}{-27}$ & $16.0 \asymErr{+1.3}{-3.9}$ \\
		AGAL305.562+00.014 & $16.56 \pm 0.09$ & $21.7 \asymErr{+5.3}{-1.5}$ & $317 \asymErr{+24}{-63}$ & $112 \asymErr{+9}{-27}$ & $4.4 \asymErr{+0.4}{-1.1}$ \\
		AGAL309.384--00.134 & $16.25 \pm 0.12$ & $23.5 \asymErr{+7.6}{-2.2}$ & $124 \asymErr{+13}{-30}$ & $103 \asymErr{+11}{-29}$ & $4.6 \asymErr{+0.5}{-1.3}$ \\
		AGAL310.014+00.387 & $16.21 \pm 0.08$ & $22.7 \asymErr{+6.5}{-1.9}$ & $149 \asymErr{+13}{-33}$ & $107 \asymErr{+11}{-28}$ & $3.8 \asymErr{+0.4}{-1.0}$ \\
		AGAL313.576+00.324 & $16.22 \pm 0.09$ & $23.8 \asymErr{+8.0}{-2.3}$ & $180 \asymErr{+19}{-45}$ & $101 \asymErr{+12}{-29}$ & $2.9 \asymErr{+0.3}{-0.8}$ \\
		AGAL316.641--00.087 & $15.81 \pm 0.08$ & $19.2 \asymErr{+2.8}{-0.7}$ & $106 \asymErr{+4}{-14}$ & $128 \asymErr{+5}{-24}$ & $3.0 \asymErr{+0.1}{-0.6}$ \\
		AGAL326.661+00.519 & $16.42 \pm 0.08$ & $21.5 \asymErr{+5.2}{-1.5}$ & $220 \asymErr{+16}{-42}$ & $113 \asymErr{+9}{-27}$ & $4.6 \asymErr{+0.4}{-1.1}$ \\
		AGAL327.119+00.509 & $16.35 \pm 0.08$ & $32.0 \asymErr{+23.2}{-5.6}$ & $157 \asymErr{+33}{-66}$ & $73 \asymErr{+16}{-29}$ & $2.4 \asymErr{+0.5}{-1.0}$ \\
		AGAL327.293--00.579 & $17.11 \pm 0.08$ & $25.6 \asymErr{+10.6}{-3.0}$ & $65 \asymErr{+9}{-19}$ & $93 \asymErr{+13}{-29}$ & $53.3 \asymErr{+7.5}{-16.7}$ \\
		AGAL327.393+00.199 & $15.92 \pm 0.08$ & $32.9 \asymErr{+25.4}{-5.9}$ & $39 \asymErr{+9}{-17}$ & $71 \asymErr{+17}{-29}$ & $3.3 \asymErr{+0.8}{-1.4}$ \\
		AGAL328.809+00.632 & $17.04 \pm 0.08$ & $25.4 \asymErr{+10.3}{-2.9}$ & $164 \asymErr{+21}{-47}$ & $94 \asymErr{+13}{-29}$ & $18.1 \asymErr{+2.5}{-5.6}$ \\
		AGAL329.066--00.307 & $15.71 \pm 0.08$ & $25.5 \asymErr{+10.5}{-2.9}$ & $29 \asymErr{+4}{-8}$ & $94 \asymErr{+13}{-29}$ & $4.8 \asymErr{+0.7}{-1.5}$ \\
		AGAL330.879--00.367 & $16.98 \pm 0.06$ & $29.9 \asymErr{+18.3}{-4.7}$ & $151 \asymErr{+28}{-58}$ & $79 \asymErr{+16}{-29}$ & $12.3 \asymErr{+2.4}{-4.6}$ \\
		AGAL330.954--00.182 & $17.24 \pm 0.07$ & $34.2 \asymErr{+28.8}{-6.5}$ & $89 \asymErr{+21}{-41}$ & $68 \asymErr{+17}{-29}$ & $28.7 \asymErr{+7.1}{-12.1}$ \\
		AGAL332.094--00.421 & $16.56 \pm 0.05$ & $28.2 \asymErr{+15.1}{-4.0}$ & $147 \asymErr{+24}{-51}$ & $84 \asymErr{+15}{-30}$ & $5.4 \asymErr{+1.0}{-1.9}$ \\
		AGAL332.826--00.549 & $17.13 \pm 0.07$ & $28.4 \asymErr{+15.4}{-4.1}$ & $137 \asymErr{+23}{-48}$ & $84 \asymErr{+15}{-30}$ & $21.6 \asymErr{+3.8}{-7.6}$ \\
		AGAL333.134--00.431 & $17.10 \pm 0.06$ & $28.5 \asymErr{+15.5}{-4.1}$ & $174 \asymErr{+29}{-61}$ & $83 \asymErr{+15}{-30}$ & $15.6 \asymErr{+2.8}{-5.5}$ \\
		AGAL333.284--00.387 & $16.81 \pm 0.09$ & $28.5 \asymErr{+15.6}{-4.1}$ & $129 \asymErr{+22}{-46}$ & $83 \asymErr{+15}{-30}$ & $10.7 \asymErr{+1.9}{-3.8}$ \\
		AGAL333.314+00.106 & $16.10 \pm 0.09$ & $28.5 \asymErr{+15.6}{-4.1}$ & $76 \asymErr{+13}{-27}$ & $83 \asymErr{+15}{-30}$ & $3.5 \asymErr{+0.6}{-1.2}$ \\
		AGAL333.604--00.212 & $16.86 \pm 0.06$ & $28.6 \asymErr{+15.7}{-4.1}$ & $82 \asymErr{+14}{-29}$ & $83 \asymErr{+15}{-30}$ & $18.9 \asymErr{+3.4}{-6.7}$ \\
		AGAL337.406--00.402 & $16.98 \pm 0.08$ & $27.9 \asymErr{+14.6}{-3.9}$ & $148 \asymErr{+24}{-51}$ & $85 \asymErr{+15}{-30}$ & $14.3 \asymErr{+2.5}{-5.0}$ \\
		AGAL337.704--00.054 & $16.65 \pm 0.08$ & $28.9 \asymErr{+16.3}{-4.3}$ & $69 \asymErr{+12}{-25}$ & $82 \asymErr{+15}{-30}$ & $13.6 \asymErr{+2.5}{-4.9}$ \\
		AGAL337.916--00.477 & $17.09 \pm 0.06$ & $27.8 \asymErr{+14.2}{-3.8}$ & $161 \asymErr{+26}{-54}$ & $86 \asymErr{+15}{-30}$ & $17.4 \asymErr{+3.0}{-6.0}$ \\
		AGAL339.623--00.122 & $16.20 \pm 0.08$ & $27.2 \asymErr{+13.2}{-3.6}$ & $131 \asymErr{+20}{-43}$ & $88 \asymErr{+14}{-30}$ & $2.9 \asymErr{+0.5}{-1.0}$ \\
		AGAL340.746--01.001 & $16.02 \pm 0.08$ & $26.3 \asymErr{+11.7}{-3.2}$ & $98 \asymErr{+14}{-30}$ & $91 \asymErr{+14}{-29}$ & $2.7 \asymErr{+0.4}{-0.9}$ \\
		AGAL341.217--00.212 & $16.24 \pm 0.07$ & $30.5 \asymErr{+19.6}{-4.9}$ & $78 \asymErr{+15}{-30}$ & $77 \asymErr{+16}{-29}$ & $4.2 \asymErr{+0.9}{-1.6}$ \\
		AGAL343.128--00.062 & $16.77 \pm 0.07$ & $27.8 \asymErr{+14.3}{-3.8}$ & $87 \asymErr{+14}{-30}$ & $86 \asymErr{+15}{-30}$ & $15.3 \asymErr{+2.6}{-5.3}$ \\
		AGAL345.003--00.224 & $16.61 \pm 0.08$ & $27.9 \asymErr{+14.5}{-3.9}$ & $63 \asymErr{+10}{-21}$ & $85 \asymErr{+15}{-30}$ & $14.5 \asymErr{+2.5}{-5.0}$ \\
		AGAL345.488+00.314 & $16.79 \pm 0.05$ & $24.4 \asymErr{+8.8}{-2.5}$ & $122 \asymErr{+14}{-32}$ & $99 \asymErr{+12}{-29}$ & $15.1 \asymErr{+1.9}{-4.4}$ \\
		AGAL345.504+00.347 & $16.72 \pm 0.08$ & $24.5 \asymErr{+8.9}{-2.6}$ & $168 \asymErr{+20}{-45}$ & $98 \asymErr{+12}{-29}$ & $9.2 \asymErr{+1.2}{-2.7}$ \\
		AGAL345.718+00.817 & $16.11 \pm 0.09$ & $21.7 \asymErr{+5.4}{-1.6}$ & $87 \asymErr{+7}{-17}$ & $112 \asymErr{+9}{-27}$ & $5.6 \asymErr{+0.5}{-1.4}$ \\
		AGAL351.161+00.697 & $16.71 \pm 0.07$ & $22.9 \asymErr{+6.8}{-2.0}$ & $48 \asymErr{+5}{-11}$ & $106 \asymErr{+11}{-28}$ & $36.2 \asymErr{+3.7}{-9.6}$ \\
		AGAL351.244+00.669 & $16.93 \pm 0.07$ & $22.9 \asymErr{+6.8}{-2.0}$ & $169 \asymErr{+16}{-39}$ & $106 \asymErr{+11}{-28}$ & $17.0 \asymErr{+1.7}{-4.5}$ \\
		AGAL351.416+00.646 & $16.99 \pm 0.10$ & $21.0 \asymErr{+4.6}{-1.3}$ & $70 \asymErr{+5}{-13}$ & $116 \asymErr{+8}{-27}$ & $56.9 \asymErr{+4.1}{-13.0}$ \\
		AGAL351.581--00.352 & $16.88 \pm 0.04$ & $54.3 \asymErr{+140.3}{-16.0}$ & $34 \asymErr{+14}{-24}$ & $41 \asymErr{+18}{-23}$ & $12.4 \asymErr{+5.5}{-6.9}$ \\
		AGAL351.774--00.537 & $17.13 \pm 0.06$ & $19.8 \asymErr{+3.4}{-0.9}$ & $105 \asymErr{+5}{-15}$ & $124 \asymErr{+7}{-25}$ & $59.3 \asymErr{+3.1}{-11.9}$ \\
		AGAL353.409--00.361 & $16.93 \pm 0.08$ & $30.8 \asymErr{+20.4}{-5.1}$ & $88 \asymErr{+17}{-35}$ & $76 \asymErr{+16}{-29}$ & $17.5 \asymErr{+3.7}{-6.7}$ \\
		\end{longtable}
	}
\clearpage
\section{Spectral energy distributions}
	\begin{figure*}
		\centering
		\includegraphics[width=0.32\textwidth]{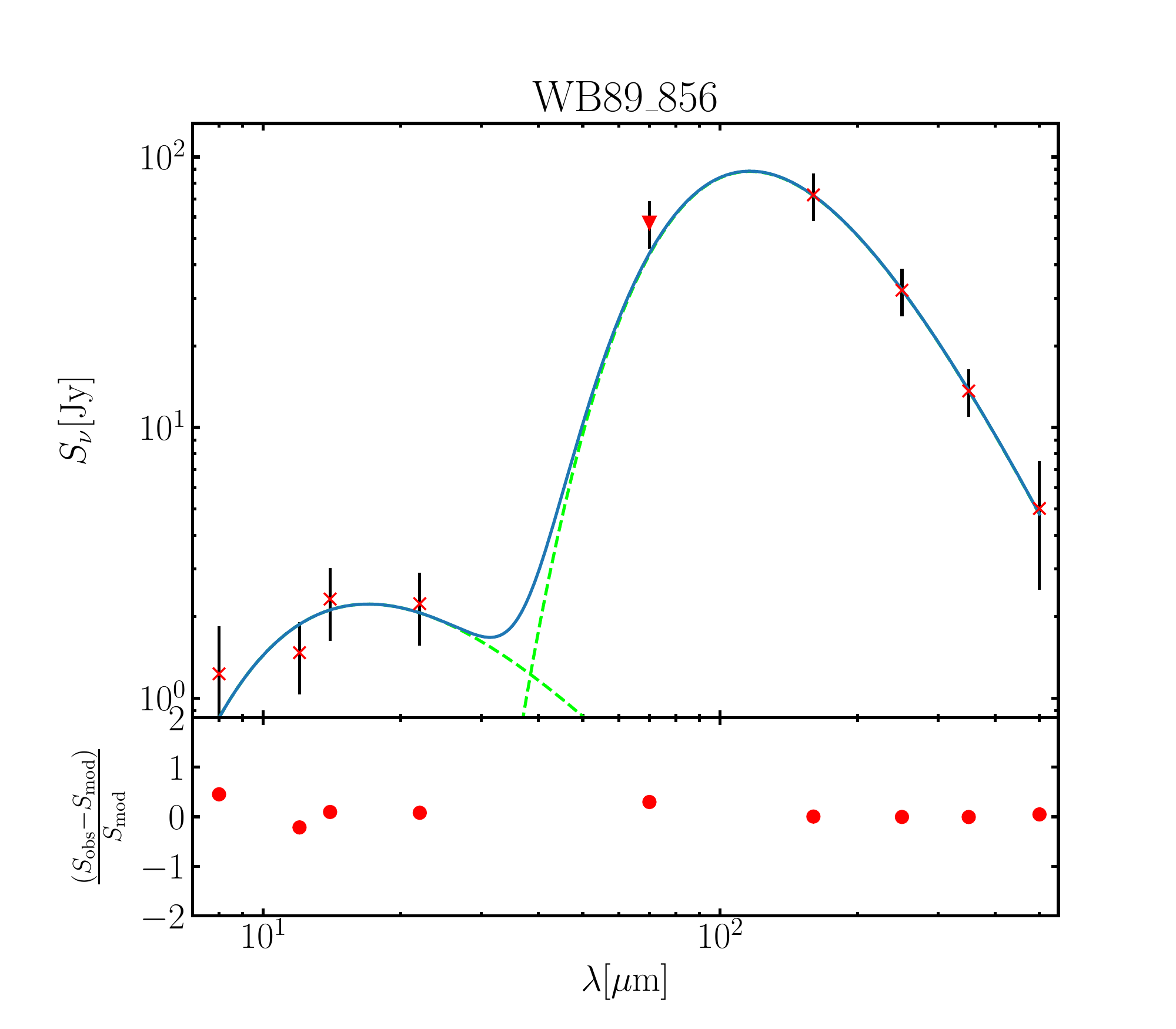}
		\includegraphics[width=0.32\textwidth]{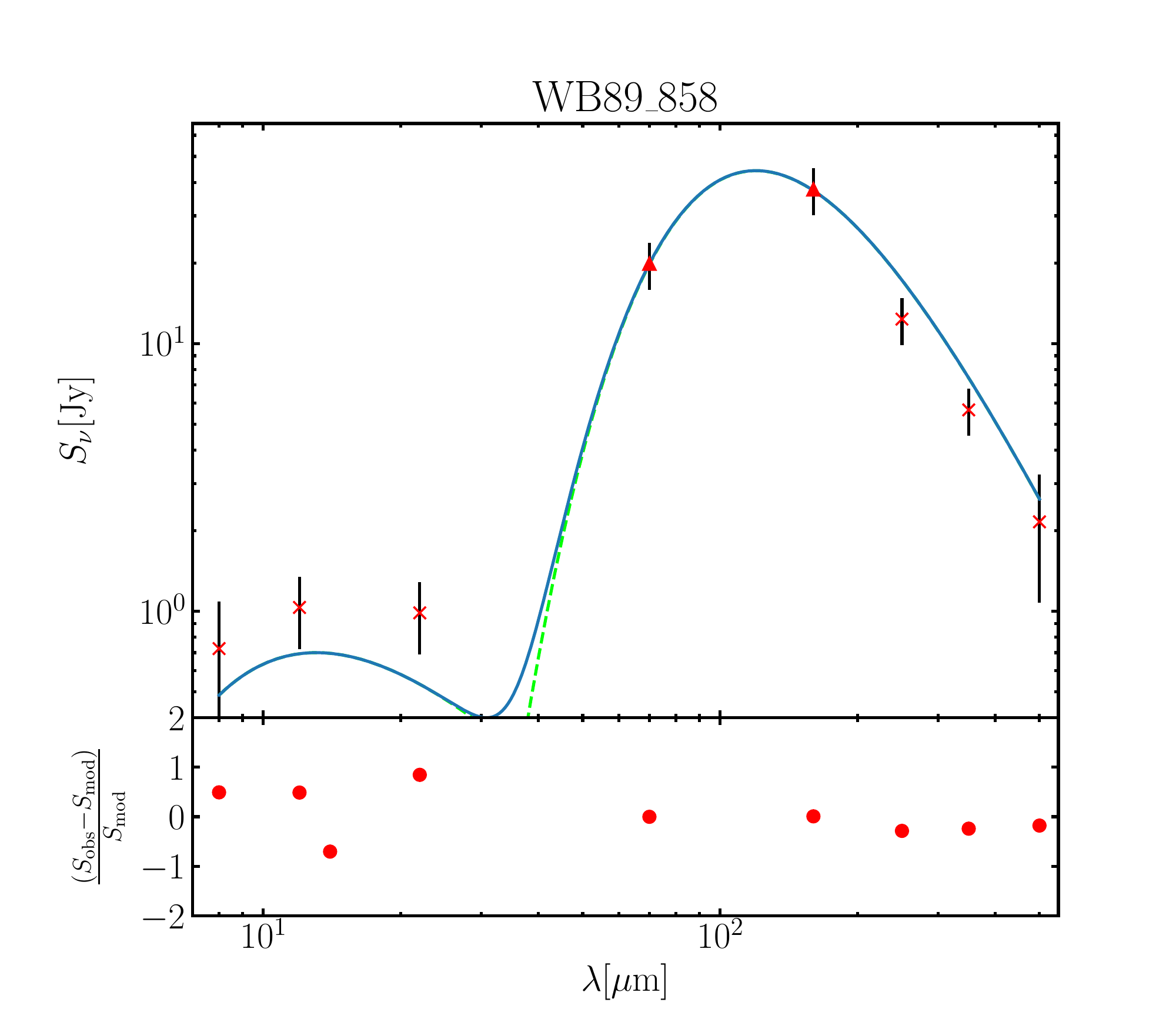}
		\includegraphics[width=0.32\textwidth]{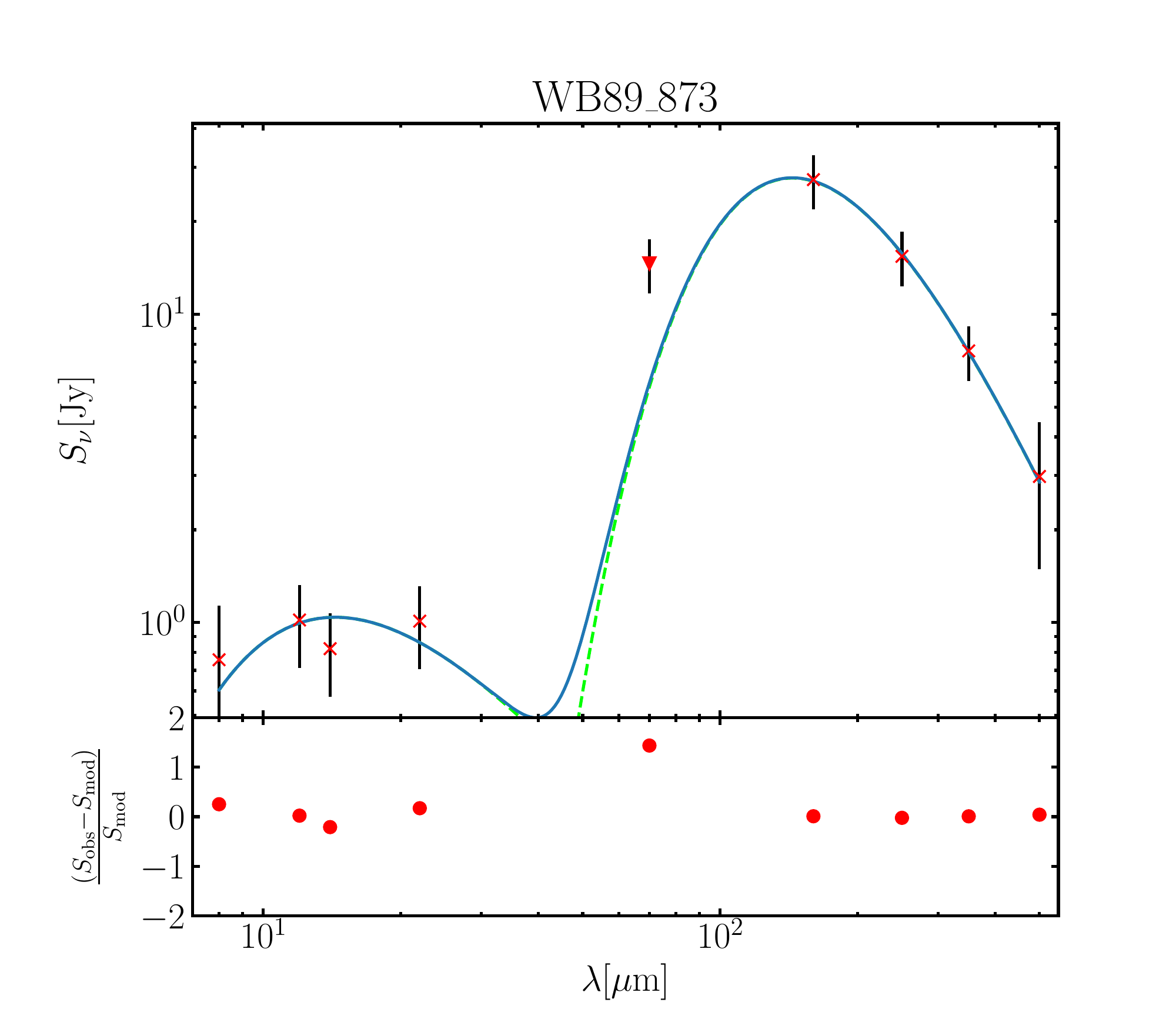}\\
		\includegraphics[width=0.32\textwidth]{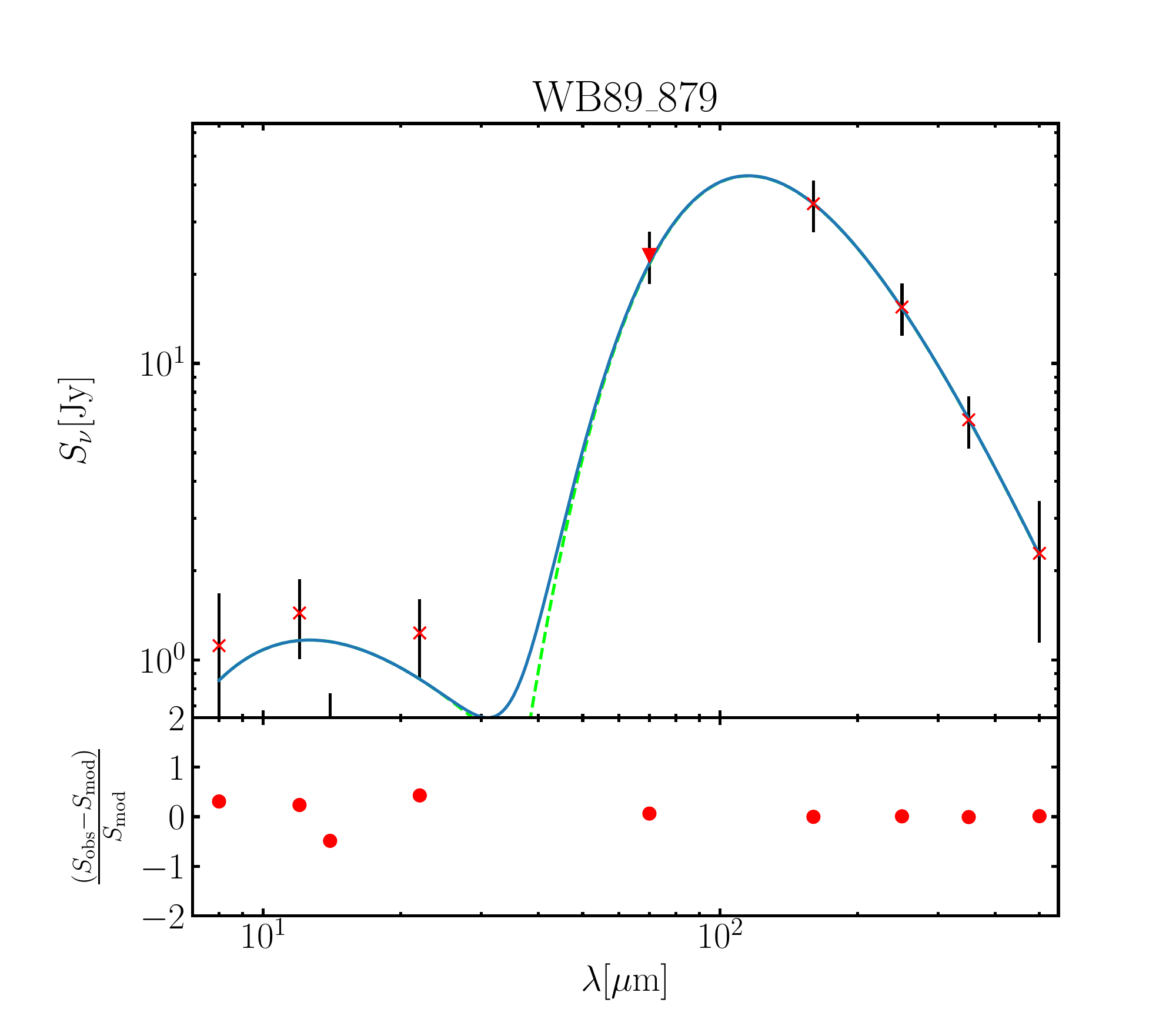}
		\includegraphics[width=0.32\textwidth]{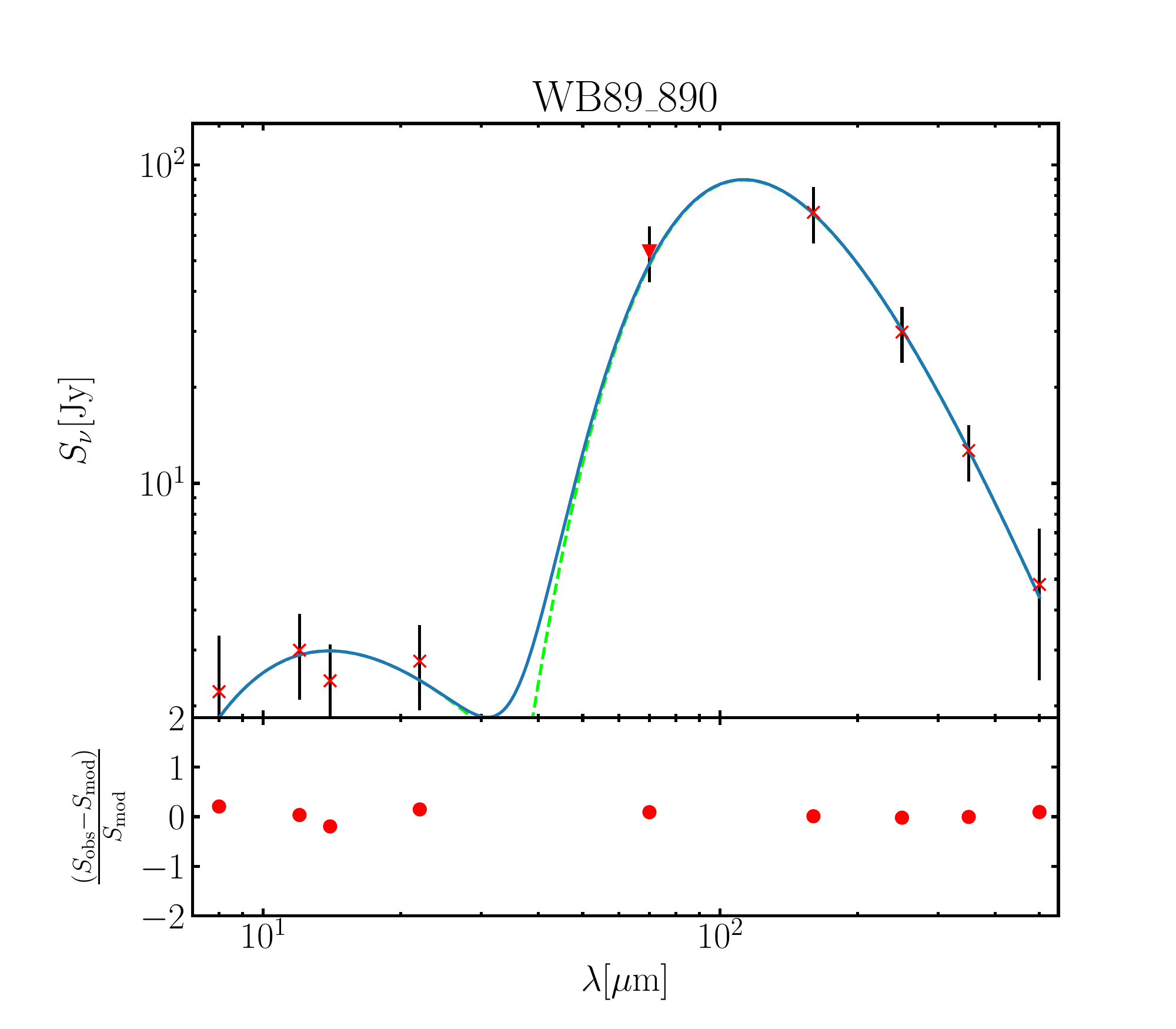}	
		\includegraphics[width=0.32\textwidth]{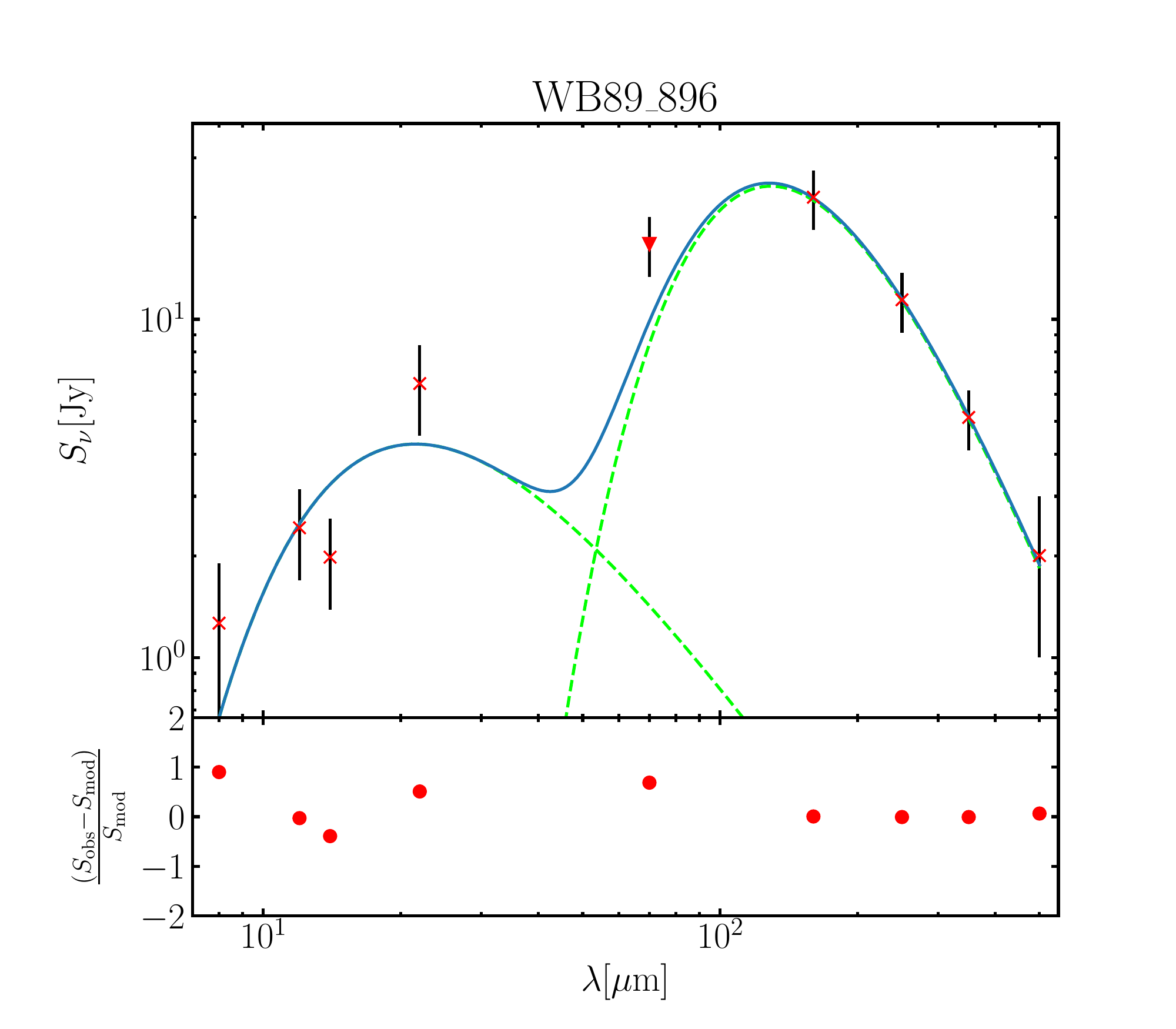}\\
		\includegraphics[width=0.32\textwidth]{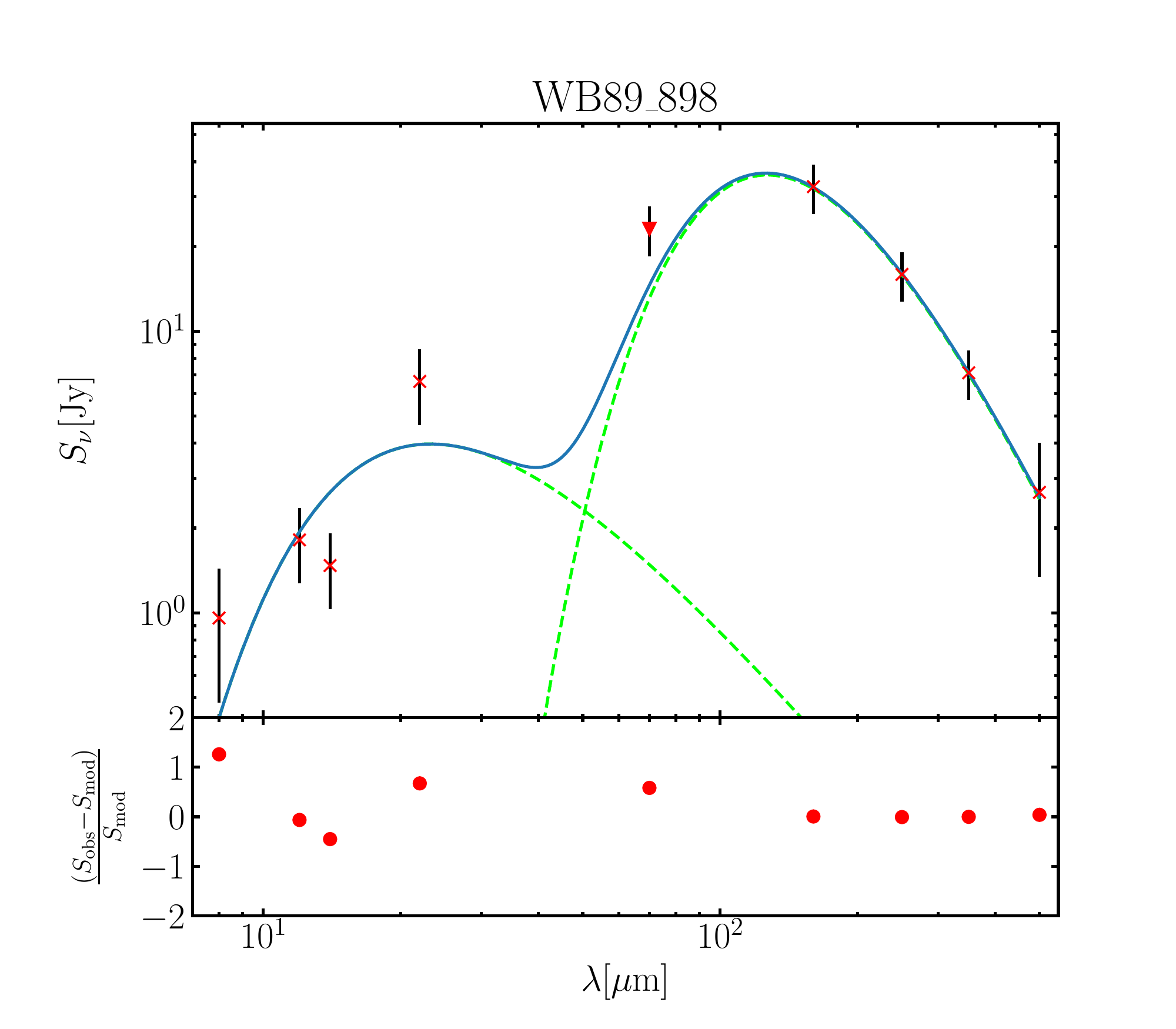}
		\includegraphics[width=0.32\textwidth]{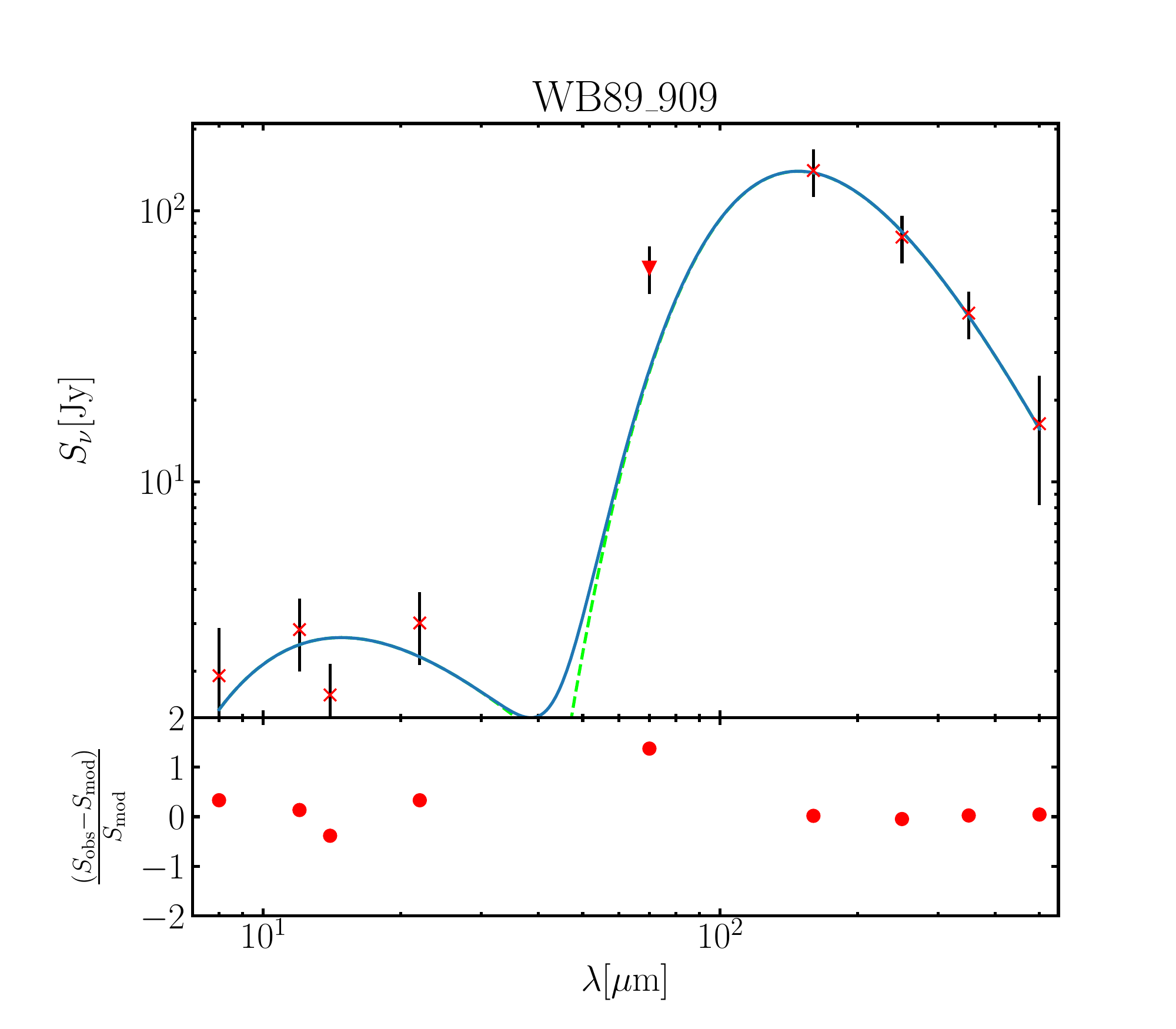}	
		\includegraphics[width=0.32\textwidth]{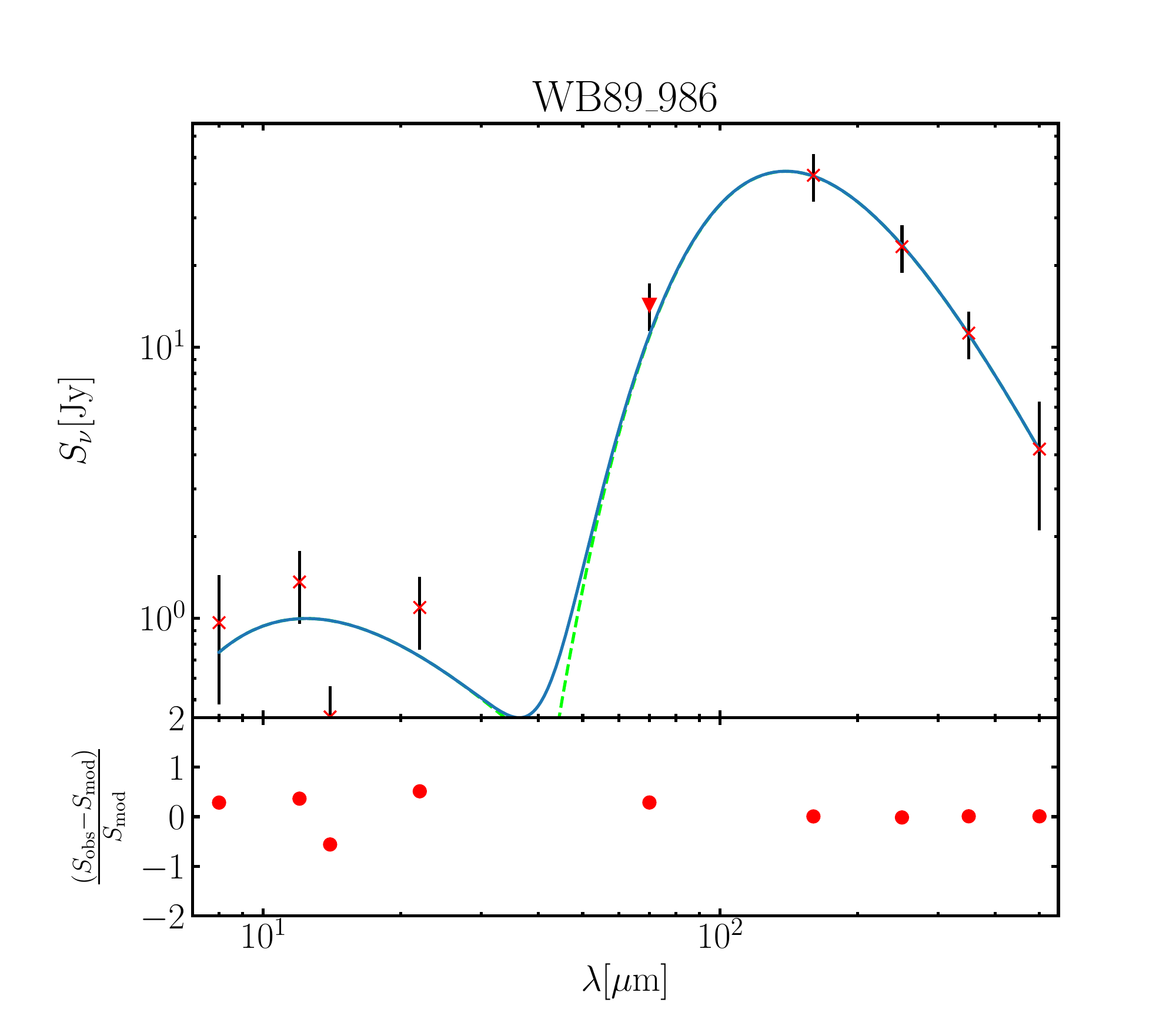}\\
		\includegraphics[width=0.32\textwidth]{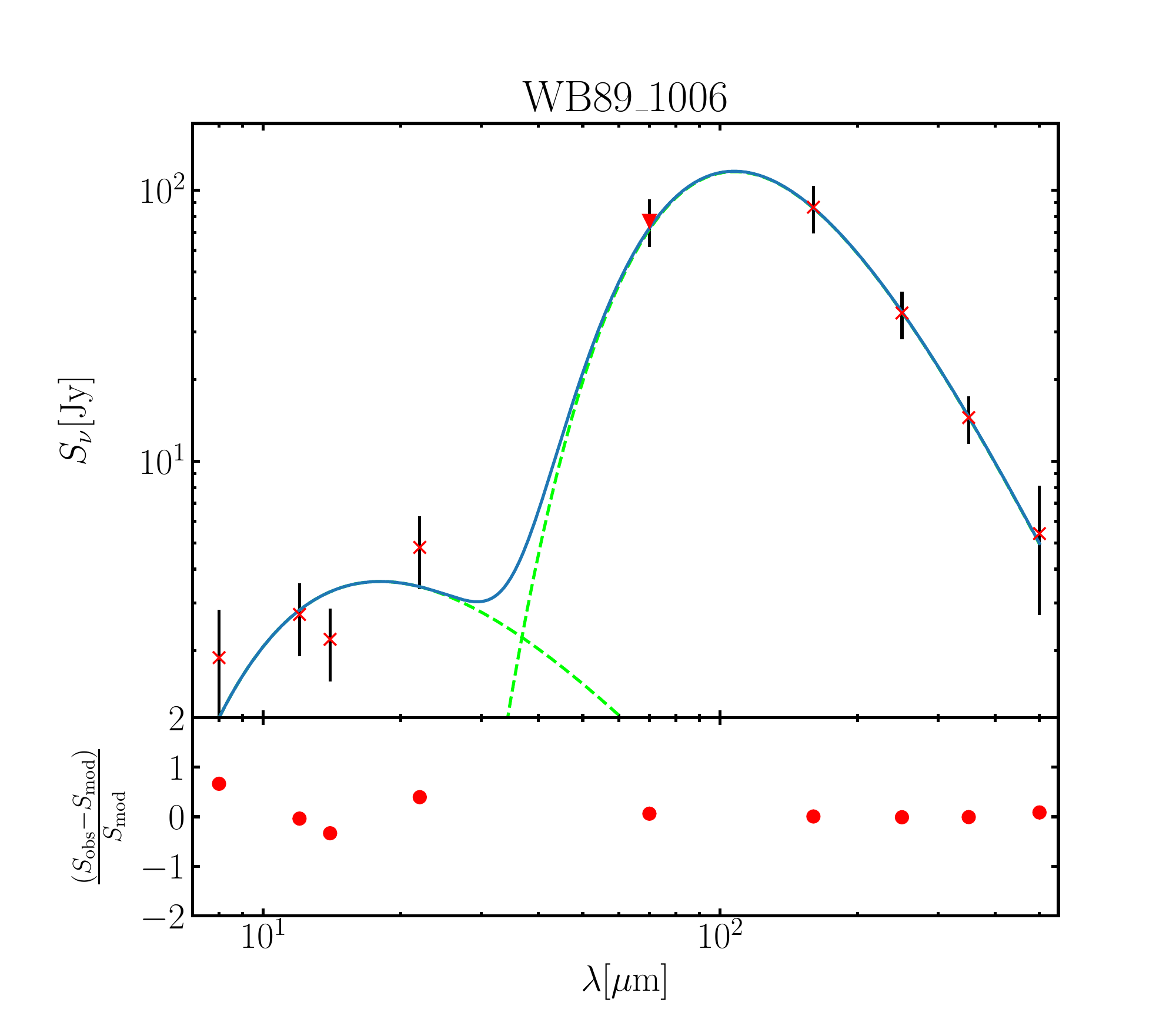}
		\includegraphics[width=0.32\textwidth]{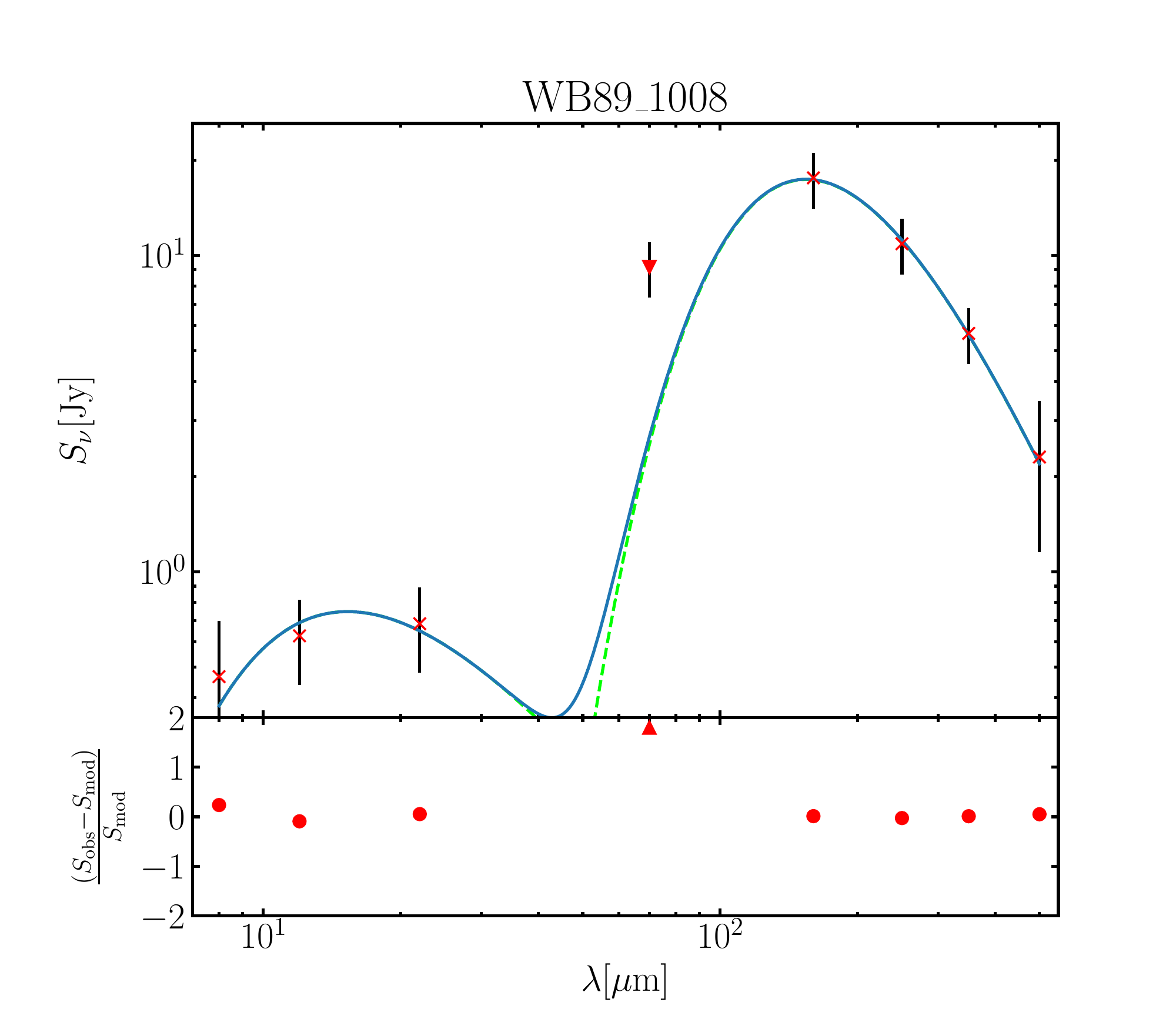}		
		\includegraphics[width=0.32\textwidth]{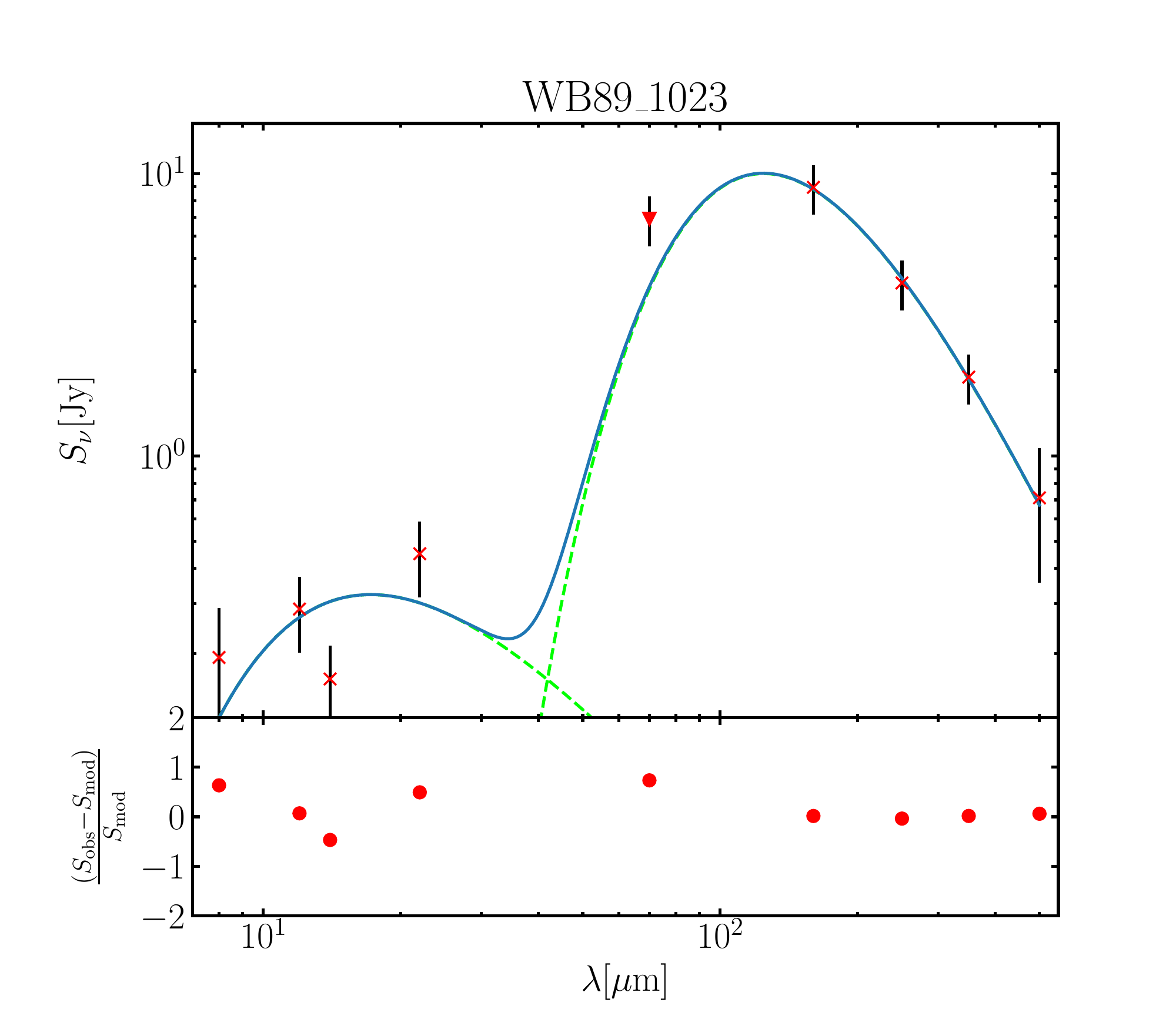}\\
		\caption{For each source we show the SED fit (top panel) and the residuals $(S_{\mathrm{obs}} - S_{\mathrm{mod}}) / S_{\mathrm{obs}}$ (bottom panel). In the top panel, the extracted fluxes are indicated by the red crosses, and upper and lower limits are indicated by triangles pointing downwards and upwards, respectively. The best fit curve is indicated in blue, and the separate contribution of the grey- and black-body is shown by the green dashed lines. 
		If the residuals in the bottom panel exceed |2|, the point is indicated by a triangle.\label{fig:seds}}
	\end{figure*}
	\begin{figure*}
		\ContinuedFloat
		\centering
		\includegraphics[width=0.32\textwidth]{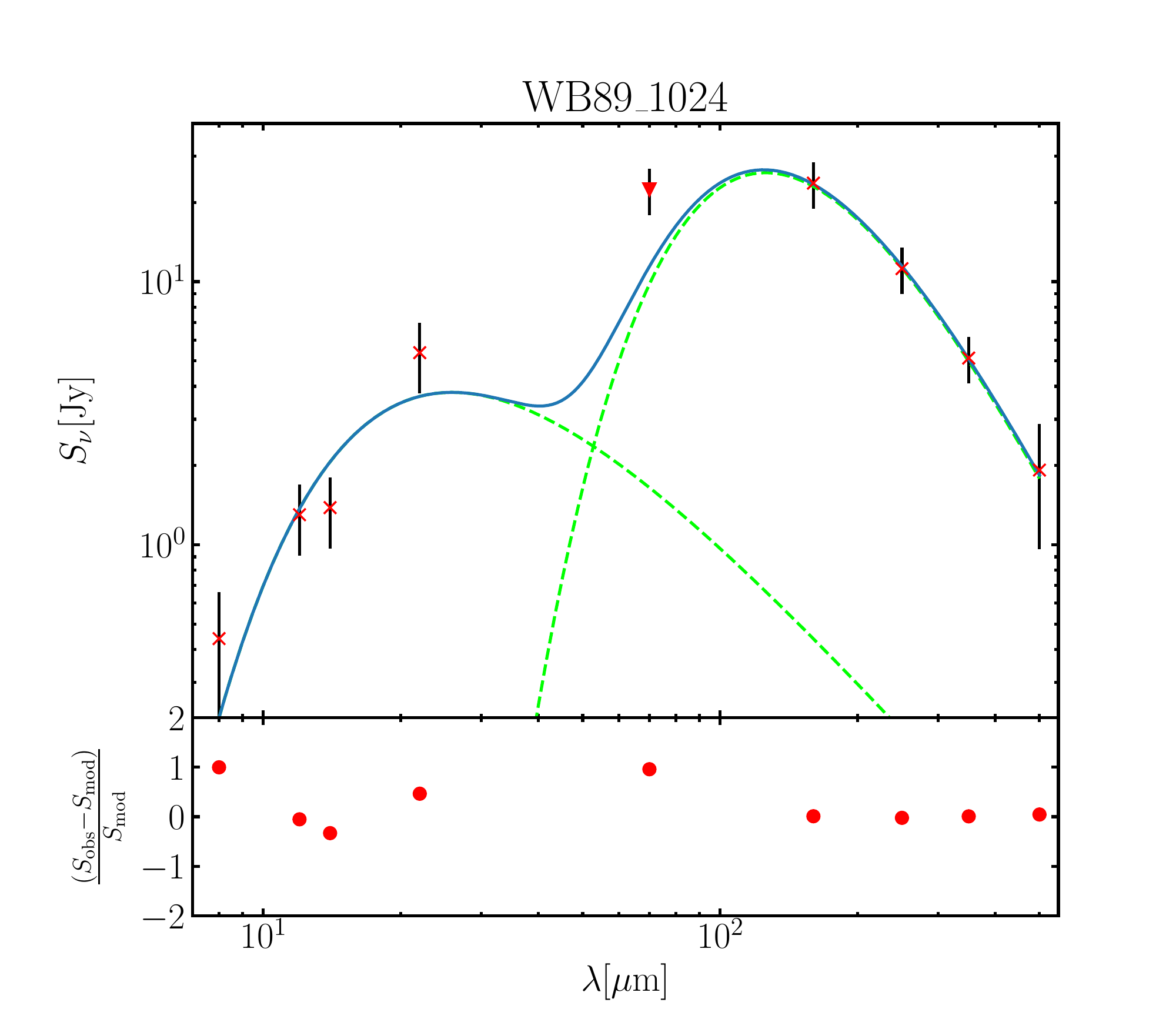}
		\includegraphics[width=0.32\textwidth]{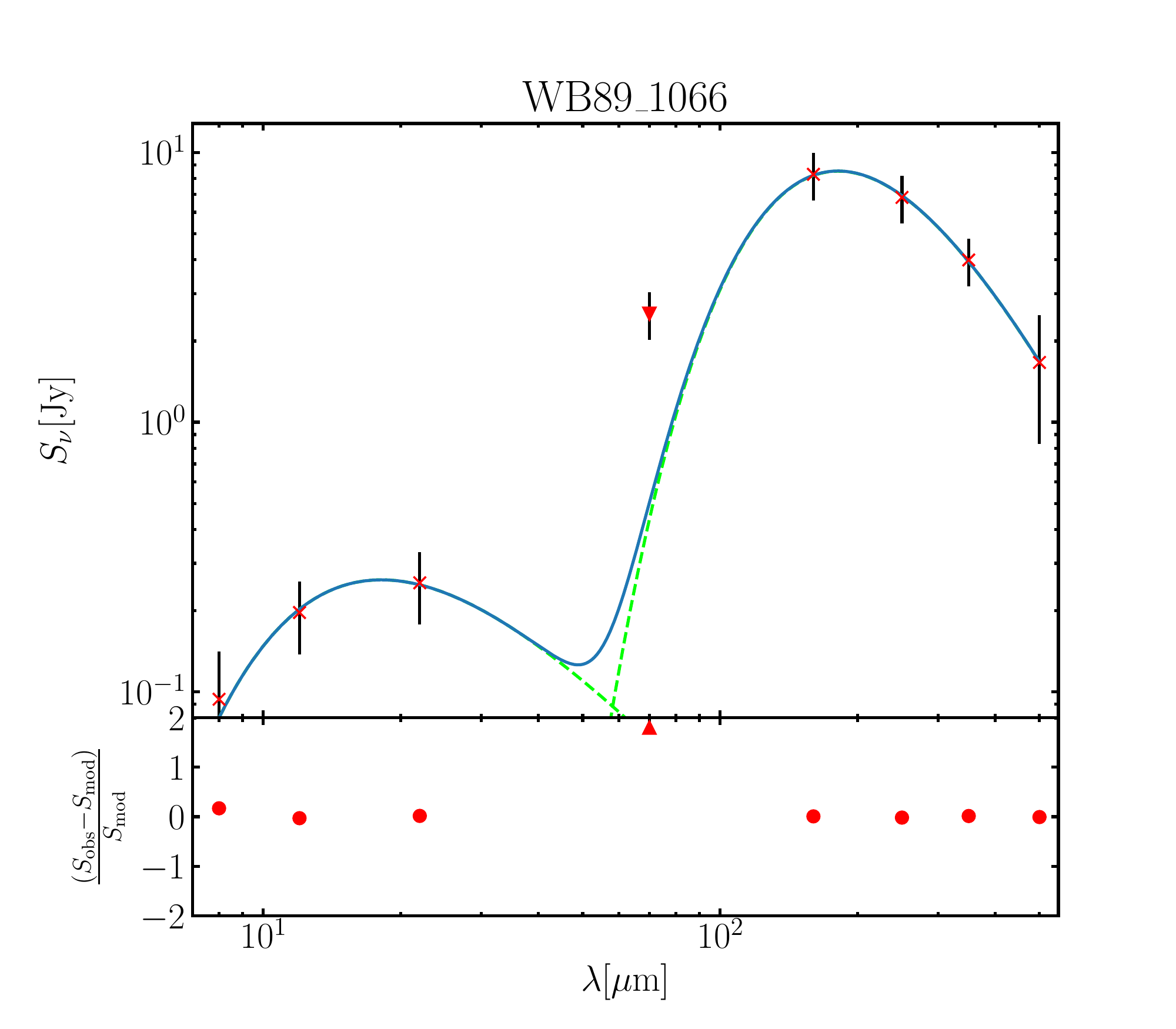}
		\includegraphics[width=0.32\textwidth]{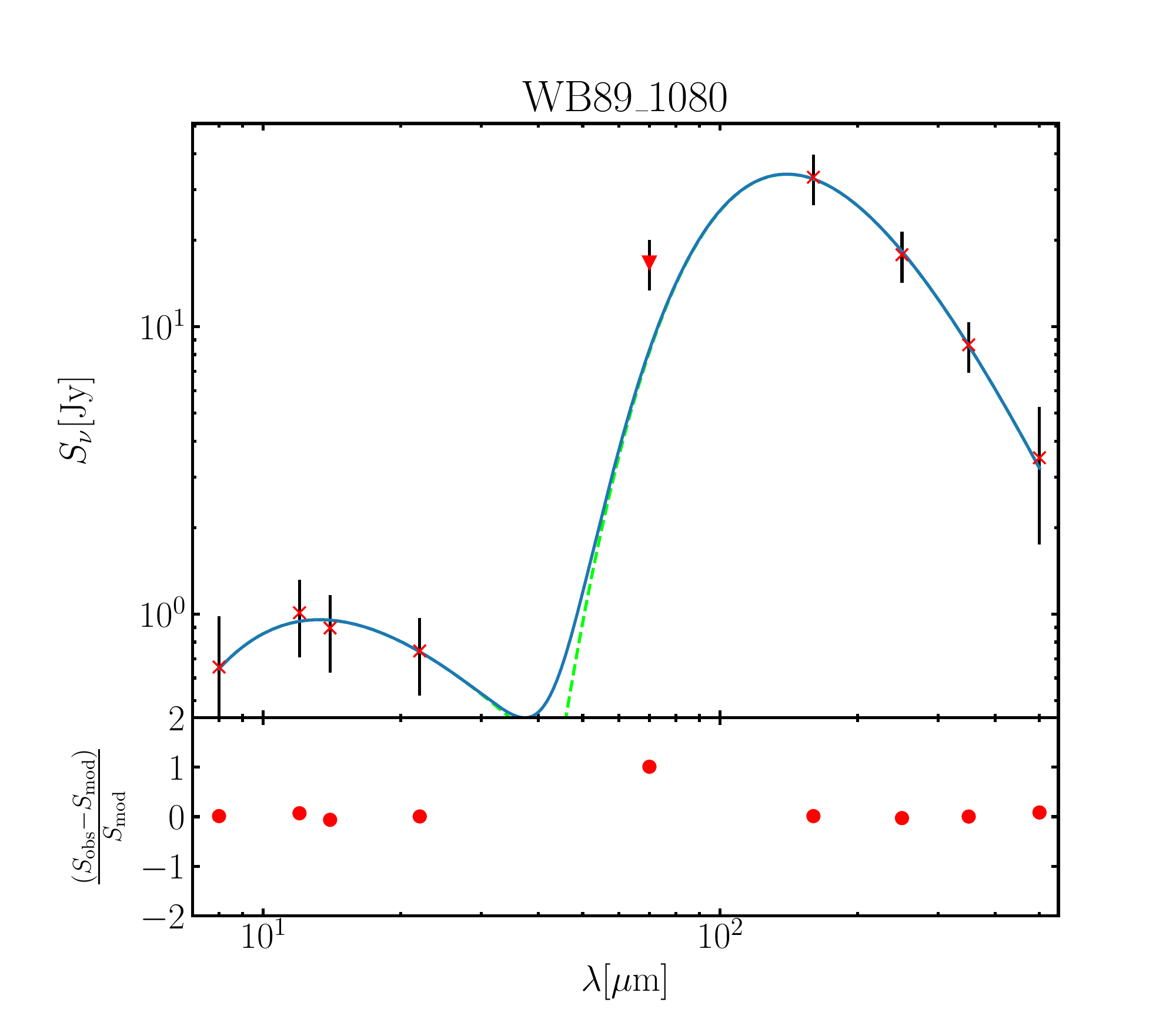}\\
		\includegraphics[width=0.32\textwidth]{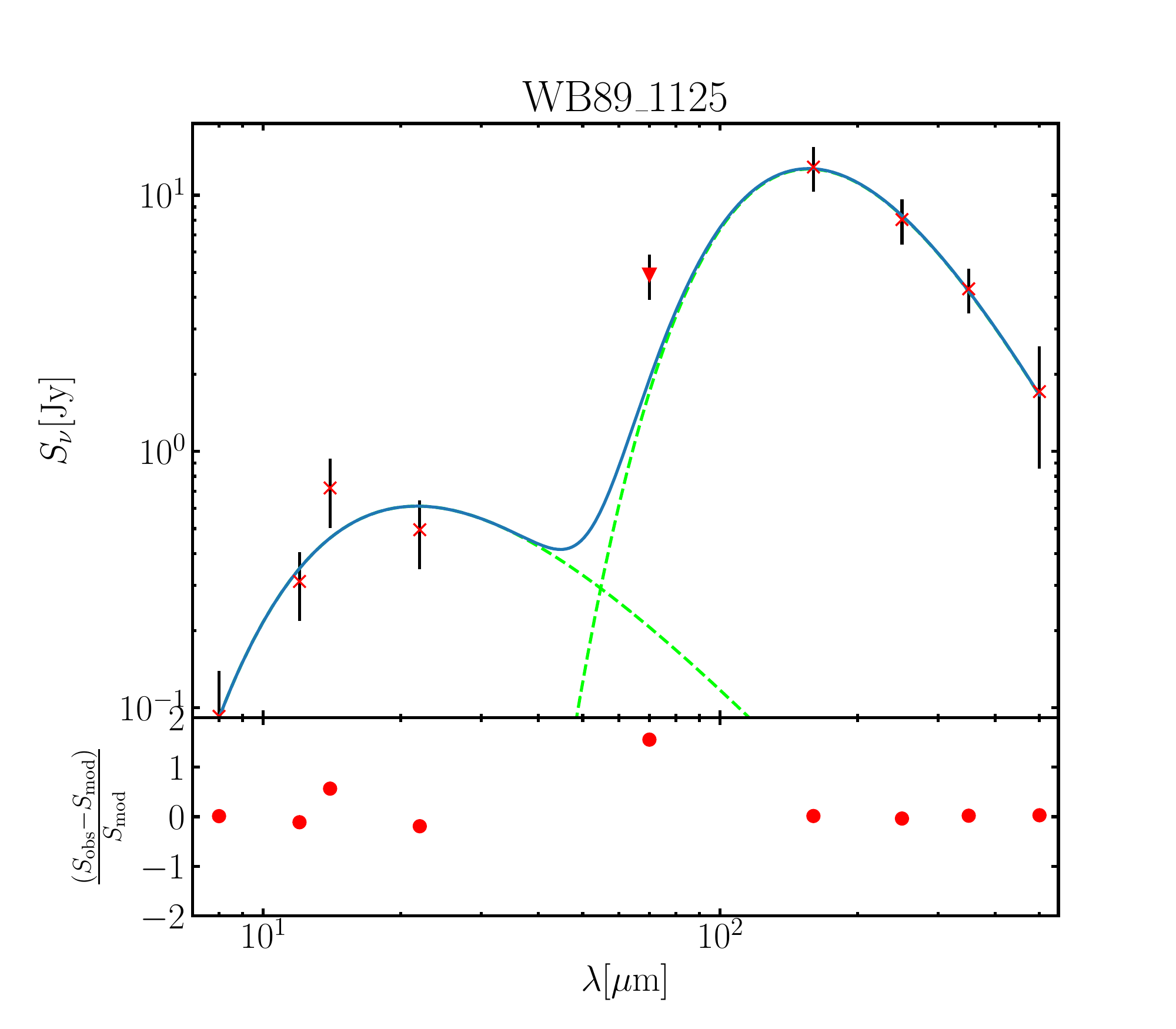}
		\includegraphics[width=0.32\textwidth]{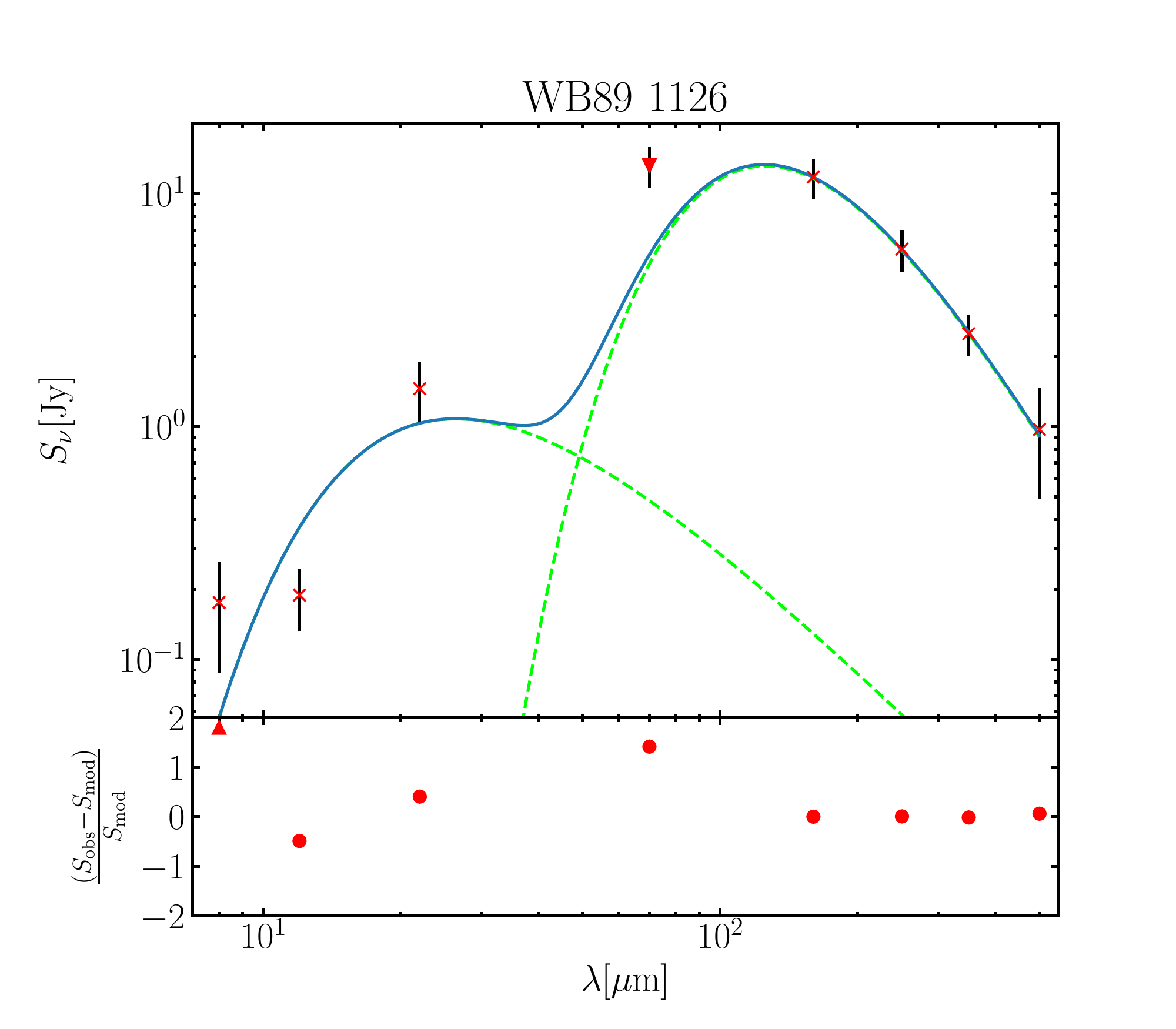}
		\includegraphics[width=0.32\textwidth]{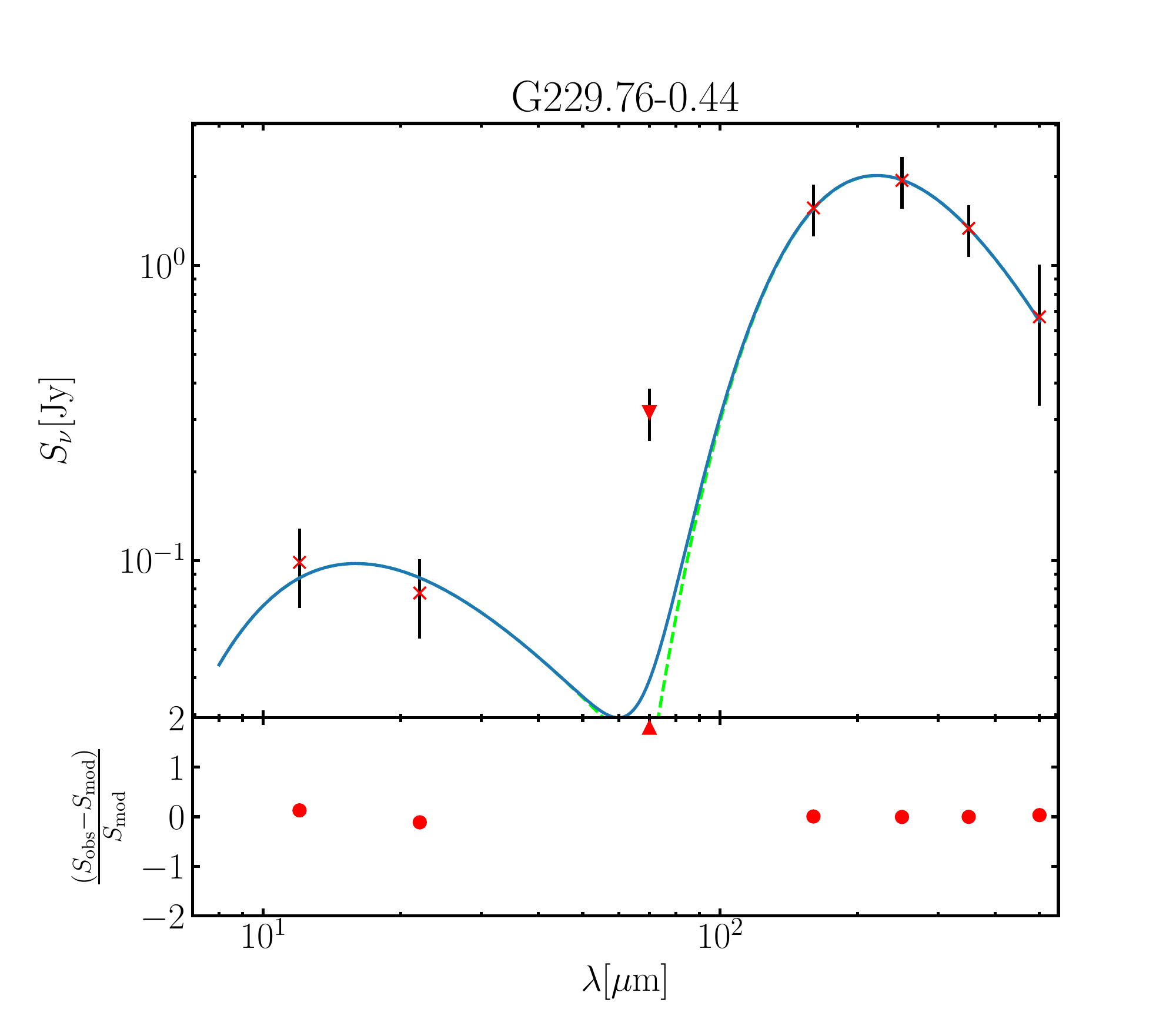}\\
		\includegraphics[width=0.32\textwidth]{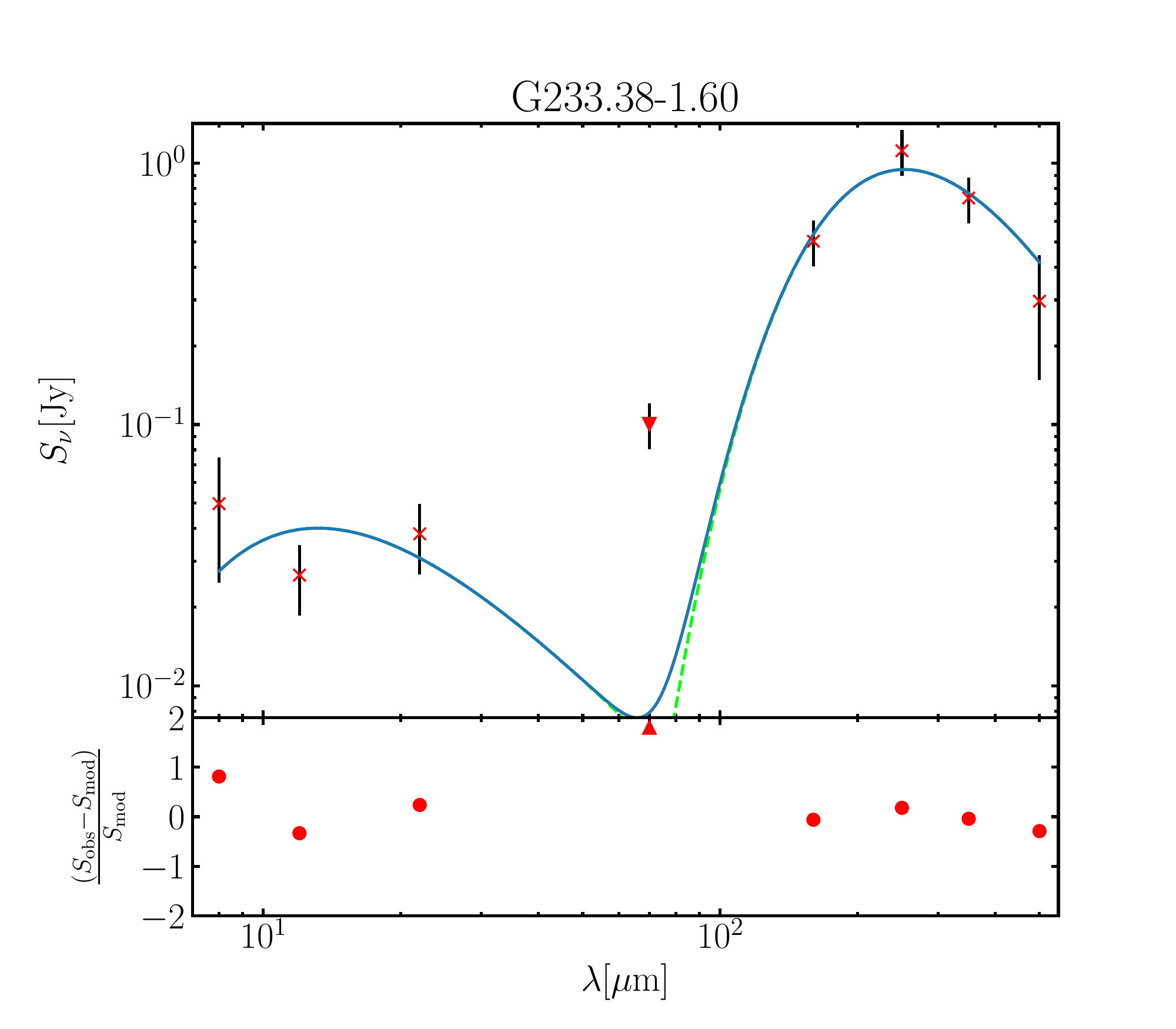}
		\includegraphics[width=0.32\textwidth]{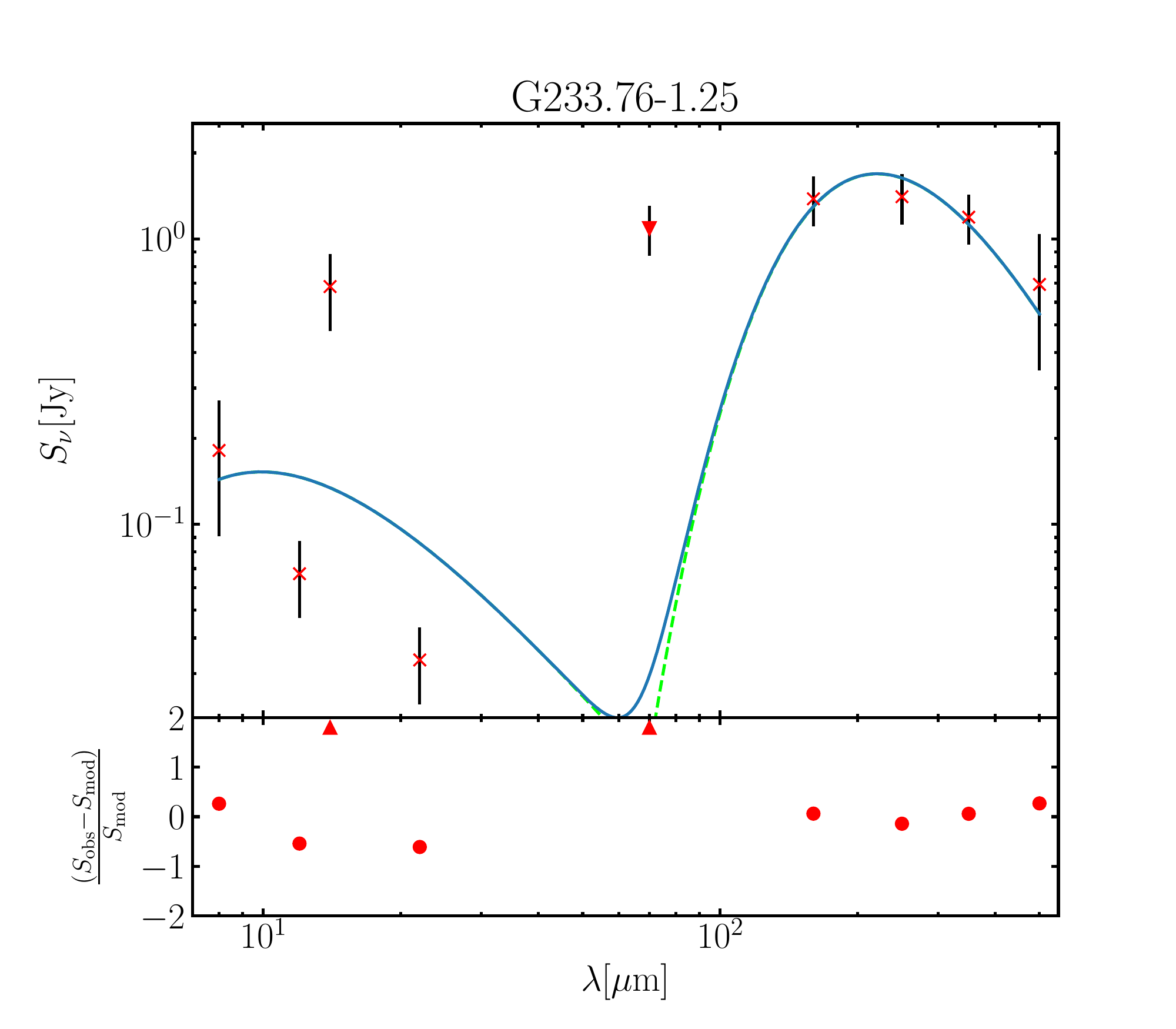}
		\includegraphics[width=0.32\textwidth]{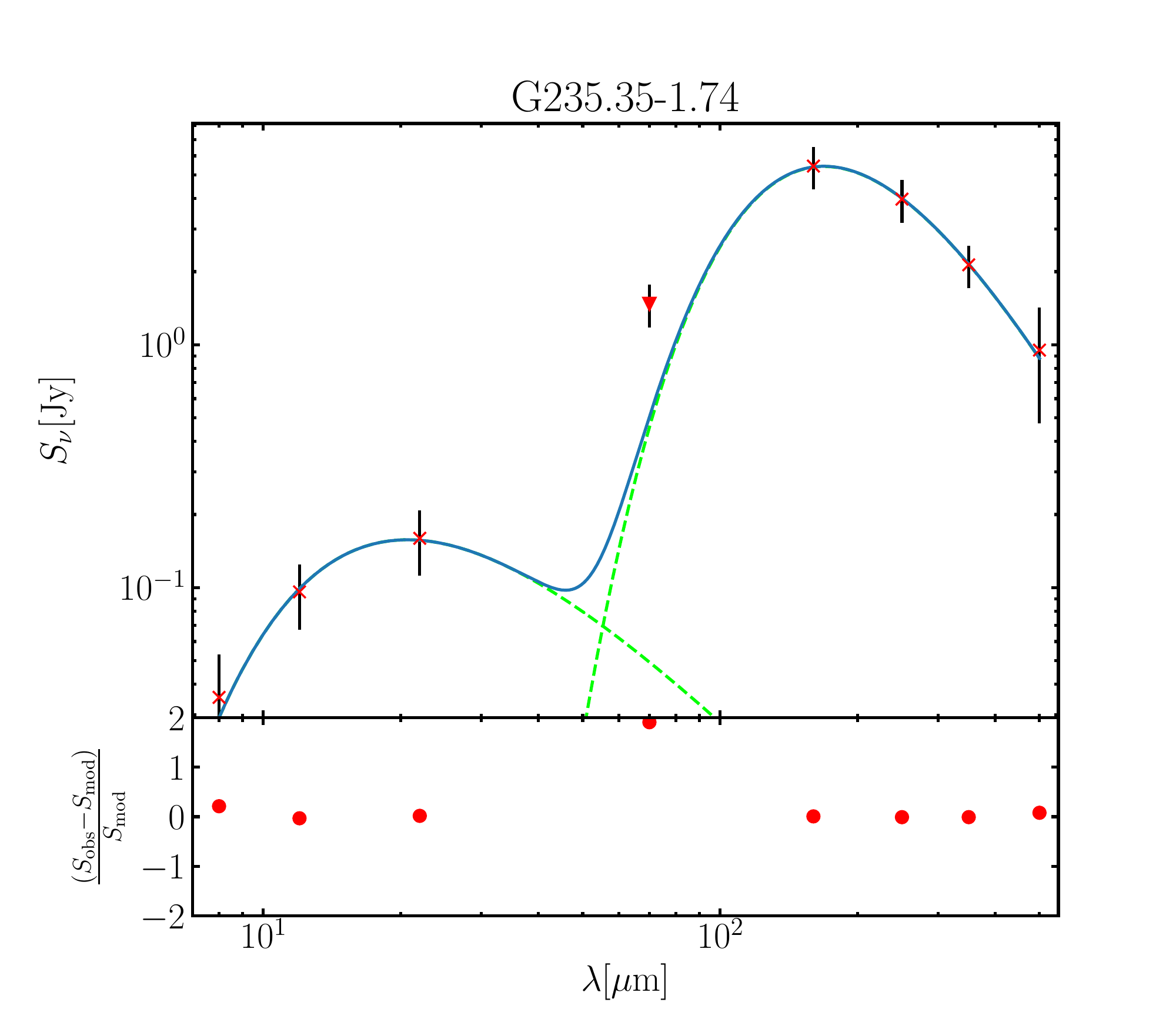}\\
		\includegraphics[width=0.32\textwidth]{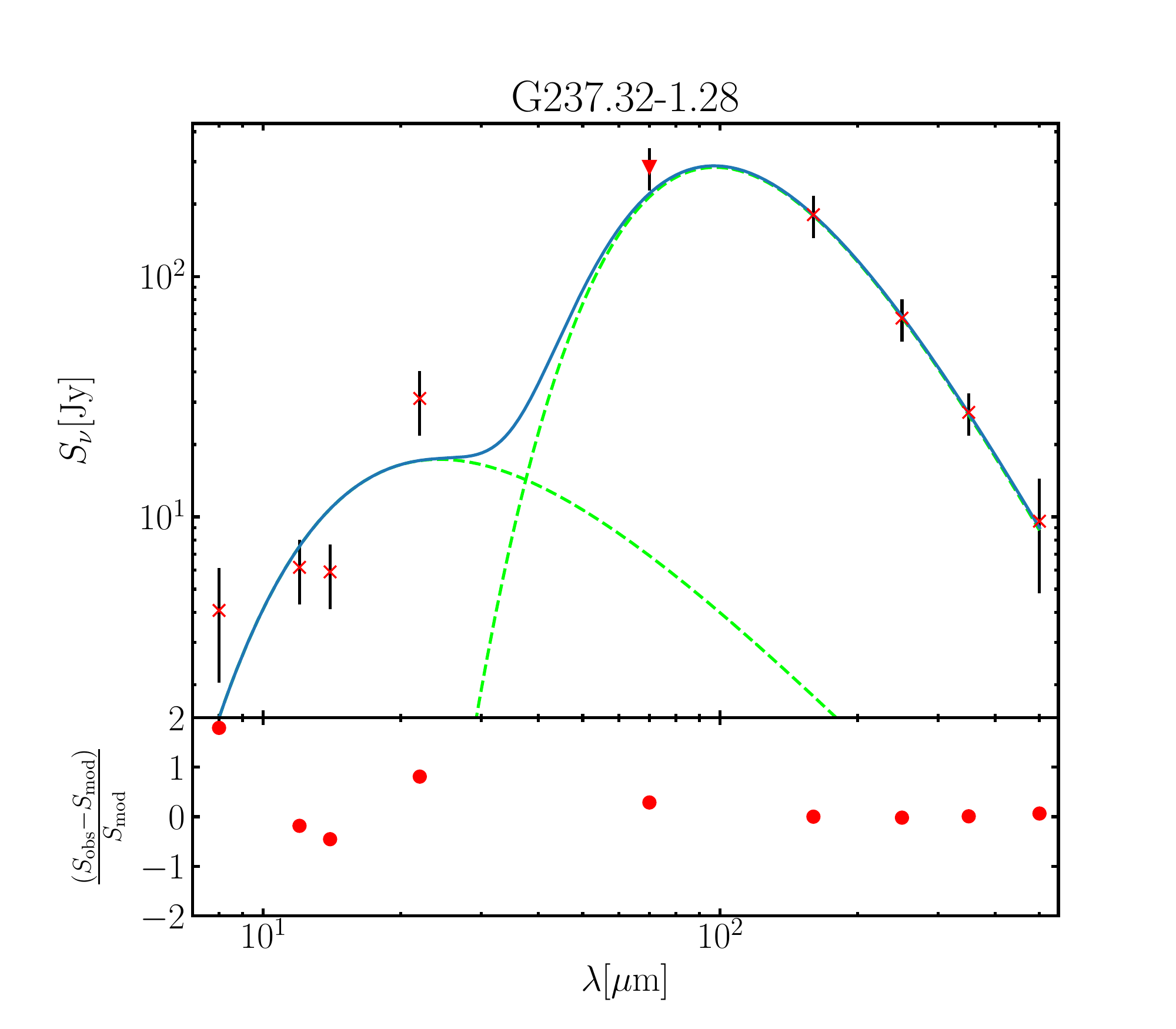}
		\caption{Continued.}
	\end{figure*}

\section{Spectra for sources in the far outer Galaxy}\label{app:spectra}
	\begin{figure*}
		\includegraphics[width=0.3\textwidth]{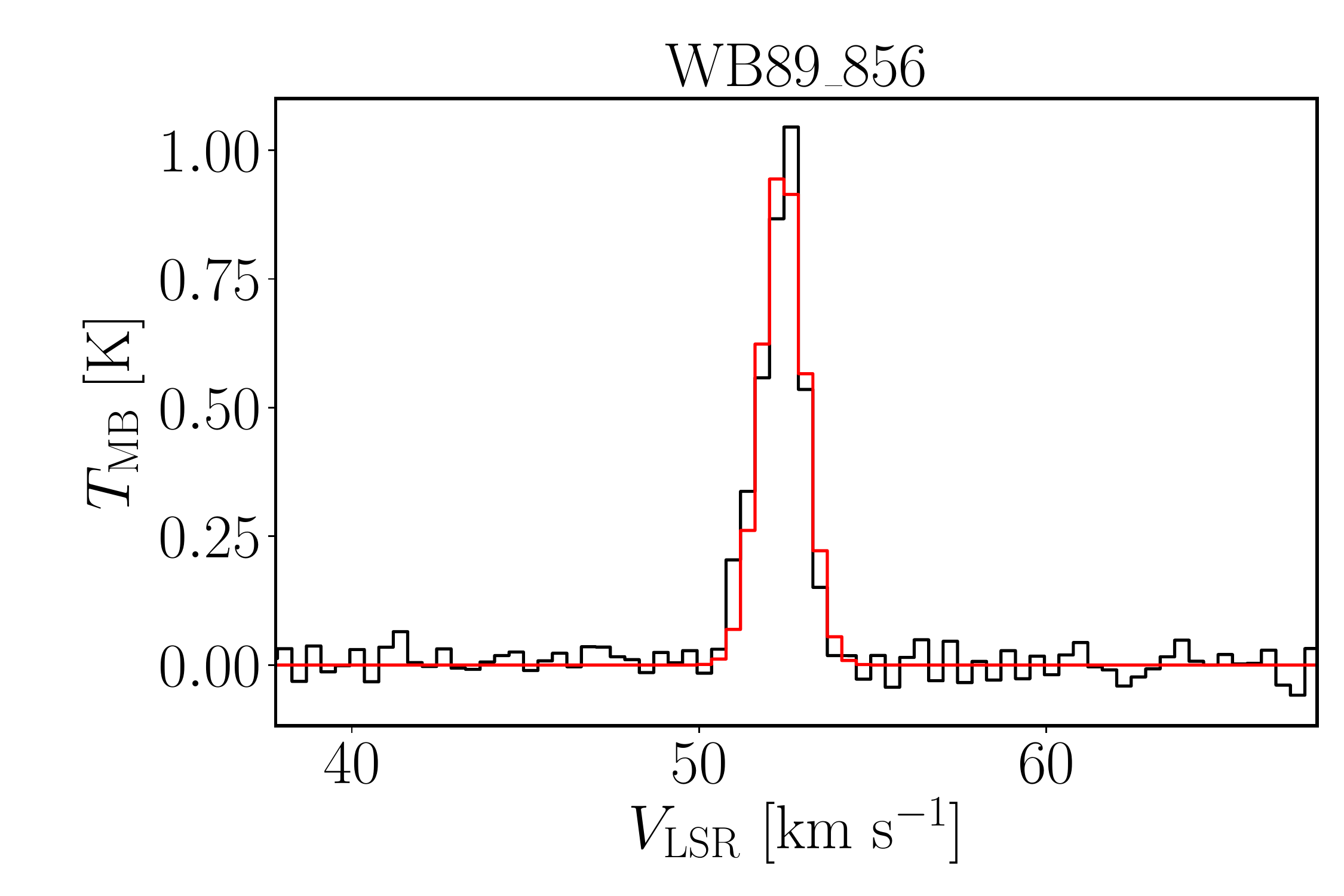}
		\includegraphics[width=0.3\textwidth]{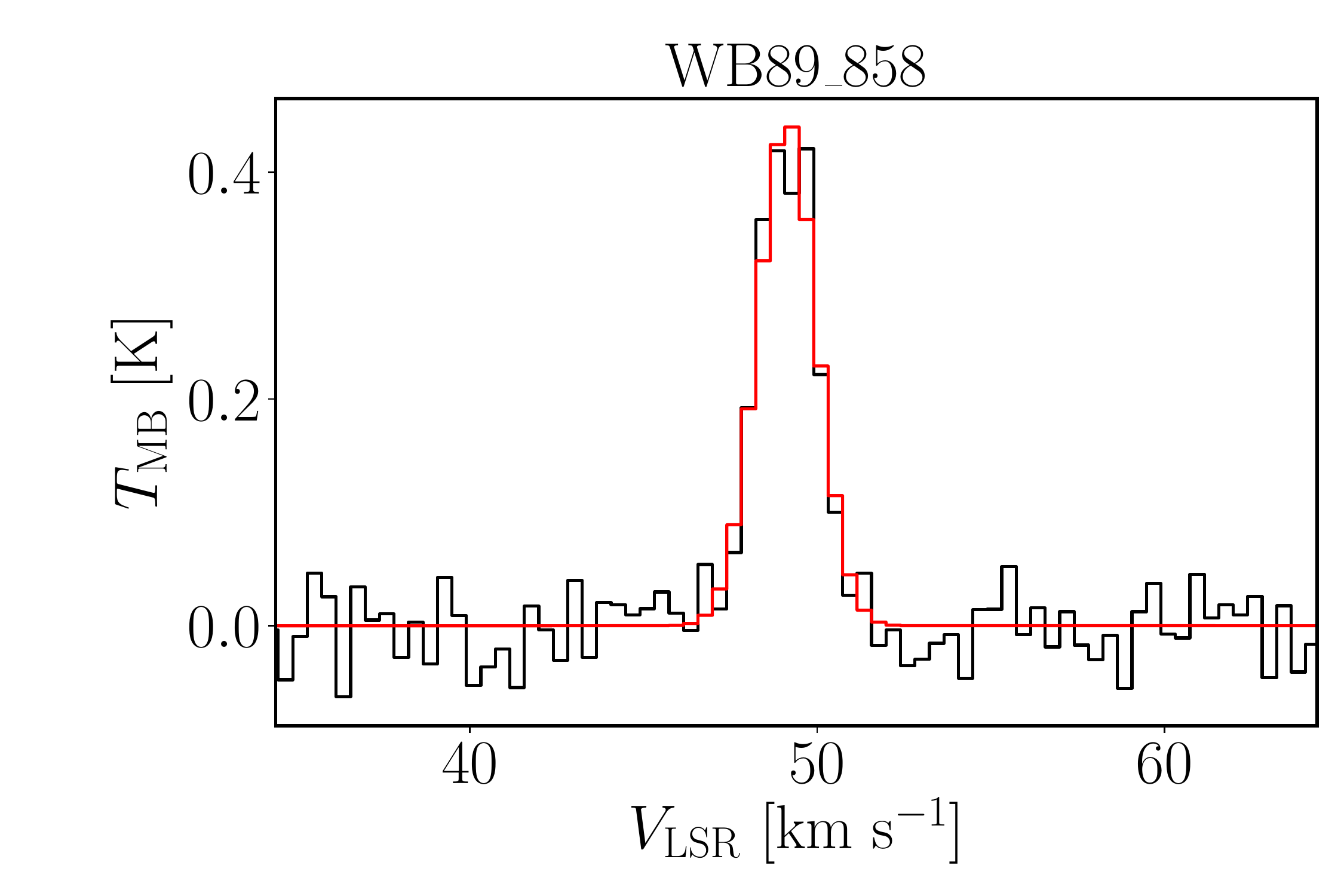}
		\includegraphics[width=0.3\textwidth]{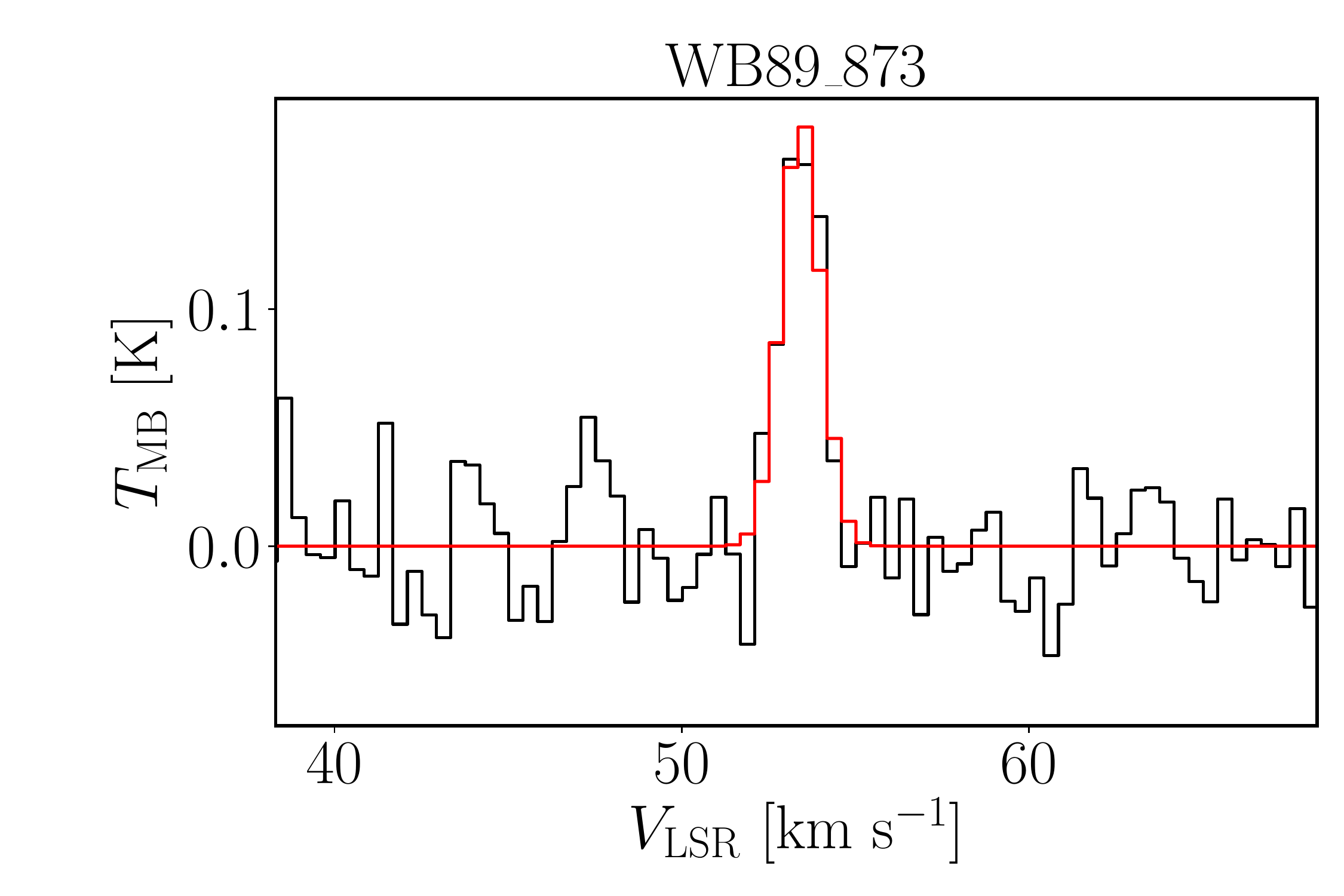}\\
		\includegraphics[width=0.3\textwidth]{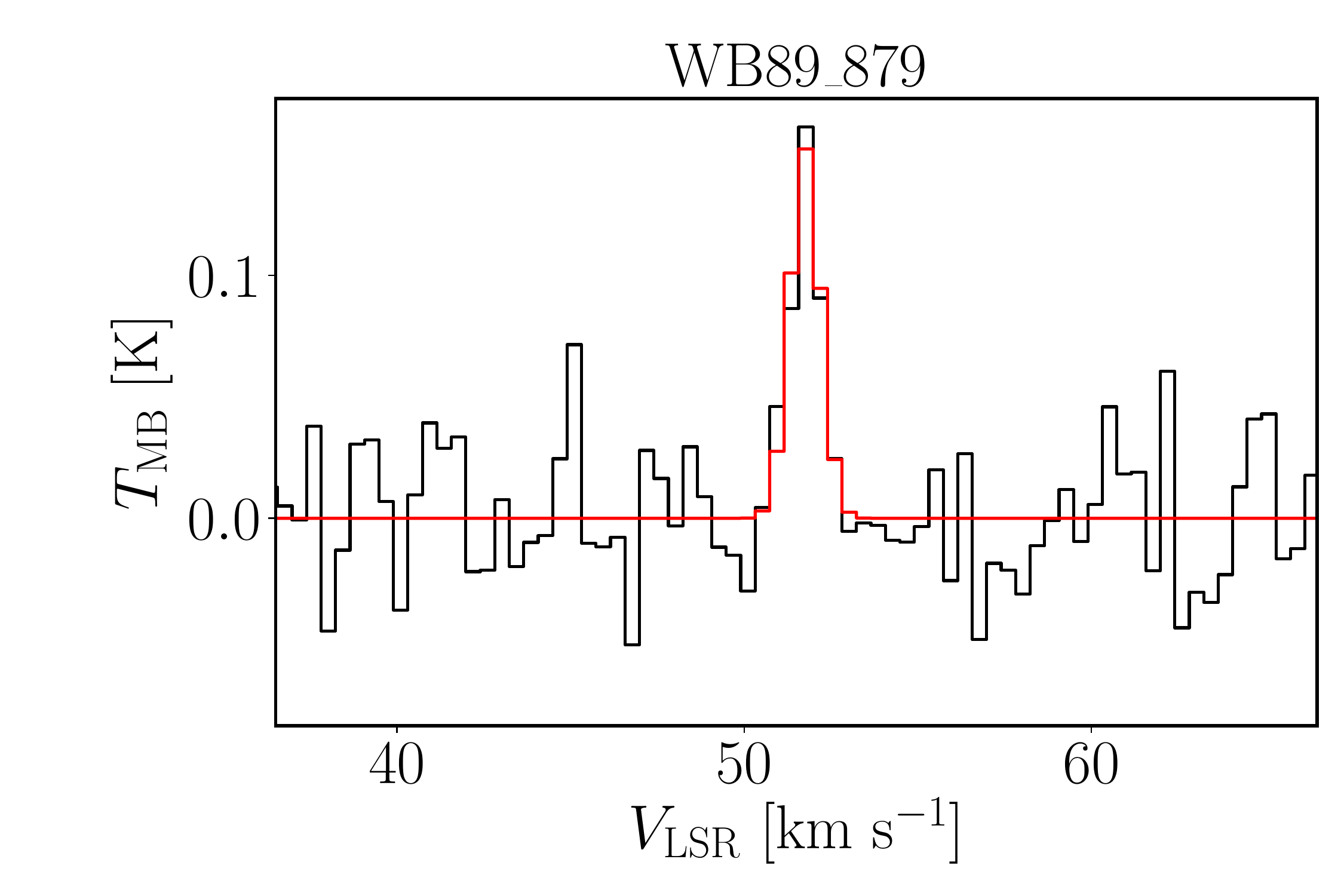}
		\includegraphics[width=0.3\textwidth]{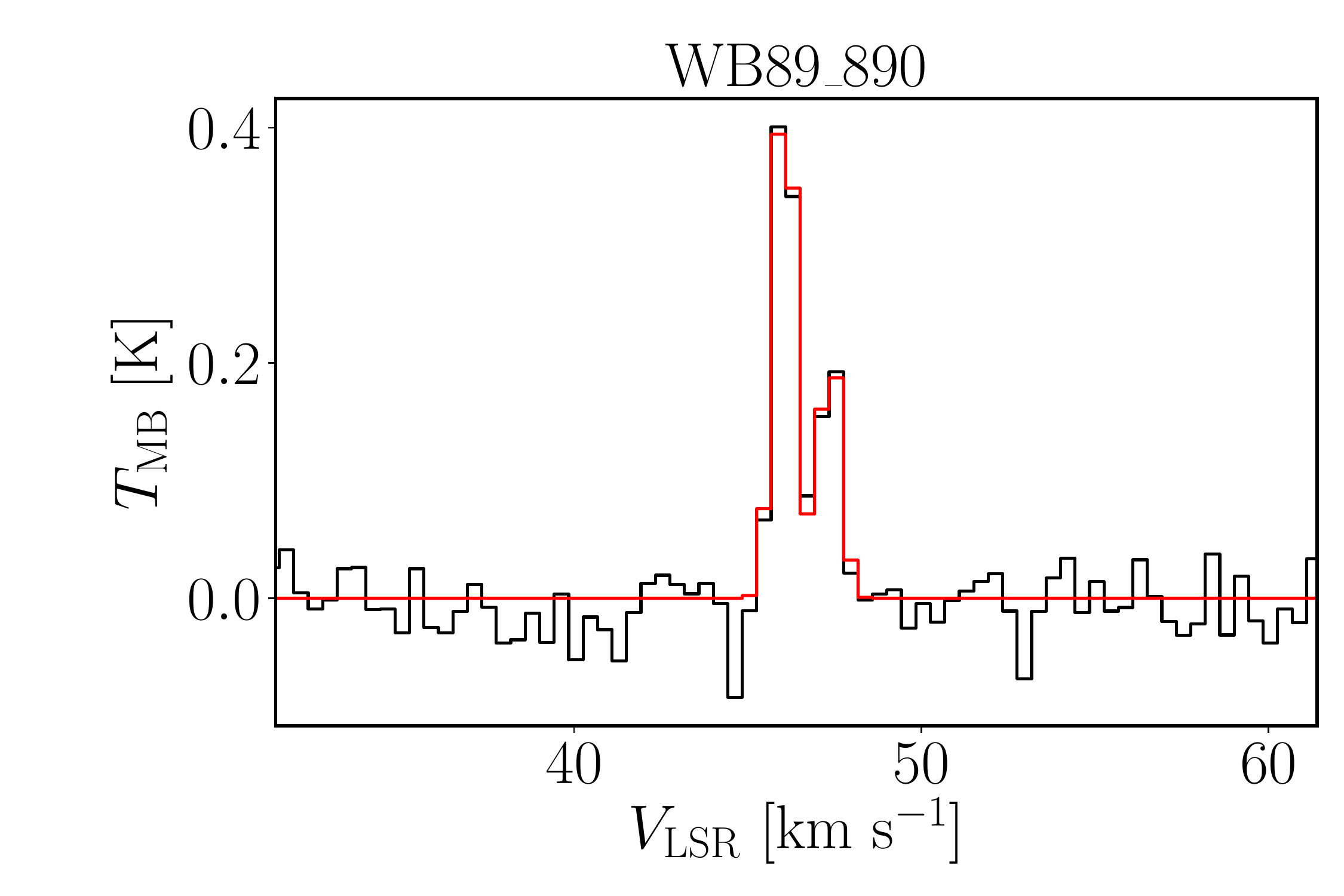}
		\includegraphics[width=0.3\textwidth]{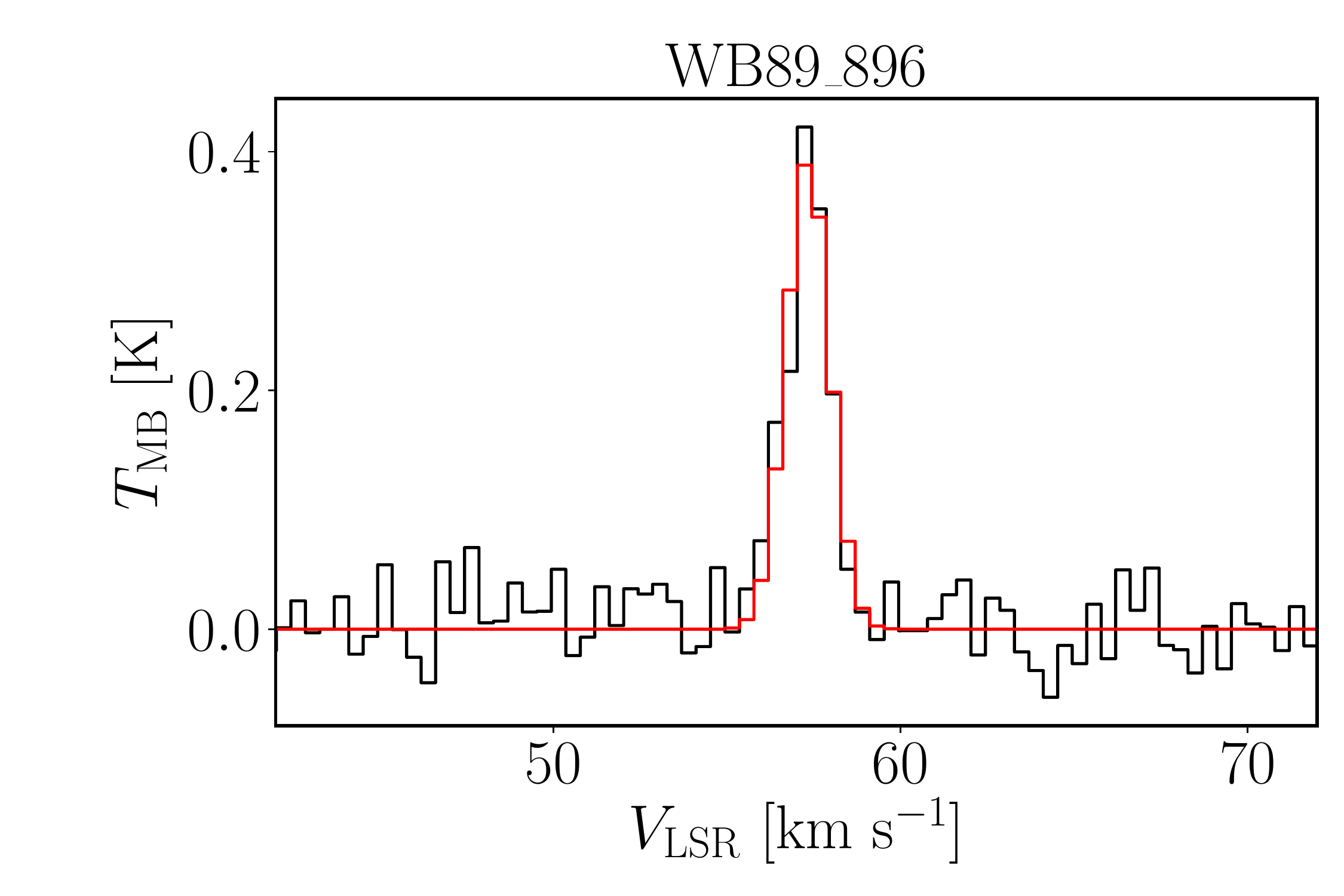}\\
		\includegraphics[width=0.3\textwidth]{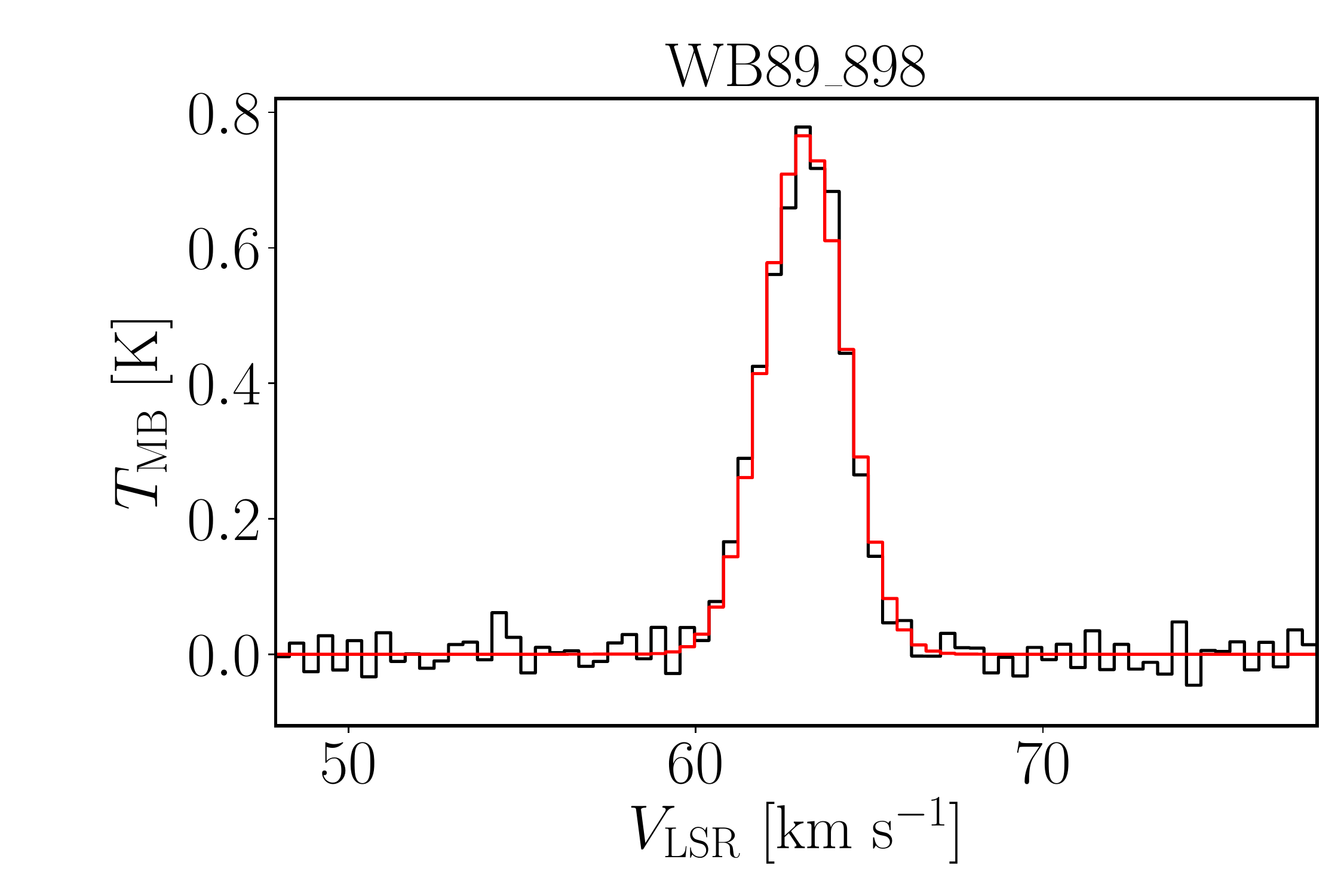}
		\includegraphics[width=0.3\textwidth]{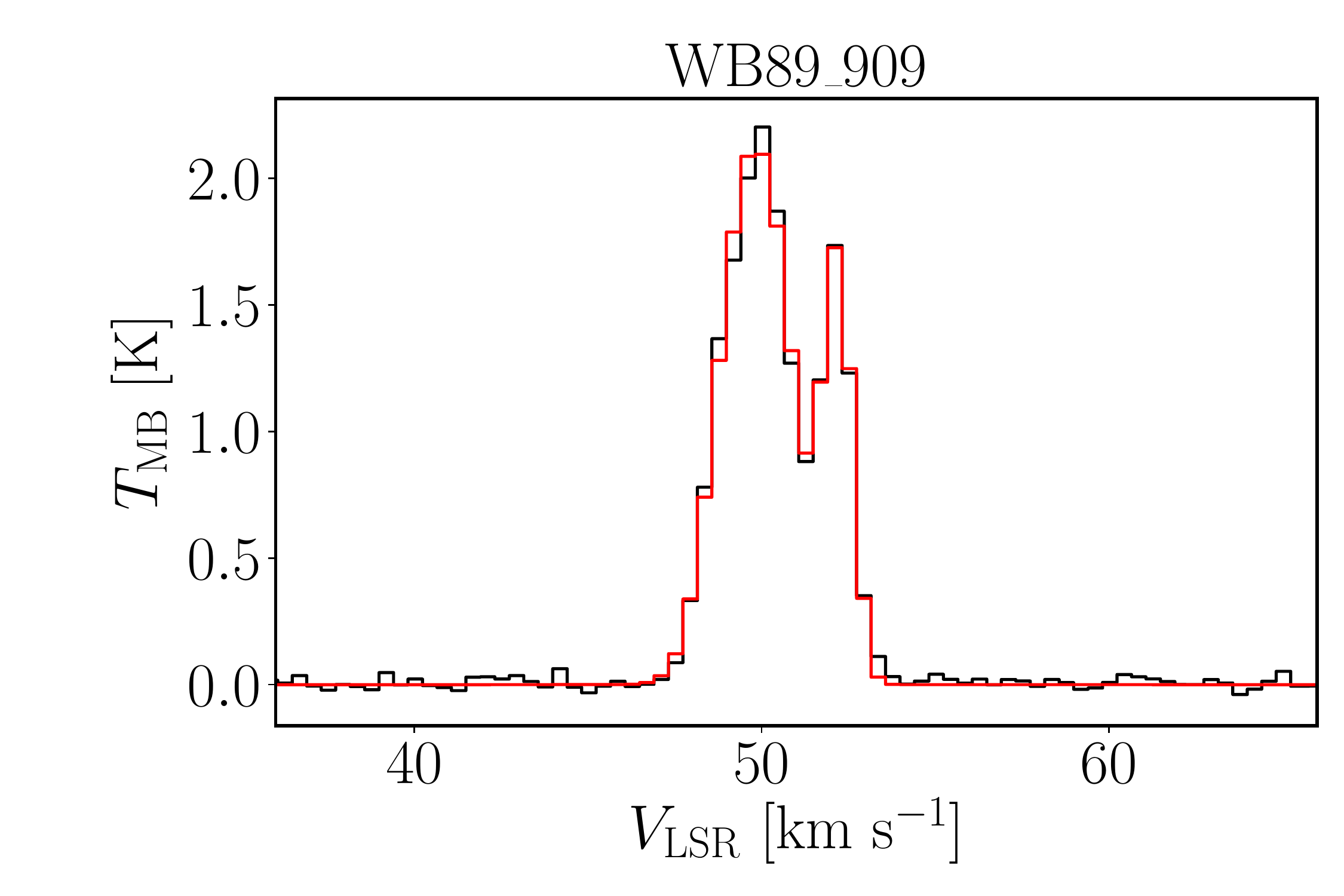}
		\includegraphics[width=0.3\textwidth]{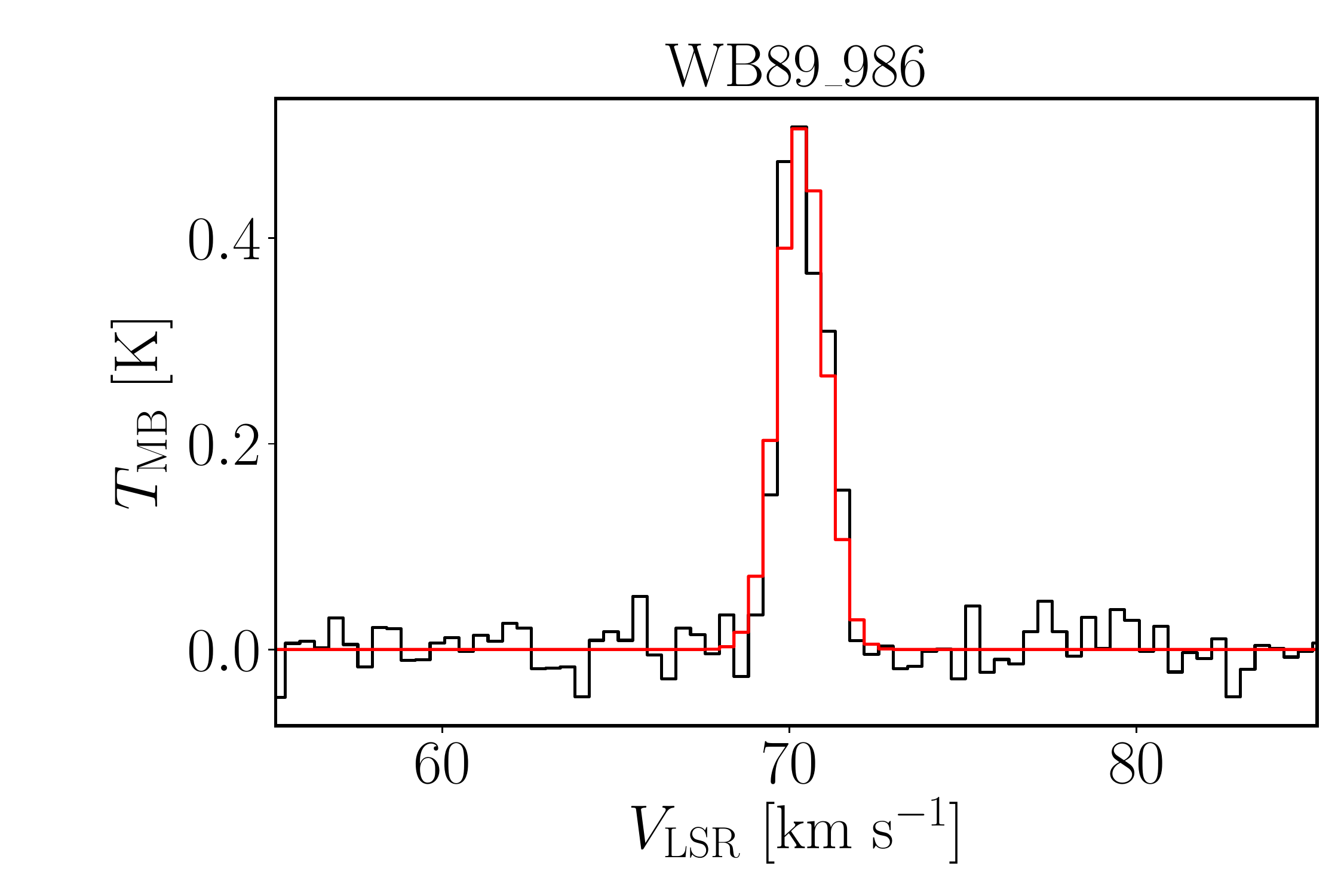}\\
		\includegraphics[width=0.3\textwidth]{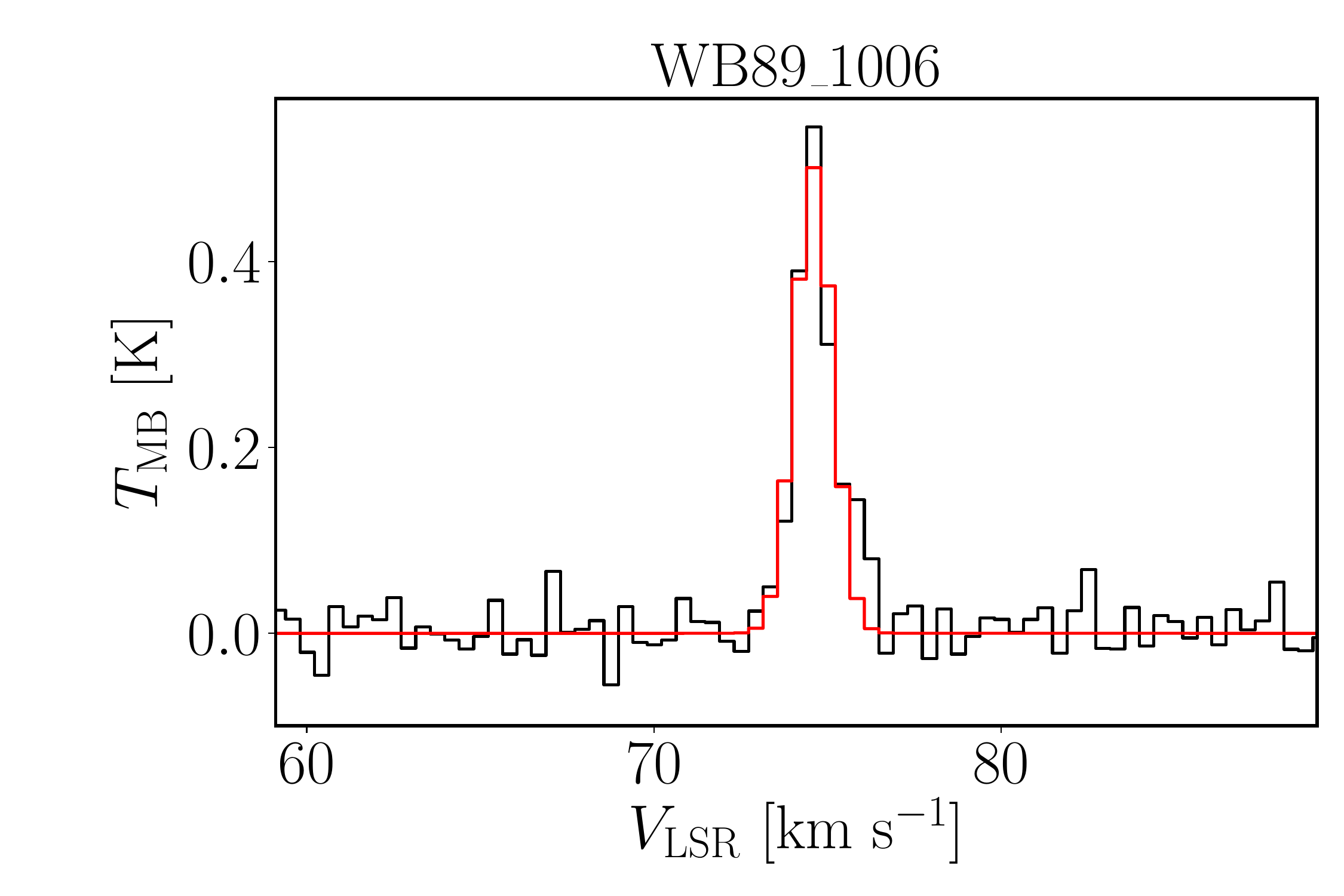}
		\includegraphics[width=0.3\textwidth]{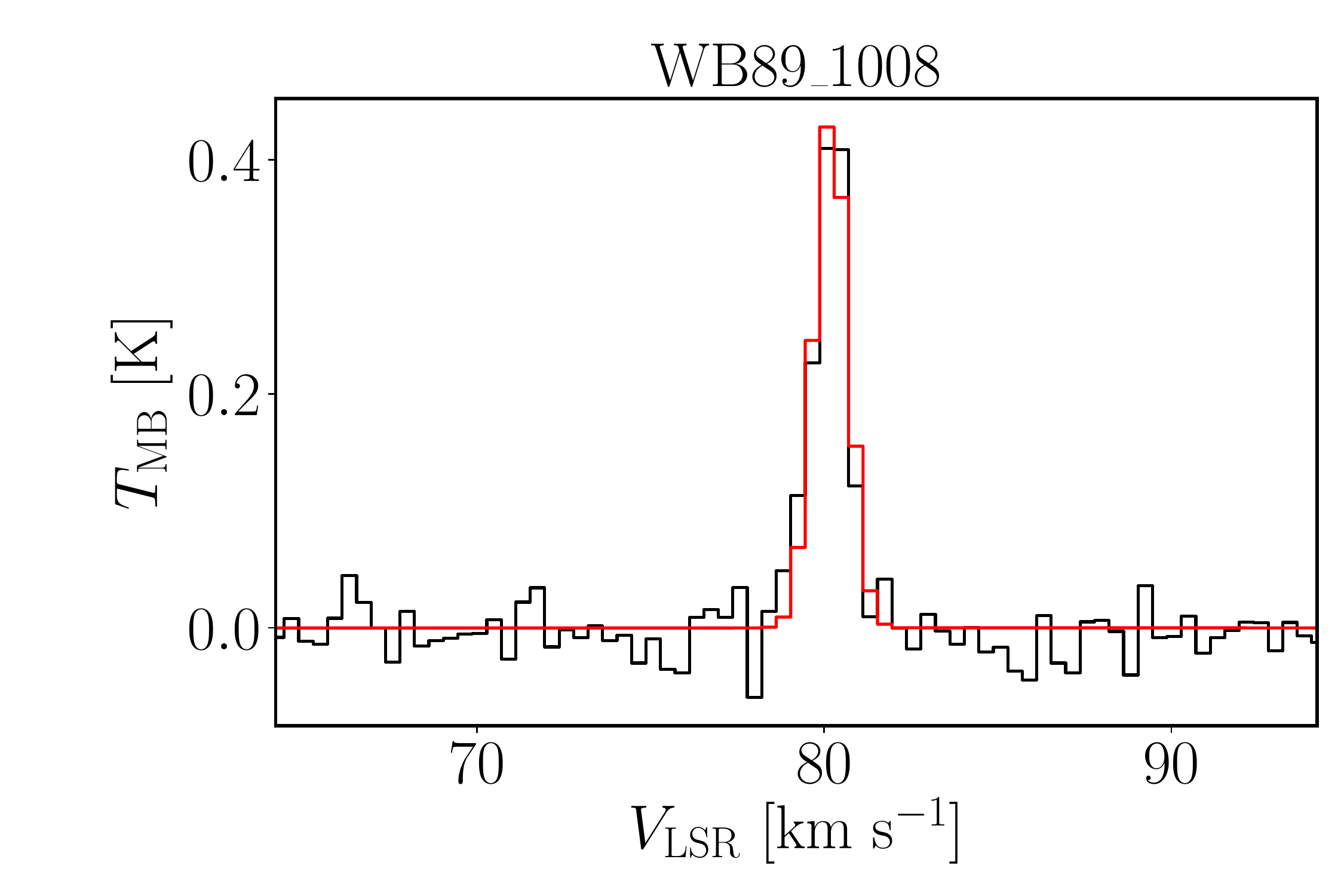}
		\includegraphics[width=0.3\textwidth]{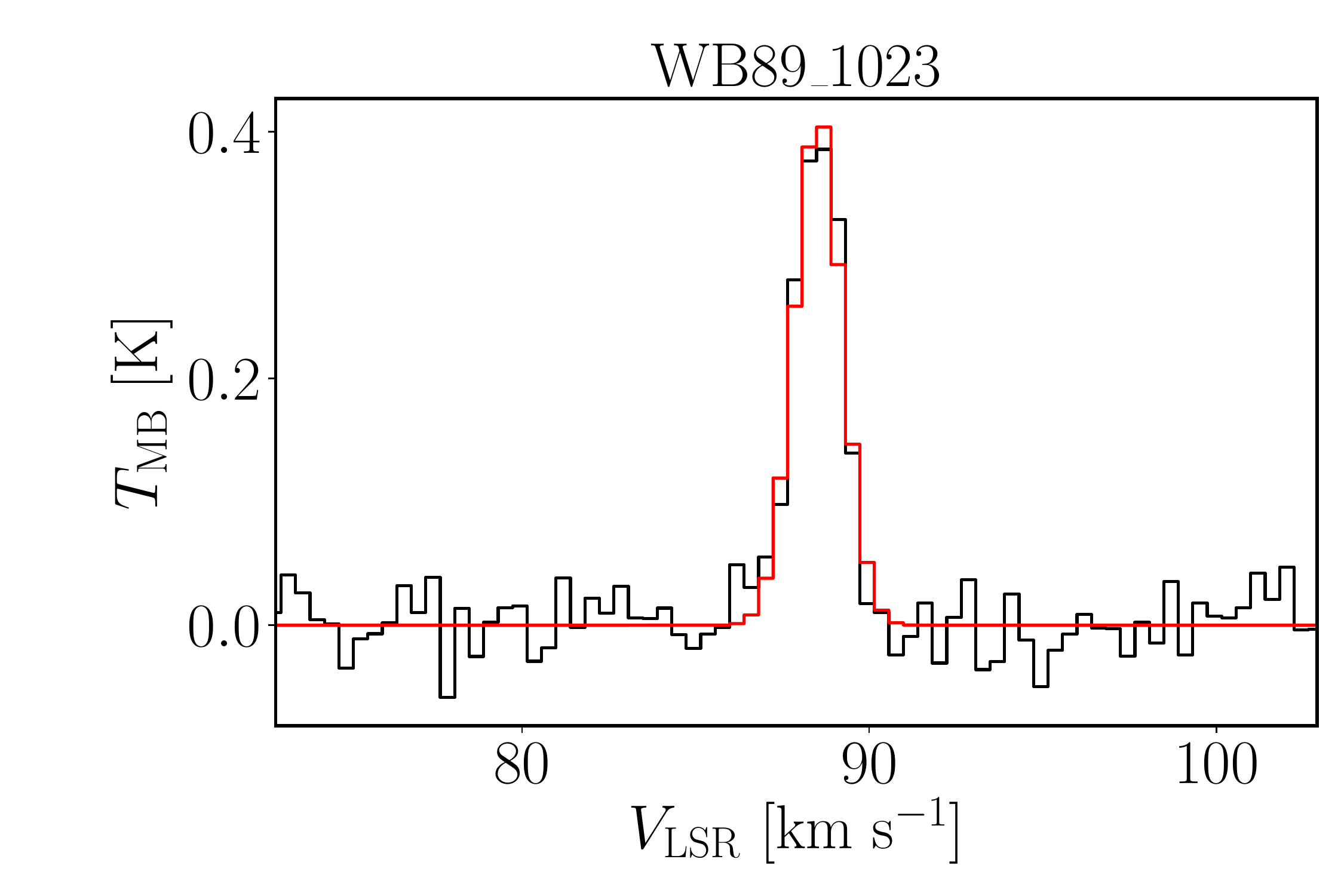}\\
		\includegraphics[width=0.3\textwidth]{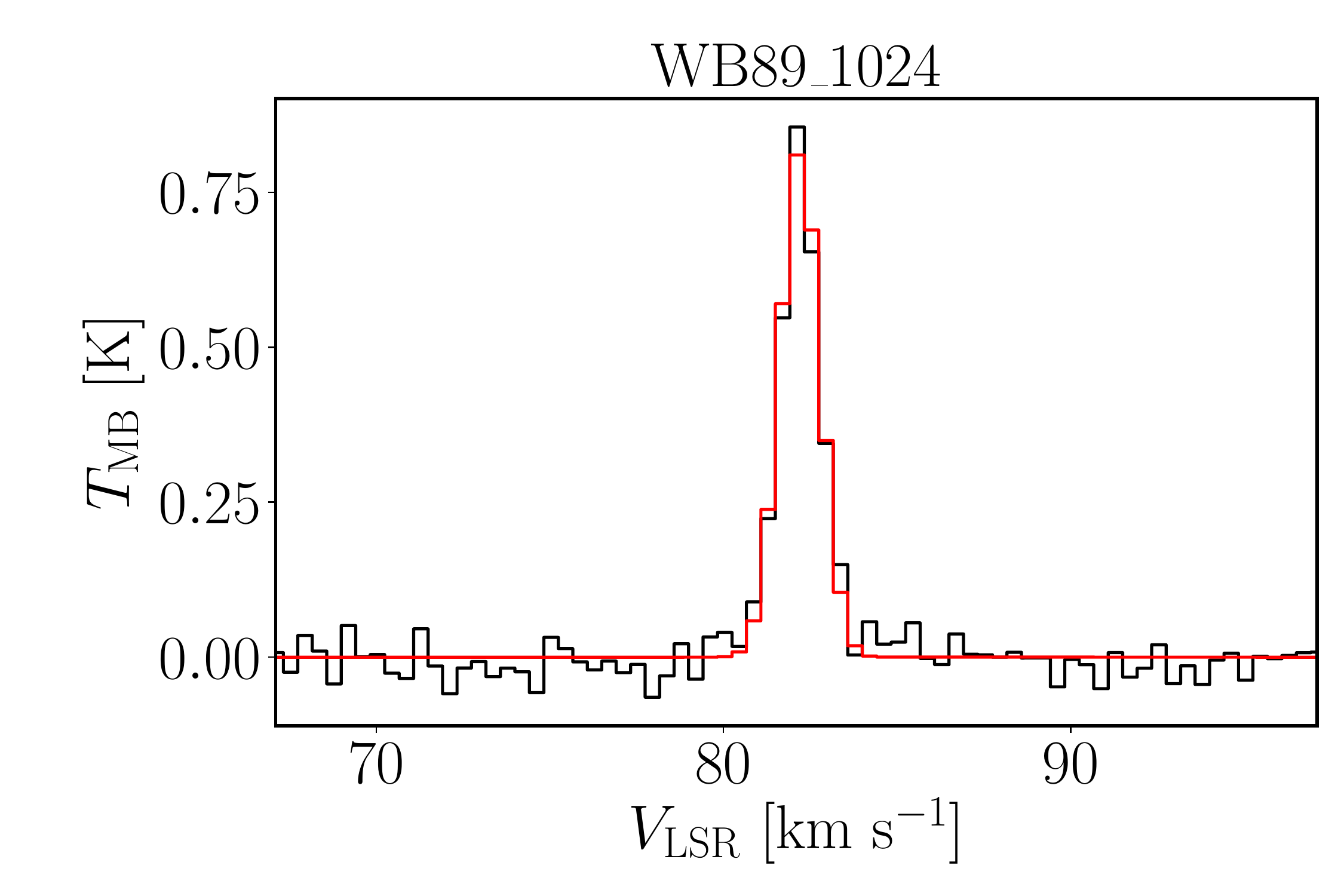}
		\includegraphics[width=0.3\textwidth]{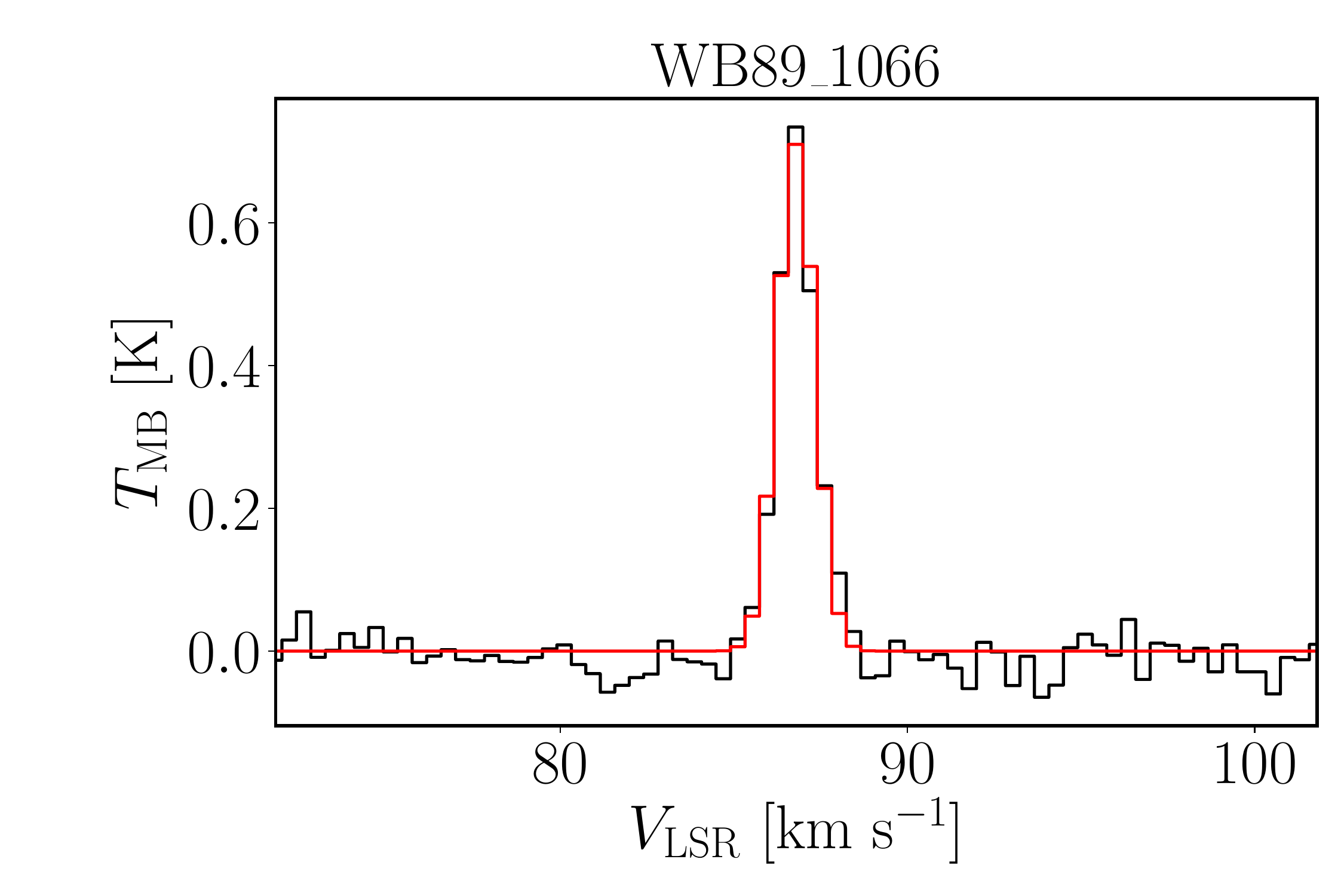}
		\includegraphics[width=0.3\textwidth]{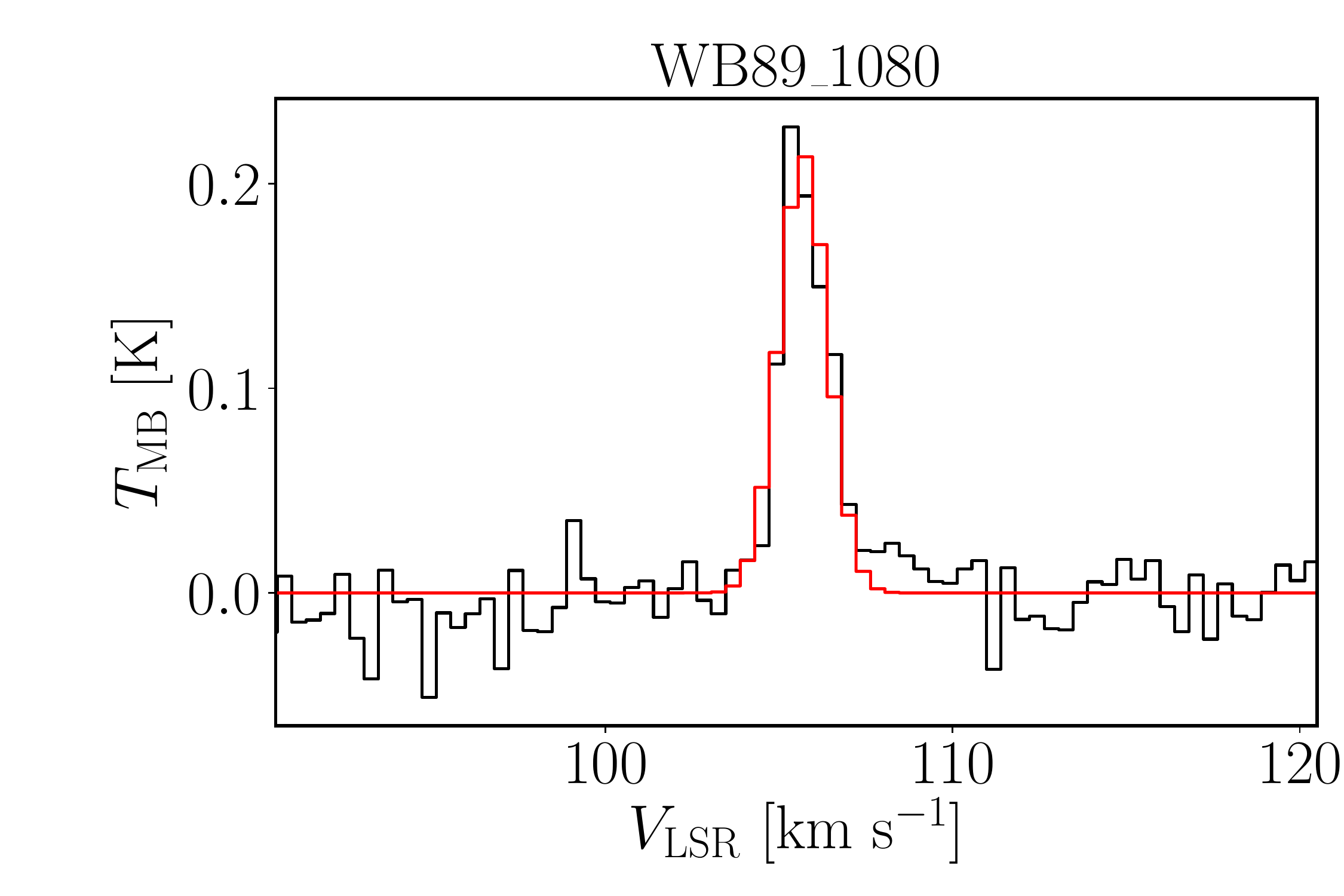}\\
		\includegraphics[width=0.3\textwidth]{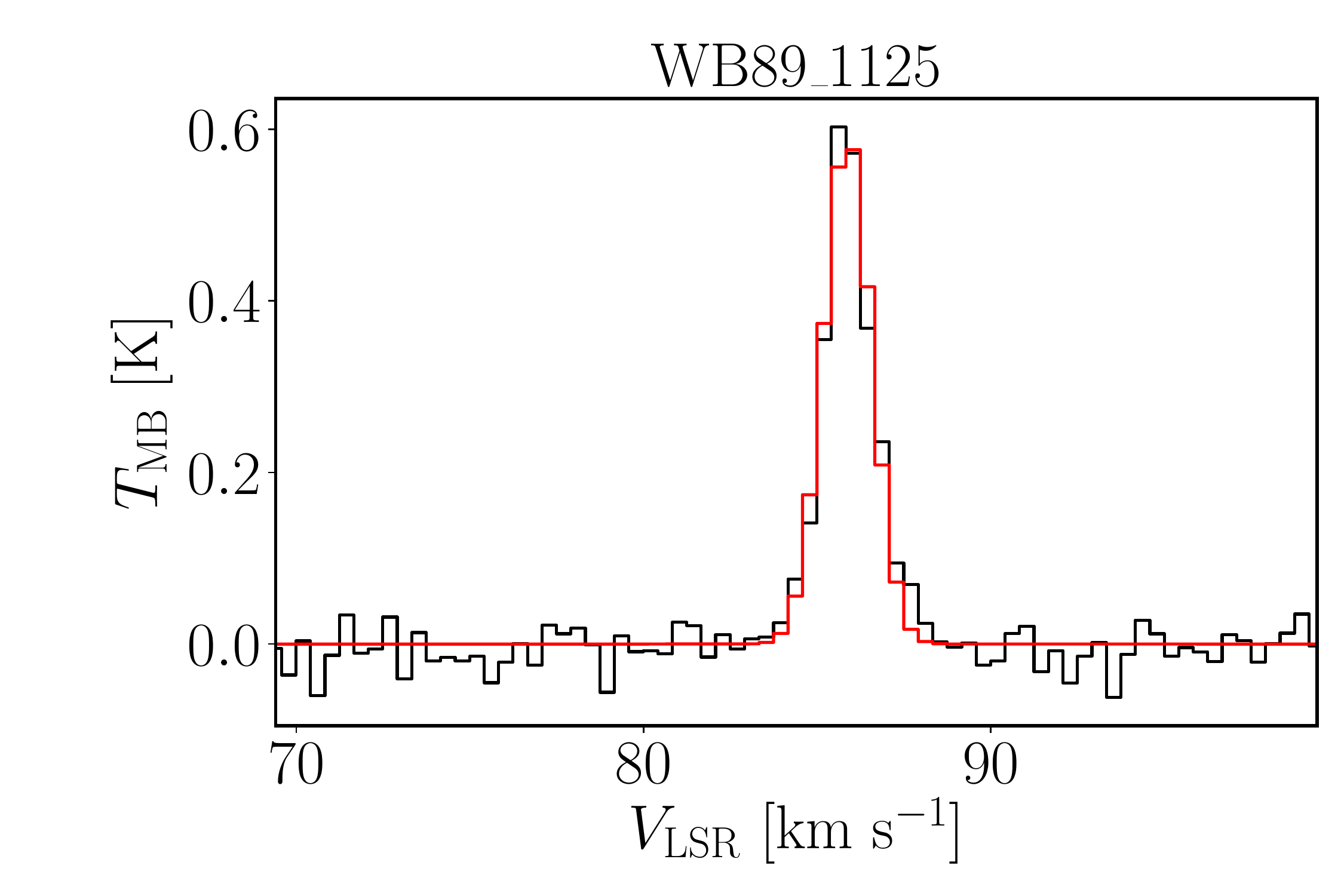}
		\includegraphics[width=0.3\textwidth]{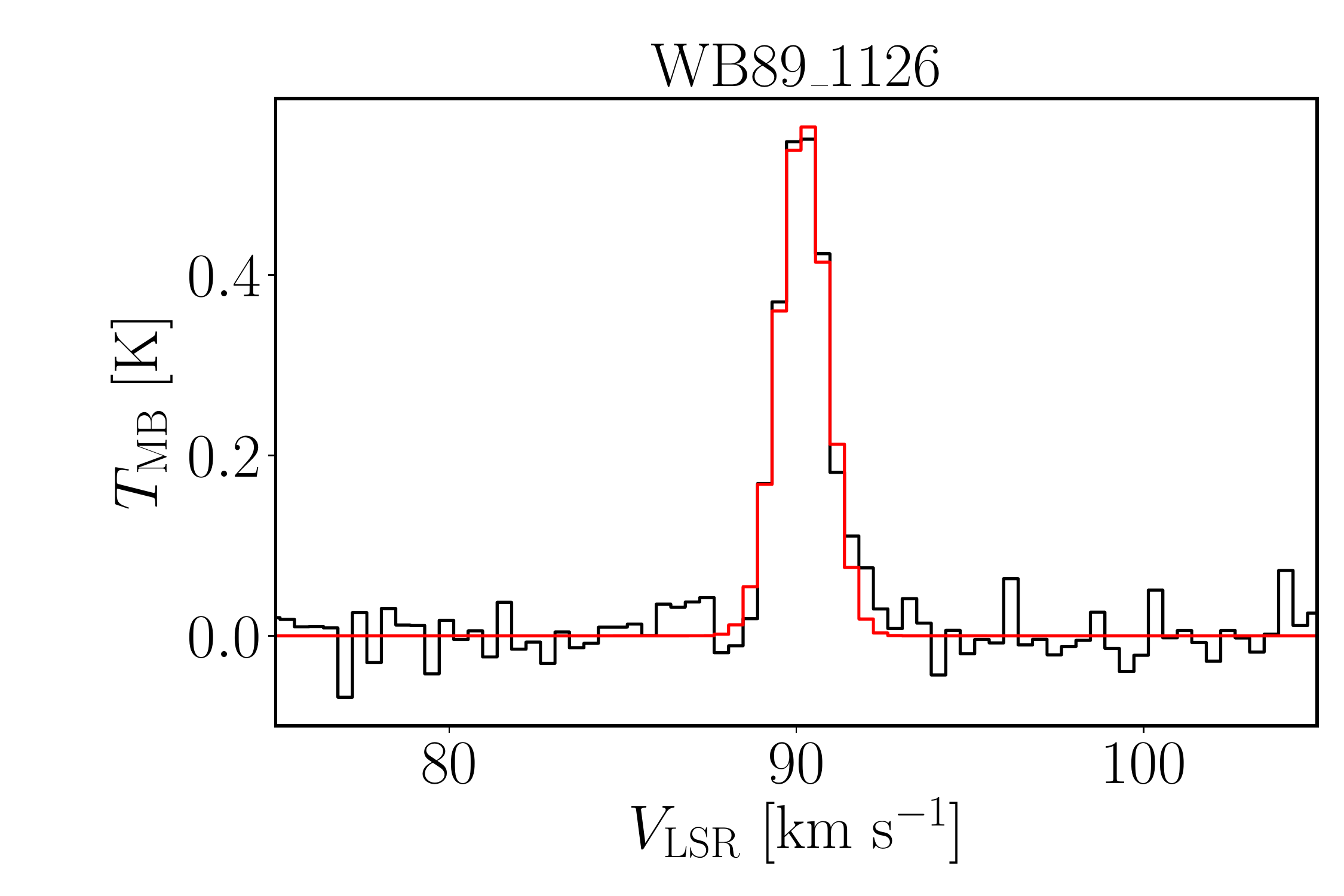}
		\includegraphics[width=0.3\textwidth]{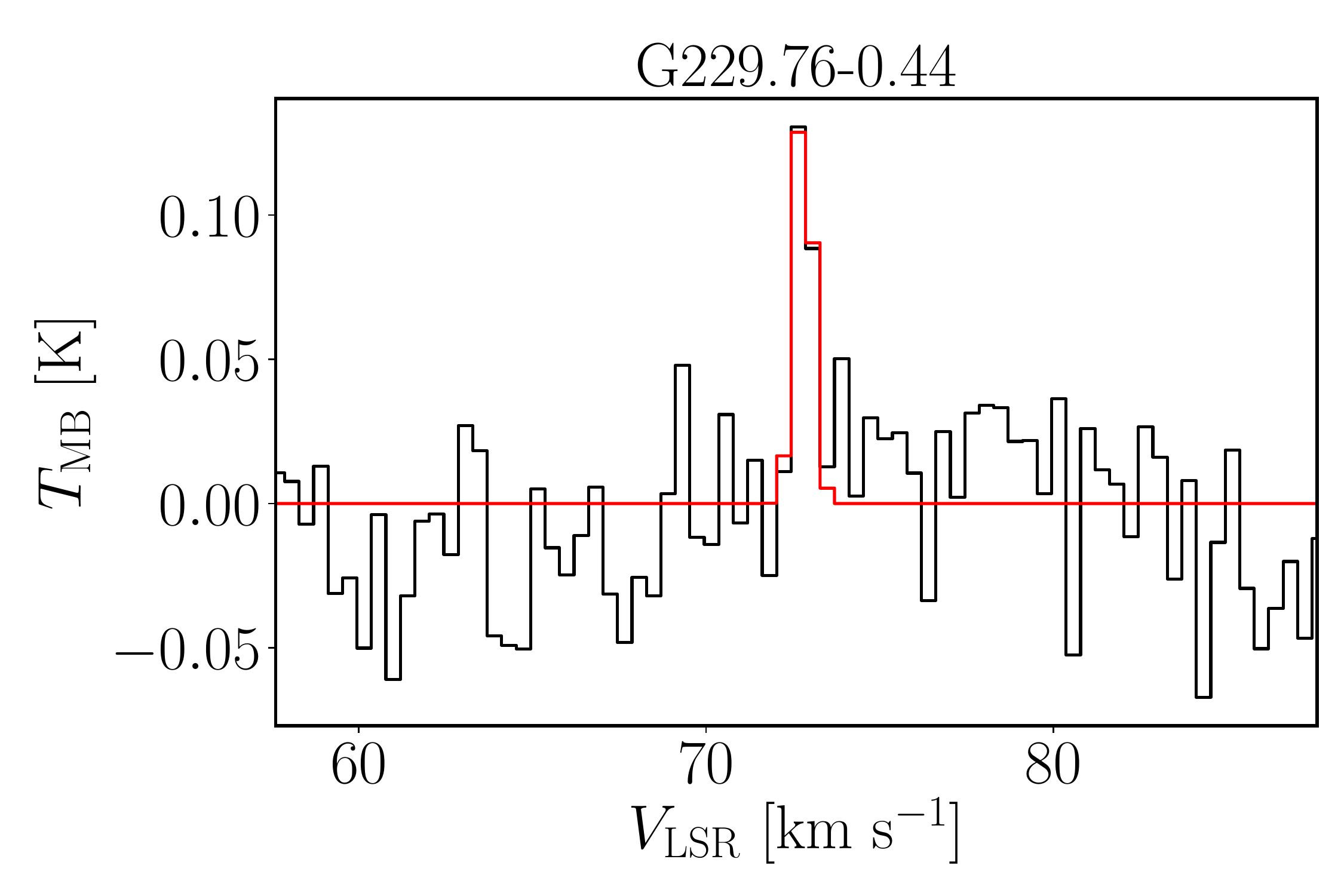}\\
		\caption{C$^{18}$O spectra of the sources in the FOG. We show in red the best fit from MCWeeds.\label{fig:spectra}}
	\end{figure*}
	\begin{figure*}
		\ContinuedFloat
		\includegraphics[width=0.3\textwidth]{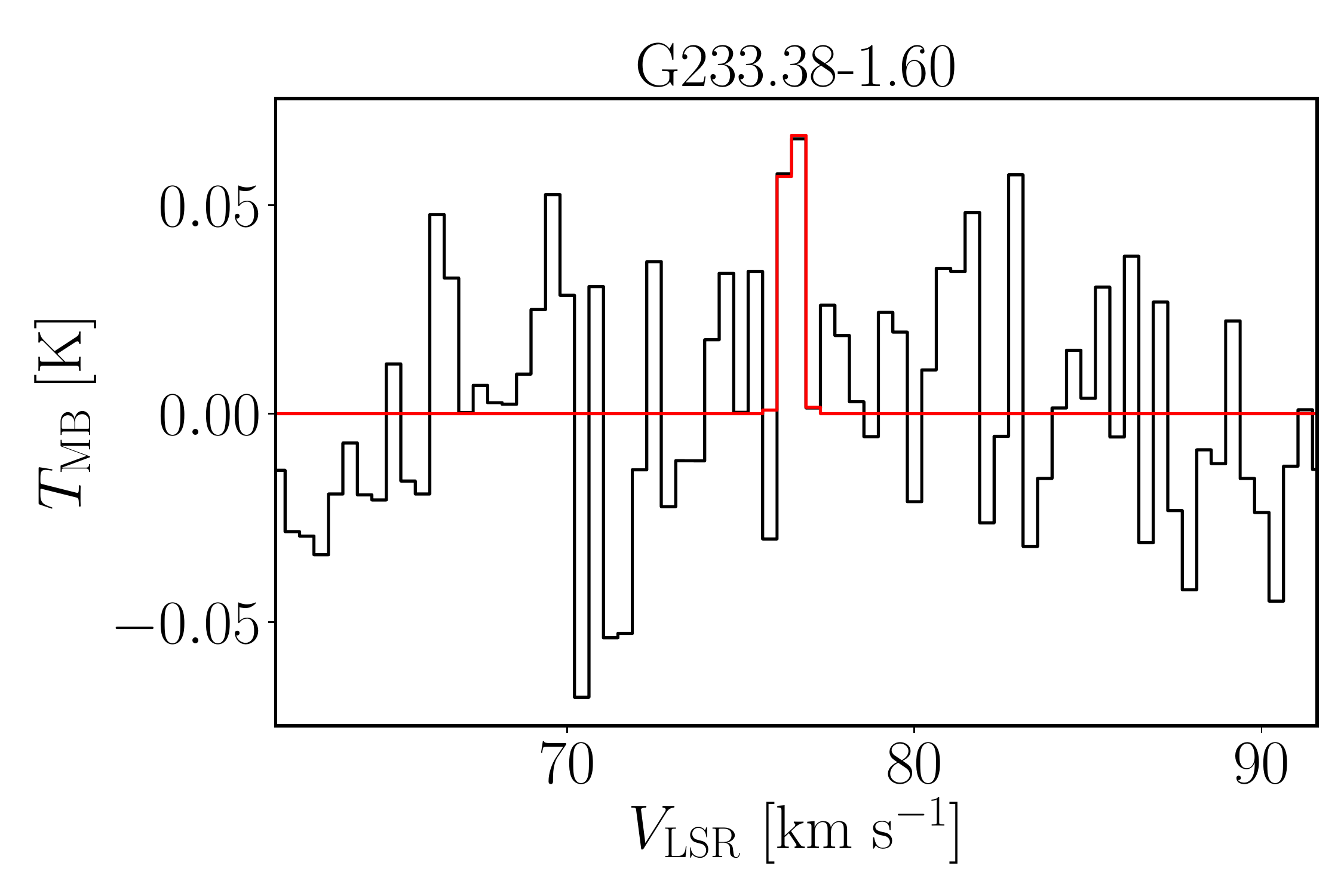}
		\includegraphics[width=0.3\textwidth]{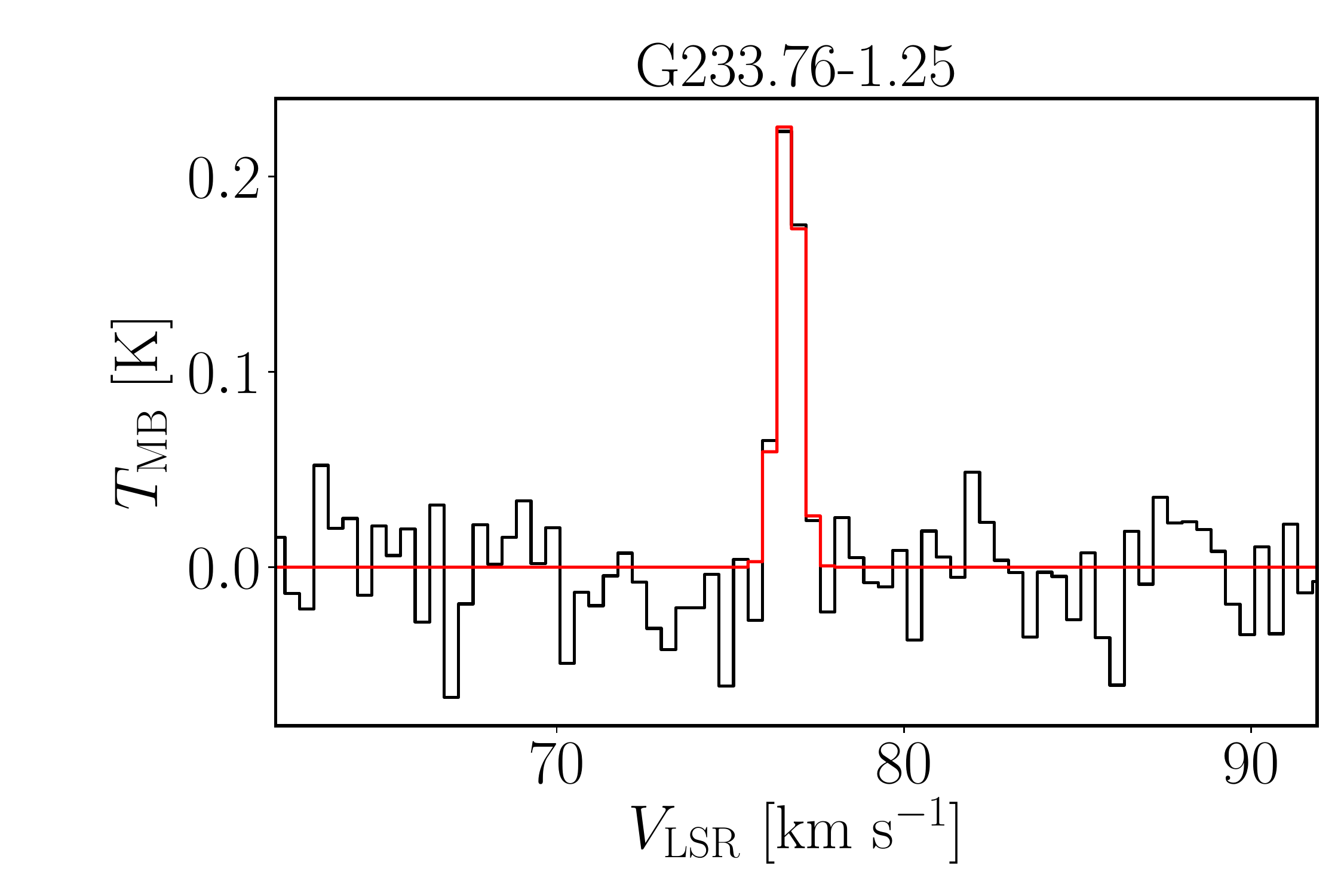}
		\includegraphics[width=0.3\textwidth]{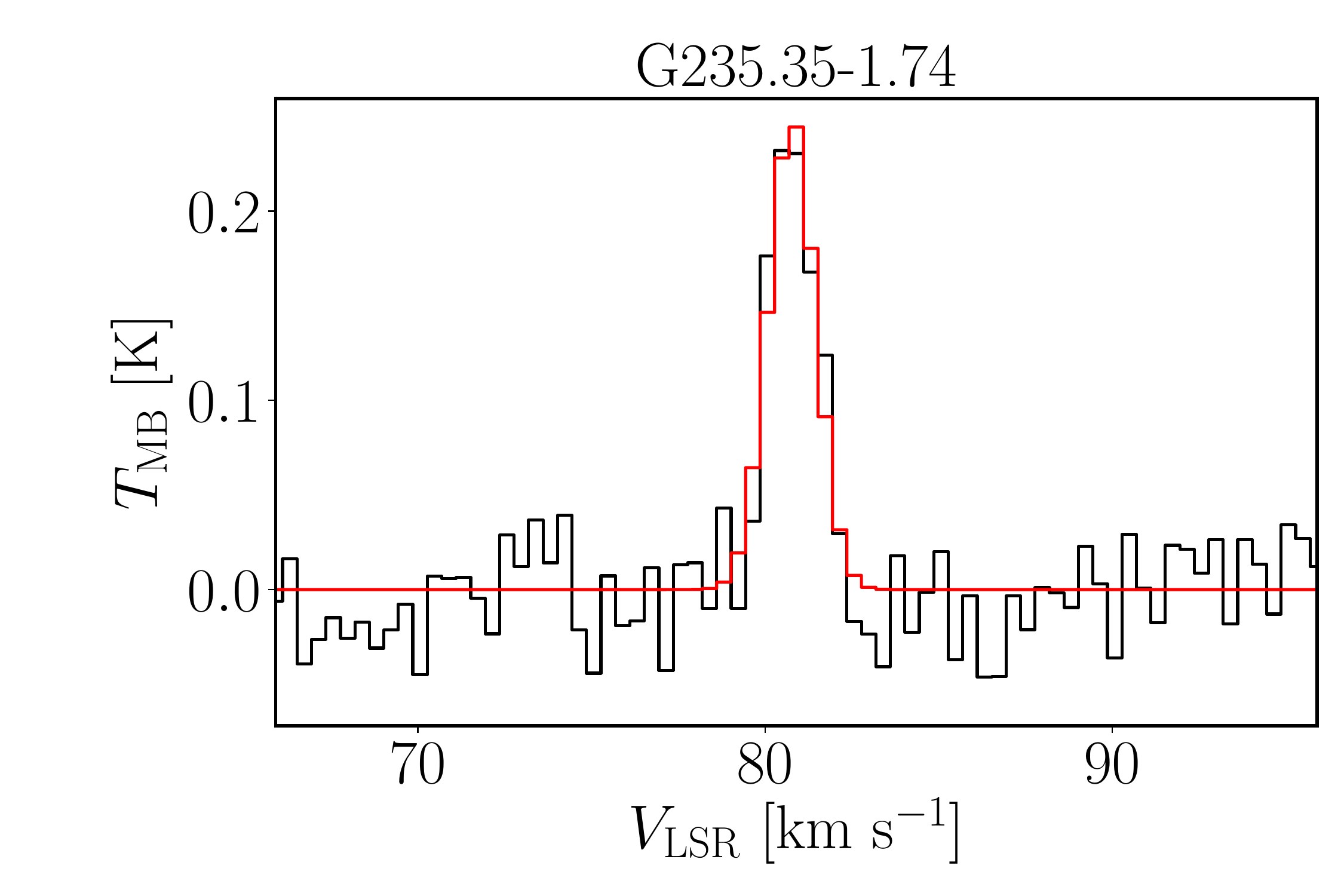}\\
		\includegraphics[width=0.3\textwidth]{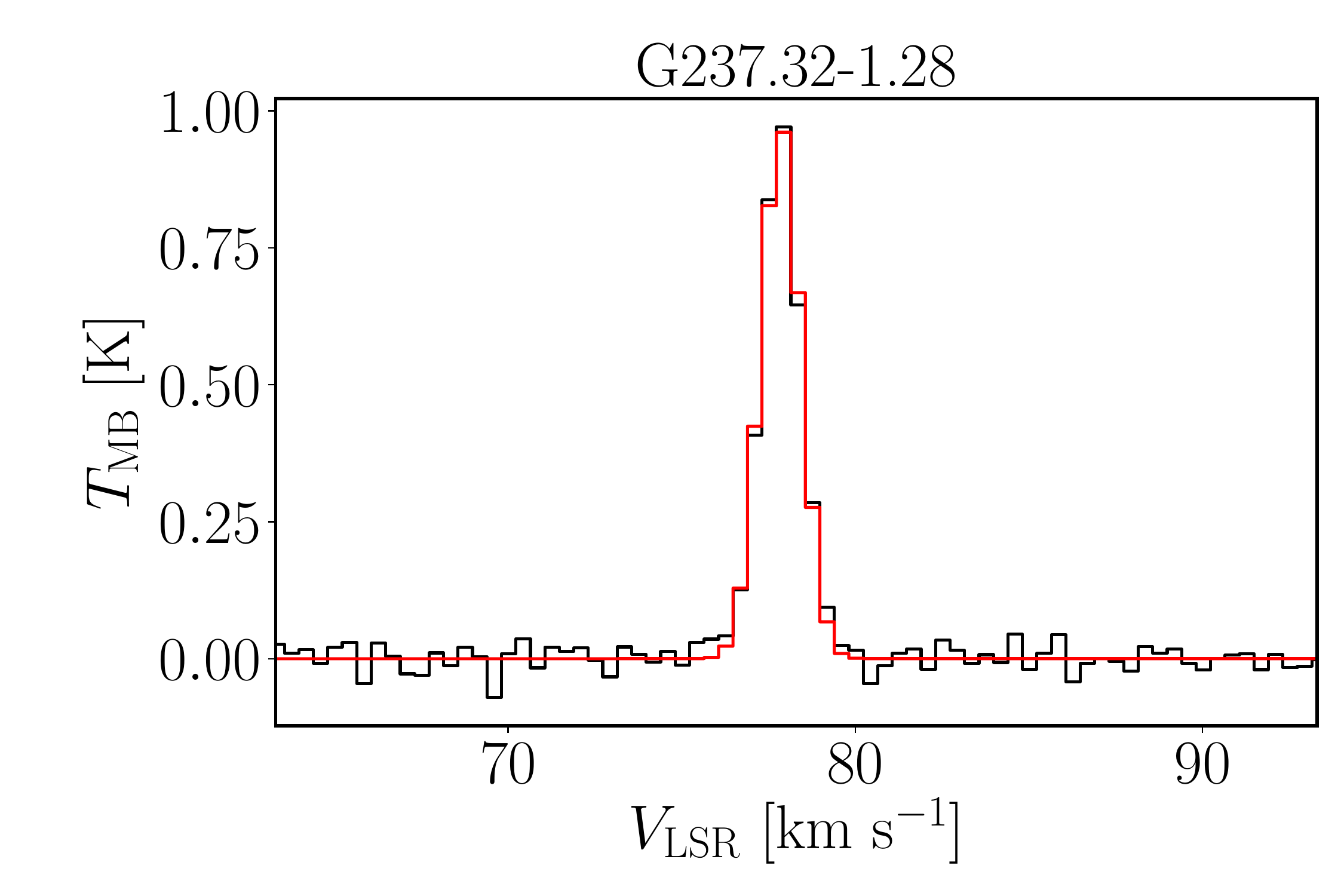}\\
		\caption{Continued.}
	\end{figure*}
	
\end{appendix}
}

\end{document}